\pdfoutput=1
\documentclass[aip,preprint]{revtex4-1}

\usepackage{xr-hyper}
\makeatletter

\newcommand*{\addFileDependency}[1]{
\typeout{(#1)}
\@addtofilelist{#1}
\IfFileExists{#1}{}{\typeout{No file #1.}}
}\makeatother

\newcommand*{\myexternaldocument}[1]{%
\externaldocument{#1}%
\addFileDependency{#1.tex}%
\addFileDependency{#1.aux}%
}

\usepackage{hyperref}
\hypersetup{
    colorlinks=true,
    linkcolor=blue,
    filecolor=magenta,      
    urlcolor=cyan,
    pdftitle={Overleaf Example},
    pdfpagemode=FullScreen,
    }

\myexternaldocument{si}

\usepackage{graphicx}
\usepackage{dcolumn}
\usepackage{bm}
\usepackage{amssymb}
\usepackage{comment}
\usepackage{xcolor}
\usepackage{amsmath}

\usepackage[utf8]{inputenc}
\usepackage[T1]{fontenc}
\usepackage{mathptmx}
\usepackage{etoolbox}
\usepackage{verbatim}

\begin{document}

\preprint{}

\title{Structural transitions of calcium carbonate by molecular dynamics simulation}




\author{Elizaveta Sidler}
\email{elizaveta.sidler@ntnu.no.}
\author{Raffaela Cabriolu}%
\email{raffaela.cabriolu@ntnu.no.}
\affiliation{Department of Physics, Norwegian University of Science and Technology (NTNU), H\o gskoleringen 5, Trondheim, 7491, Norway.}


\begin{abstract} 

Calcium carbonate (CaCO$_3$) plays a crucial role in the global carbon cycle, and its phase diagram has always been of significant scientific interest. In this study, we used molecular dynamics (MD) to investigate several structural phase transitions of calcium carbonate. Using the Raiteri potential model, we explored the structural transitions occurring at a constant pressure of 1 bar with a temperature ranging from 300 K to 2500 K, and at a constant temperature of 1600 K with a pressure ranging from 0 to 13 GPa.
With increasing temperatures, the transitions among calcite, CaCO$_3$-IV, and CaCO$_3$-V are observed and characterized. In the calcite structure, CO$_3^{2-}$ ions are ordered in a planar triangular arrangement, alternating with layers of Ca$^{2+}$ ions. As temperature increases, the calcite to CaCO$_3$-IV transition occurs, determining the partial disordering of CO$_3^{2-}$ ions. At a higher temperature, CaCO$_3$-IV transforms into CaCO$_3$-V. By applying free energy analysis, we have classified the last transition as a second-order order-disorder phase transition. 
At a temperature of 2000 K, it appears a `disordered CaCO$_3$' structure, characterized by low order within the calcium and carbonate sublattices and the free rotation of CO$_3^{2-}$ ions. 
With increasing pressures, two calcium carbonate transformations were observed. At $P=$ 2 GPa, the CaCO$_3$-IV phase undergoes a phase transition into CaCO$_3$-V, demonstrating that the model can describe the transition between these two phases as pressure and temperature-driven. Another phase transition was detected at $P=$ 4.25 GPa -- CaCO$_3$-V transits into the recently discovered CaCO$_3$-Vb phase. This transition is classified as a first-order phase transition by structural analysis and free energy-based arguments.  

\end{abstract}


\maketitle

\section{Introduction}

Calcium carbonate is one of the most abundant minerals on Earth.\cite{deer2013introduction} It is involved in a variety of natural processes, such as the carbon dioxide sequestration process via aqueous wollastonite carbonation, \cite{Wouter_wollastonite, CO2_seqestation} and the biomineralization of exoskeletons and shells of marine organisms in water.\cite{Biomin_mollusc}  

Calcium carbonate is characterized by rich polymorphism and a complex temperature-pressure phase diagram. \cite{Bayarjargal_phase_diagr_VII} Its high-temperature and pressure phase transitions are important for understanding geological formation, the global carbon cycle, \cite{carbonateGeology} climate change phenomena, \cite{hanein2021decarbonisation} and other processes. Insight into CaCO$_3$ structural transformations can benefit various industrial processes, including cement production, \cite{amato2013concrete} the development of optical devices, \cite{DeviceCalcitic} ceramics, coatings, and novel composite materials. \cite{Composites}

High-pressure polymorphs of calcium carbonate are of crucial importance as host structures for carbonate materials. Specifically, these polymorphs exhibit significant influence in the upper-mantle pressure range, spanning from 23 to 135 gigapascals (GPa), and extend their relevance into the lower-mantle pressure regions. \cite{Merlini_III_VI_xray} 
 
The CaCO$_3$ phase diagram is particularly complex, and yet, only partially explored. 
CaCO$_3$ occurs naturally in three anhydrous allotropic forms: calcite, aragonite, and vaterite, with calcite being the most stable at ambient conditions. Its structure was first identified by Bragg in 1914 \cite{bragg1914analysis} as belonging to R${\bar3}$c space group symmetry, with coplanar carbonate groups and calcium ions sixfold coordinated. The calcite trigonal unit cell has parameters of $a=$ 4.991(2) \AA{}, $c=$ 17.062(2) \AA{} at $T\simeq$ 293 K, $P\simeq$ 1 bar. \cite{calcite_str} Aragonite is a high-pressure CaCO$_3$ phase, and is stable up to $\simeq$ 40 GPa in ambient temperatures. \cite{aragon_to_post_aragon} Finally, vaterite, of which the detailed structure is still under debate, is a thermodynamically metastable form of CaCO$_3$. \cite{vater_struct}


Many polymorphic modifications of calcium carbonate have been discovered using theoretical and experimental methods. \cite{redfern2000structural, Bayarjargal_phase_diagr_VII, martirosyan2021effect} Already in 1912, Boeke reported a phase transition of calcite into a new phase called calcite V (hereafter CaCO$_3$-V or V) at 1243 K. \cite{boeke1912schmelzerscheinungen} 
More recently, several experimental studies have referred to this transition as an order/disorder orientational transition. \cite{order_disorder_2, order_disorder_2_2009} 
Another calcium carbonate metastable phase, CaCO$_3$-IV, was also identified as a separate phase in 1976, \cite{calcite_4_first} again without any accurate determination of its structure.

The crystal structures of both phases were finely resolved in 2013 by Ishizawa et al. \cite{Ishizawa_2013, Ishizawa_2014} with X-ray diffraction (XRD) experiments under a carbon dioxide atmosphere at 0.4 MPa pressure. Compared to calcite, phases IV and V present different distributions of oxygen positions in the carbonate ions. In calcite, perpendicular to the $c$-axis, planar triangular CO$_3^{2-}$ ions vibrate around fixed positions. Additionally, these ions are oriented in opposite directions within the alternating layers perpendicular to the $c$-axis. After the reversible transition into phase IV at $T=$ 985 K, the ions are partially disordered while the structure still belongs to the same symmetry group as calcite. In phase V, the carbonate groups become orientationally disordered around their 3-fold axes, all equivalent in R$\bar{3}$m space group symmetry with unit cell parameters of $a=$ 4.9697(9) \AA{}, $c=$ 8.9295(2) \AA{} at $T=$ 1275 K.

Previous MD research implemented by Kawano et al. \cite{Kawano_2009} revealed that the temperature-induced structural transition from calcite to phase V exhibits a first-order behavior at low-pressure, i.e. $ P\simeq$ 1 bar, without the appearance of an intermediate phase IV. Conversely, the same transition from calcite to phase V appears to be the second order at pressures of 1 and 2 GPa, characterized by the appearance of phase IV as an intermediate phase.
Several metastable calcium carbonate structures appear during the compression of calcite crystals at room temperature. At 1.7 GPa, calcite transforms into monoclinic CaCO$_3$-II with lattice symmetry equal to $P2_1/c$, as determined by Merrill et al. in 1975.\cite{Merrill_II} 
CaCO$_3$-III and CaCO$_3$-IIIb can emerge sequentially at pressures between 2.5 and 15 GPa, as Pippinger et al., \cite{ Pippinger_III_IIIb_phase_diagr} Koch-Müller et al., \cite{KochMüller_IIIb_III_IIIb_VI} and  Merlini et al. reported. \cite{Merlini_III_IIIb_coexist} 
These triclinic polymorphs were resolved by both Raman scattering \cite{Pippinger_III_IIIb_phase_diagr} and single-crystal synchrotron XRD experiments. \cite{Merlini_III_VI_xray} 
Numerous Density Functional Theory (DFT)-based calculations have shown the stability of those polymorphs with different functionals. \cite{Belkofsi}  
Beyond pressures of 15 GPa, various structures such as CaCO$_3$-VI and CaCO$_3$-VII have been reported in experimental works of Koch-Müller et al., \cite{KochMüller_IIIb_III_IIIb_VI} Bayarjargal et al..\cite{Bayarjargal_phase_diagr_VII} 

At pressures above 40 GPa, the post-aragonite structure was found experimentally by Ono et al.. \cite{post-aragonite} A recent study by Druzhbin et al. \cite{Druzhbin} identified yet another high-temperature, high-pressure calcium carbonate phase, calcite-Vb (hereafter Vb or CaCO$_3$-Vb). This phase was defined with symmetry $P2_1$/m and lattice parameters $a=$ 6.284(5) \AA{}, $b=$ 4.870(3) \AA{}, $c=$ 3.991(4) \AA{}, $\beta=$ $107.94(3)^{\circ}$ at $T\simeq$ 1573 K and $P\simeq$ 4 GPa. CaCO$_3$-Vb is stable at pressures from 4 to 9 GPa and temperatures between 1200 K and 1800 K, and is considered to be an intermediate phase between CaCO$_3$-V and aragonite. Although no triple points were observed, Druzhbin et al.'s experimental study revealed points of IV, V, and Vb in the proximity of the sequence V to Vb to aragonite as pressure increases. \cite{Druzhbin} 
Furthermore, their XRD patterns confirm that Vb is a distinct phase from the \textit{disarg} or aragonite-like phase predicted for the same temperature-pressure range by the previous MD works. \cite{Gavryushkin_2020} 

Significant progress has been made in understanding the polymorphism and phase transitions of CaCO$_3$. However, persistent disagreements are due to the system's complexity and the limitations of current experimental and theoretical models. For instance, debates continue regarding the stability and characteristics of high-temperature and high-pressure phases, suggesting the possible existence of undiscovered polymorphs. Furthermore, inconsistency between experimental observations and modeling predictions regarding the properties of these phases enhances the challenges in accurately characterizing them. The precise locations of phase boundaries at elevated pressures and temperatures, especially the conditions under which calcite transforms into high-pressure phases, remain unresolved. Experimental determination of these boundaries is hindered by kinetic barriers that influence phase transformations. Moreover, the kinetics of transformations under non-equilibrium conditions, compounded by factors such as impurities and defects, pose additional obstacles to comprehensive study. In modeling, discrepancies often arise in predicting transition conditions and phase stability, highlighting the limitations of simplistic assumptions that may not fully capture the real conditions. In solid-solid transformations, the kinetics are particularly challenging because these transformations occur with very limited diffusion. The rate of transformation depends on how effectively the new phase can form from the starting phase without inducing significant strain.

In this paper, we investigated high-temperature and high-pressure phases of calcium carbonate—namely calcite, CaCO3-IV, -V, and -Vb—using molecular dynamics (MD) simulations \cite{MolecularSim} with a Raiteri potential model.\cite{Raiteri_2015} We employed structural analysis and metadynamics to explore the phase transitions. Our simulations estimate the transition of calcite into IV at $T=$ 1000 K and $P=1$ bar, aligning well with experimental data. We observed carbonate ion disorder, thermal expansion, and IV symmetry that match experimental and theoretical findings. \cite{Ishizawa_2013,Ishizawa_2014,Kawano_2,Kawano_2009} A peak at $T=$1540 K in the heat capacity was identified as a second-order phase transition based on free energy and thermodynamic analysis. Synthetic X-ray diffraction (XRD) patterns from our MD data are consistent with diffraction peaks for the model unit cell of the V structure.\cite{gsas_II,Ishizawa_2013,Ishizawa_2014} We also identified a first-order phase transition to a disordered CaCO$_3$ structure, previously unreported resulting from the potential model. Our pressure-driven studies revealed a first-order transition to the Vb phase, supported by free energy, synthetic XRD patterns, and symmetry analysis.\cite{gsas_II,Druzhbin} Finally, we discuss preliminary findings on domain and defect formation in incommensurate CaCO$_3$ phase transitions.


\newpage 
\section{Models and Methods}\label{sec:method}

\subsection{Molecular dynamics simulations}


The MD technique based on an empirical force field model was used to study bulk calcium carbonate at different pressures and temperatures. We employed Raiteri's full-atomistic potential, which has been specifically designed and parameterized for calcium carbonate to replicate numerous experimental properties, such as lattice constants, density, and compressibility \cite{Raiteri_2015} -- all of which are closely related to the phenomena we have investigated. Preliminary tests ensured that our implementation of the Raiteri potential for the calcite crystal phase was consistent with previous literature (see Section \ref{ImpleTest} in the supplementary material for more details). All MD simulations were conducted using the Large-scale Atomic/Molecular Massively Parallel Simulator (\textsc{lammps}) package. \cite{LAMMPS} The Parrinello-Rahman thermostat and barostat coupling, \cite{parrinello1981polymorphic} in conjunction with periodic boundary conditions, were utilized to simulate the isobaric-isothermal ensemble (NPT). 

We have studied three sample dimensions to highlight possible size effects, especially in the kinetics of the phase transitions. Specifically, a small sample with 3360 particles (672 CaCO$_3$ units), a medium sample with 9900 particles (1980 CaCO$_3$ units), and a large sample containing 14529 particles (2904 CaCO$_3$ units) were investigated. The initial configurations for our simulations were built from the unit cell of the calcite structure, as determined by the X-ray investigations of Maslen et al.. \cite{calcite_str}

All the investigated samples underwent a preliminary equilibration in the NPT ensemble following the same protocol: at each $P$ and $T$, the sample was relaxed for 1 to 3 ns, depending on its size. Then, this relaxed configuration was used as a starting point for production simulations, which spanned from 3 to 8 ns -- with the simulation time also depending on the sample size. 
Changes of $P$ and $T$ over broad intervals were always made in small steps, to keep the system as close as possible to the equilibrium state. 


Here we consider two avenues for driving phase transitions: transitions at constant pressure driven by temperature changes and transitions at constant temperature driven by changes in pressure. 


To investigate temperature-driven transitions, the three samples (small, medium, and large) were equilibrated at a temperature of 300 K and a pressure of 1 bar and used as starting configurations. 
These starting samples were kept at a constant pressure of 1 bar, while being gradually heated from $T=$ 300 K up to $T=$ 2500 K with a temperature step of $\Delta T=$ 100 K. The temperature step was reduced to $\Delta T=$ 10 K around suspected transition points to increase the resolution of transition temperatures. 
The samples were heated with rates from 1 to 30 K ns$^{-1}$. Different heating rates allowed us to create several independent trajectories and did not result in any significant differences in the observed phase transitions.


To investigate pressure-driven transitions, the three samples (of the same three sizes) were initially equilibrated at $T$ = 1600 K and $P$ = 1 bar within the standard NPT protocol.
These structures were compressed up to 13 GPa with a pressure step of $\Delta P$ = 1 GPa. Close to the transition points, the step size was reduced to 0.25 GPa for better resolution of transition pressures. 
The samples were compressed with rates from 0.05 to 0.3 GPa ns$^{-1}$. Again, different rates were selected to create several independent trajectories. The choice of compression rate was not found to have a significant impact on the phase transitions we found.
For more details on the simulation protocol, see Section~\ref{sec:results}.

\subsection{Analysis}

For an isothermal-isobaric ensemble, the heat capacity per particle can be recovered from the fluctuations of total enthalpy, i.e. $H$, according to the formula: \cite{allen2017computer, MolecularSim} 

\begin{equation}
\begin{aligned}
C_p &= \left(\frac{\partial H}{\partial T}\right)_P = \frac{ \left < H^2 \right >- \left <H \right >^2}{k_bT^2N_p} \\
H   &= E + PV 
\label{Eq_HeatCapacity}
\end{aligned}
\end{equation}

where $N_p$ is the number of particles, $\left < H^2 \right >-\left <H \right >^2$ is the standard deviation of total enthalpy, $T$ is the temperature, and $k_b$ is the Boltzmann constant. We recover the instantaneous enthalpy from our simulations by summing the total energy $E$ with the product of the pressure, $P$, and the instantaneous volume, $V$. We calculated the (linear) thermal expansion coefficient using the following formulas: 

\begin{equation}
\alpha = \frac{1}{V_0} \frac{dV}{dT}
\label{eq:therm_exp}
\end{equation}

\begin{equation}
\alpha_i = \frac{1}{L_{i0}} \frac{dL_i}{dT}
\label{eq:lin_therm_exp}
\end{equation}

where $\alpha$ is the thermal expansion coefficient, $V_0$ is the volume at the initial temperature $T = T_0$, $\alpha_i$ is the linear thermal expansion coefficient along one of the three possible directions ($i = a, b, c$ are the symmetry axes, parallel to $x, y, z$ in the simulation box for calcite symmetry), and $L_i$ is the length of the simulation box along a chosen direction ($L_{i0}$ = $L_i$ at $T = T_0$). The differentiation in the equations \ref{eq:lin_therm_exp},\ref{eq:therm_exp} was implemented via the central finite-difference method.

Structural analysis was performed using a suite of different tools. The radial total (pair) distribution functions (RDFs, PDFs) were calculated with \textsc{lammps}. The local Steinhardt's bond orientational order parameter $\bar{q}_l$ \cite{q} were determined with \textsc{plumed}. \cite{plumed_q} Finally, the synthetic XRD patterns were found by applying the \texttt{pymatgen} python package. \cite{pymatgen} To determine the unit cells of the discovered phases, we analyzed the XRD data using \textsc{gsas-ii}. \cite{gsas_II} The details of the analysis are presented in the supplementary material.

The centrosymmetry or central symmetry parameter (CSP), introduced by Kelchner et al., \cite{CSP} was calculated using the \textsc{ovito} visualization tool, \cite{ovito} according to the formula: 

\begin{equation}
P_{CSM} = \sum_{i=1}^{\frac{N}{2}}|\vec{r_i}+\vec{r}_{i+\frac{N}{2}}|^2
\label{Eq_CSP}
\end{equation}

where $\vec{r_i}$ and $\vec{r}_{i+\frac{N}{2}}$ are vectors between a central atom and spatially-opposite neighbors and $N$ is the number of neighbors within a chosen coordination sphere. As it is defined CSP is a valuable tool for studying the central symmetry of various crystal phases. A CSP of 0 corresponds to the ideal centrosymmetric crystal, where all the contributions from all neighbor pairs in Eq.~\ref{Eq_CSP} cancel out. However, atomic sites within a defective region or a crystal structure lacking centrosymmetry result in a non-centrosymmetric neighborhood with a positive CSP. We have used the CSP to analyze crystal symmetry and locate possible defects as it has been proven to be a powerful tool for defect identification. \cite{def_param_1, def_param_2} All images and analyses from \textsc{ovito} were performed using the ``smoothed'' trajectories option.

Together with the CSP, the coordination number (CN), calculated within the first and second coordination spheres, was used to provide a precise characterization of the structure during its evolution, as well as identification and analysis of crystal defects.

The diffusion and mobility of the particles were studied by evaluating the mean squared displacement (MSD) as a function of time with the MDAnalysis python package. \cite{MDA_1, MDA_2}

\subsection{Free energy calculations}

To comprehensively study the phase transitions, an in-depth analysis of the free energy ($G$) landscape has been conducted by employing the metadynamics method through \textsc{plumed} version 2.8. \cite{plumed} This methodology involves mapping the free energy in terms of carefully chosen collective variables (CV), dependent on the positions of $N$ particles within the system. The metadynamics approach involves incorporating a biased time-dependent Gaussian potential, also known as $V(s, t)$, which is periodically added to the Hamiltonian of the system and explicitly defined by the following formula: 

\begin{equation}
    V(s, t) = h\sum_{t'<=t}\exp \left[-\frac{(s-s(t'))^2}{2\sigma^2} \right]
\end{equation}

where $s$ is a collective variable, $t$ and $t'$ are simulation times, and $h$ and $\sigma$ are chosen Gaussian parameters. 

We employed these biased Gaussian potentials to fill minima in the free energy landscape, in order to accelerate the search for metastable structures and overcome energy barriers between states. This metadynamics approach of integrating the history-dependent potential allows us to fully explore the system's free energy landscape as a function of chosen CVs. \cite{metad}

In the extended simulation time limit, which here is defined as the time where the biased potential has explored all the energy minima within the studied range of CVs, the free energy $G(s,t)$ can be recovered as follows:

\begin{figure*}
    \centering
    \includegraphics{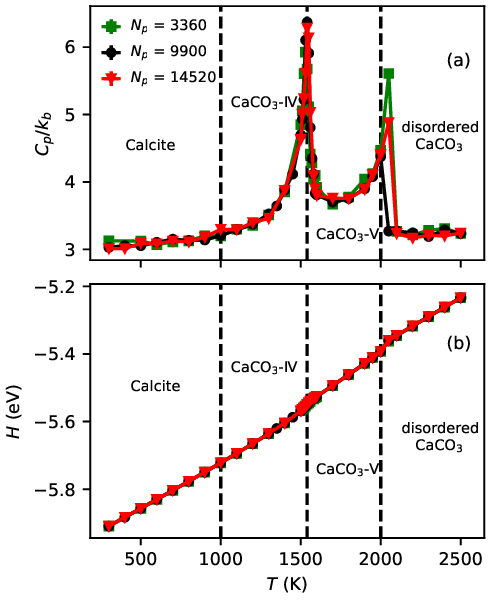}
    \includegraphics{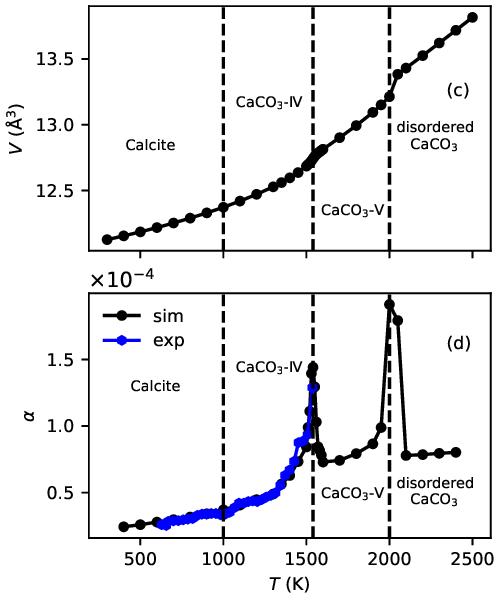}
    \caption{Heat capacity, $C_p$ (a), and enthalpy, $H$ (b), calculated at different temperatures for the small, medium, and large samples in green, black, and red, respectively. Volume, $V$ (c), and thermal expansion coefficient, $\alpha$ (d), from the simulation of the medium sample (black lines) and the experimental work of Dova et al. \cite{order_disorder_2_2009} (blue line). The vertical black dashed lines signify the temperatures of the three temperature-driven phase transitions. } 
    \label{fig:cp_4_5}
\end{figure*}

\begin{equation}
    G(s, t) = - \lim_{t\to\infty} V(s, t)
\end{equation}

The free energies for both pressure- and temperature-driven phase transitions were studied using metadynamics. We used $\bar{q}_6$ parameter between oxygen atoms (O-O) and $\bar{q}_6$ parameter between calcium atoms (Ca-Ca) as CVs to explore free energy in the vicinity of particular phase transitions.
The preliminary analysis from which we selected these CVs, and the details of their application are presented in Section~\ref{sec:results} and the supplementary material.

In all the metadynamics simulations, we added a Gaussian term with $\sigma$ = 0.01 and $h$ = 0.14 eV every 500 steps and eventually collected 10000-18000 additions.

\section{Results and Discussion}\label{sec:results}

We divide this Section into two parts, considering first the temperature-driven phase transitions, and second the pressure-driven phase transitions. 

\subsection{Temperature-driven CaCO\texorpdfstring{$_3$}{} phase transitions}\label{subsec:temp}

The calcium carbonate samples were simulated at pressure $P$ = 1 bar and temperature $T$ ranging from 300 K to 2500 K. From these simulations, we observed the fluctuations of
the enthalpy and collected the heat capacity of the systems for the three sample sizes. The corresponding plots are presented in Fig.~\ref{fig:cp_4_5}~(a), (b). No size effects were detected for the three samples within statistical variations. 
Consequently, our following investigation focused on the medium-sized sample, which produced results equivalent to the large sample but with lower computational costs. As such, we performed three independent simulations of the medium sample and averaged the results for heat capacity, enthalpy, volume, and thermal expansion coefficient as presented in Fig.~\ref{fig:cp_4_5}.

In Fig.~\ref{fig:cp_4_5}~(a), two distinct peaks are evident. The first peak, occurring at $T = T_V$ = (1540 $\pm$ 13) K, appears to be associated with the CaCO$_3$-IV to CaCO$_3$-V structural phase transition, as we will demonstrate below. We will show that the second peak, at $T = T_m $ = (2000 $\pm$ 17) K, corresponds to the transformation of the CaCO$_3$-V crystal into a more disordered phase, which we refer to as ``disordered CaCO$_3$''. Besides these two peaks, we have also marked the temperature $T = T_{IV}$ = (1000 $\pm$ 13) K at which our calcite sample assumes the features that experimental studies associate with the CaCO$_3$-IV phase. 

\begin{figure*}
    \centering
    \includegraphics{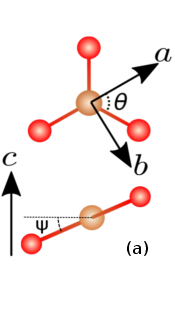}
    \includegraphics{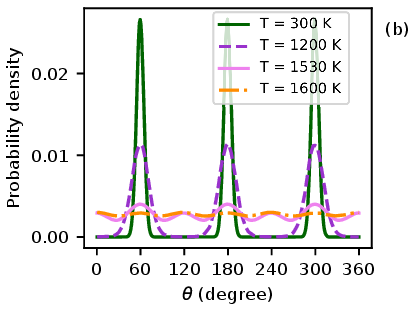}
    \includegraphics{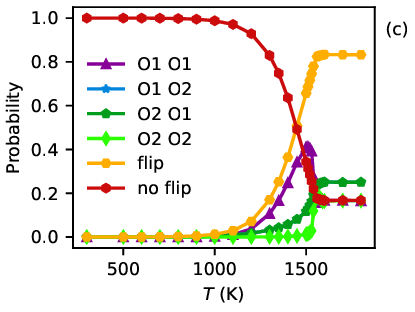}\\
    \includegraphics{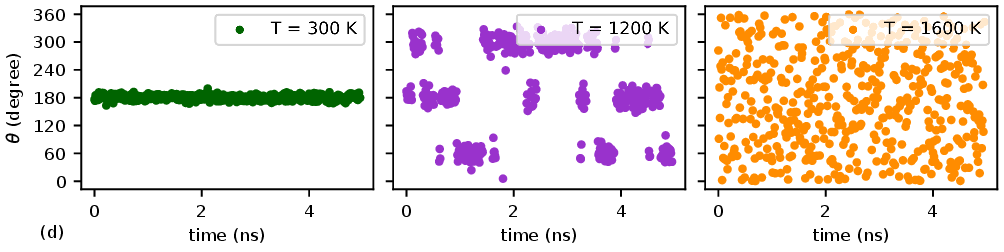}
    \caption{(a) Illustration of the $\theta$ and $\psi$ angles for one CO$_3^{2-}$ ion alongside the $abc$ symmetry axes. (b) $\theta$-angle probability density at different temperatures as given in the legend. (c) Temperature dependence of the probability of $\theta$ angles in the odd layer to: flip $120^{\circ}$ in the O1 state (purple line); flip $60^{\circ}$ from O1 to O2 state (blue line); flip $60^{\circ}$ from O2 to O1 state (dark green line); flip $120^{\circ}$ in the O2 state (light green line); flip overall, i.e. the sum of all the previous probabilities (orange); or remain in the same orientation, i.e. 1 minus the probability to flip (red line). (d) Time evolution of $\theta$ relative to the same C-O covalent bond at the three different temperatures indicated in the legends.}
    \label{fig:angles}
\end{figure*}

The CaCO$_3$-IV and CaCO$_3$-V phases are well documented in the literature, and our investigations are consistent with the previous experimental and modeling results. The experimental studies of Ishizawa et al.\cite{Ishizawa_2013, Ishizawa_2014} report the transition temperature of the IV$\leftrightarrow$V phase change as 1240 K, whereas our estimation differs by 240 K. This discrepancy is most likely due to the nature of the approximated potential we have used.

Regarding the second peak in our heat capacity graph, experimental investigations \cite{KARUNADASA201921, TUPITSYN2024102106} have demonstrated that at high temperatures and low pressure, calcite decomposes into crystalline CaO and gaseous CO$_2$. This reaction cannot be reproduced by our empirical (non-reactive) force field model, suggesting that the observed ``disordered'' phase may be an artifact of the potential model we are using. Nevertheless, the formation of this phase is still interesting to discuss.

Previous experimental literature \cite{Ishizawa_2013, Ishizawa_2014} refers to the transition from calcite to the metastable phase CaCO$_3$-IV as an order/disorder orientational transition, revealed by the variability of the distribution of oxygen orientations in carbonate triangle units at $T\simeq$ 934 K.
In Fig.~\ref{fig:cp_4_5}~(a), the heat capacity does not show any peak or delta function spike corresponding to this transition. The enthalpy plotted in Fig.~\ref{fig:cp_4_5}~(b) and volume plotted in Fig.~\ref{fig:cp_4_5}~(c) also do not exhibit any abrupt changes in the region around $T = $1000 K; however, peaks in heat capacity are not a universal feature of all types of transitions. Nonetheless, we will show that our simulation results after $T = T_{IV}$ replicate the substantial variability in the distribution of oxygen orientations seen in experiments -- thus, for this reason, we marked $T_{IV}$ as the transition temperature into phase IV in the graphs of Fig.~\ref{fig:cp_4_5}. 

In Fig.~\ref{fig:cp_4_5}~(d), we compare the thermal expansion coefficient, $\alpha$, calculated from our simulation data with that derived from the experimental data. We have shifted the experimental thermal expansion data to a higher temperature by 330 K to align them with our simulation results and better facilitate their comparison. 
The shifted experimental results show excellent agreement with our simulation results, suggesting that this order/disorder mechanism is not readily apparent in standard thermodynamic variables.

In what follows, we consider the variability of the orientation of oxygen in carbonate ions to investigate phases IV, V, and the `disordered' phase in our simulations.

Consistent with the previous literature, we describe in Fig.~\ref{fig:angles}~(a) the orientation of the carbonate ions using the angle $\theta$ as the rotation angle between the C-O covalent bonds and $a$-axis of symmetry, and the angle $\psi$ as the libration angle of the carbonate group out of the $ab$ symmetry plane. Based on experimental findings, the calcite unit cell has a distinctive arrangement of carbonate ions in alternating odd and even layers around the three-fold axis of symmetry along the $c$-axis. 
The $\theta$ angles formed for the three atoms of oxygen in each ion are discrete and assume the values $\theta=60^{\circ}, 180^{\circ}$, and $300^{\circ}$ in the odd layers, and $\theta=0^{\circ}, 120^{\circ}$, and $240^{\circ}$ in the even layers, with equal probability density.
This means that, in adjacent layers, the $\theta$ angles are rotated by $60^{\circ}$ relative to each other.

Referring to the notation in Refs. \cite{Ishizawa_2013, Ishizawa_2014}, we consider the oxygen atoms in the carbonate ions within the alternating layers of calcite to exist in two different states, either in the O1 state, as in the odd layers, or the O2 state, as in the even layers. 

The probability distribution of the $\theta$-angles recovered from our simulation data, Fig.~\ref{fig:angles}~(b), captures changes in the rotational mobility of carbonate ions within the odd layers at elevated temperatures. 
The three peaks at $60^{\circ}, 180^{\circ}$, and $300^{\circ}$, which are most prominent at $T=$ 300 K (green line), correspond to the O1 configuration of the calcite structure within odd layers. At $T =$ 1200 K ($ T > T_{IV}$, purple line), the probability distribution becomes broader, suggesting increased rotational mobility of the CO$_3^{2-}$ ions, characteristic of the CaCO$_3$-IV phase. Six peaks of different heights appear in the graph at $T=$ 1530 K (pink line), indicating that a critical condition for $\theta$ is reached. At this temperature, the ordering of carbonate triangles breaks down, as the probabilities of finding the O1 state within even layers and the O2 state within odd layers become nearly equal. Finally, at $T=$ 1600 K($T > T_{V}$, orange dashed line), the probability densities of the O1 and O2 states within the same layer become identical. At this point, carbonate ions rotate freely in the $ab$-plane, which is typical of the CaCO$_3$-V phase. 

We used the probability of $\theta$ flipping into a different orientation, of either the O1 or O2 state, as a criterion to define the transition temperature $T_{IV}$. As shown in Fig.~\ref{fig:angles}~(c), the `flip probability' (orange line) is $\simeq0$ at low temperatures while the sample assumes the calcite symmetry. At $T=$ 1000 K, this probability exceeds 0.01, from which we define the transition temperature between calcite and phase IV, $T_{IV}$. After the transition to phase IV, at $T > T_{IV}$, the flip-probability increases rapidly with rising temperature until T=$T_V$= 1540 K, where the transition into phase V occurs. At this point, the probability assumes a constant value of 0.8. The individual O1 and O2 transition probabilities, from which the overall flip probability was derived are also presented in Fig.~\ref{fig:angles}~(c). 

The disordering of carbonate ions' orientation with temperature is also illustrated in the time evolution of the $\theta$-angle of one single C-O bond, as presented in Fig.~\ref{fig:angles}~(d). At $T=$ 300 K, $\theta$ remains constant at $180^{\circ}$ for 5 ns (green dots), which is representative of the calcite structure. At $T=$ 1200 K ($ T > T_{IV}$, purple dots), $\theta$ flips between one of the three discrete orientations characteristic of its layer (central panel in Fig.~\ref{fig:angles}~(d)), indicating the onset of disorder within the layer. Finally, at $T =$ 1600 K $T > T_{V}$, the transition into CaCO$_3$-V is completed, and the same $\theta$-angle assumes all the possible values, including those characteristic of the O2 configuration. 
Additional plots demonstrating the disordering process between odd and even layers are presented in the supplementary material. 

Through XRD pattern analysis of our MD results, we identified two distinct unit cells. The unit cell at $T=$ 1200 K (phase IV) has a R$\bar{3}$c symmetry -- the same as calcite -- and parameters of $a=$ 4.931 \AA{}, $c=$ 17.622 \AA{}. The unit cell at $T=$ 1600 K (phase V) has a R$\bar{3}$m symmetry and parameters of $a=$ 4.922 \AA{}, $c=$ 9.166 \AA{}. The calculated XRD patterns from our simulations and the matching with the model unit cell, as determined by GSAS II, are presented in Fig.~\ref{fig:XRD_GSAS} of the supplementary material, along with a detailed explanation of the analysis.

The fact that the point group symmetry remains unchanged at the transition from calcite to phase IV, and that it changes to $R\bar{3}m$ symmetry with the transition from phase IV to phase V, is consistent with Ishizawa's study.\cite{Ishizawa_2013}
This further confirms the identification of the calcite-IV and IV-V transitions.
 
It is worth mentioning that rapidly heating an equilibrated sample from $T$ = 500 K to $T$ = 1700 K induces an instantaneous transition of calcite into the CaCO$_3$-V symmetry phase, bypassing the intermediate CaCO$_3$-IV phase. From this observation, we conclude that in this model the appearance of phase IV is not a necessary condition for the transition into phase V.


\begin{figure}
    \centering
    \includegraphics[width=0.45\textwidth]{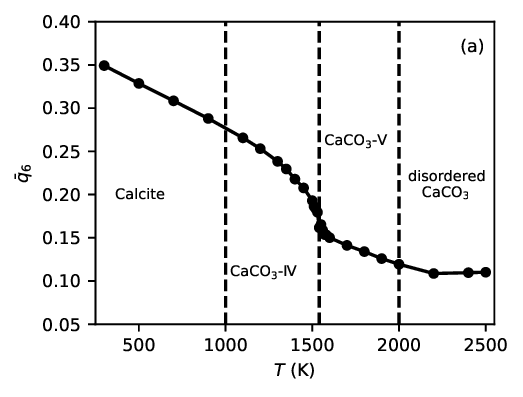}\\
    \includegraphics[width=0.45\textwidth]{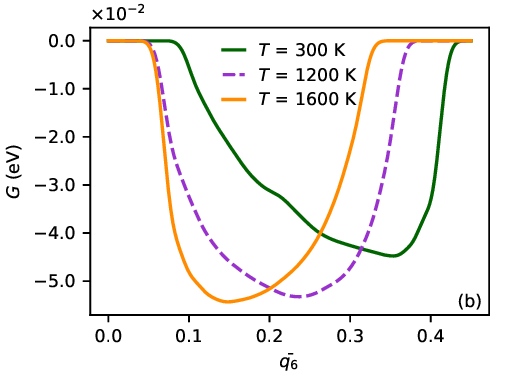}
    \caption{ (a)(O-O) averaged $\bar{q}_6$ parameter as a function of temperature. (b) $G$ as a function of the $\bar{q}_6$ parameter at three temperatures given in the legend.}  
    \label{fig:F_m_T}
\end{figure}

To complete our discussion on the temperature-driven transitions, we used metadynamics to explore the free energy landscape at several temperatures. We compared several order parameters and found that the averaged Steinhardt's orientational order parameter $\bar{q_6}$, calculated between oxygen atoms (O-O), was the most suitable CV. 
For the remainder of this section, we use $\bar{q_6}$ to refer to the parameter calculated using the distances between oxygen atoms and averaged over the number of particles. 
In Fig.~\ref{fig:F_m_T}~(a), the $\bar{q_6}$ parameter plotted against temperature clearly distinguishes between the three aforementioned phases. It decreases continuously and monotonically from $\bar{q_6} = 0.35$ to $ \bar{q_6} = 0.10$ across the temperature range covering the calcite, IV, and V phases. 

The relation between the variability of $\theta$-angles and $\bar{q_6}$ is discussed in Section~\ref{supp-q_6} and in Fig.~\ref{supp-fig:theta_q} of the supplementary material.   

\begin{figure}[h!]
    \centering
    \includegraphics{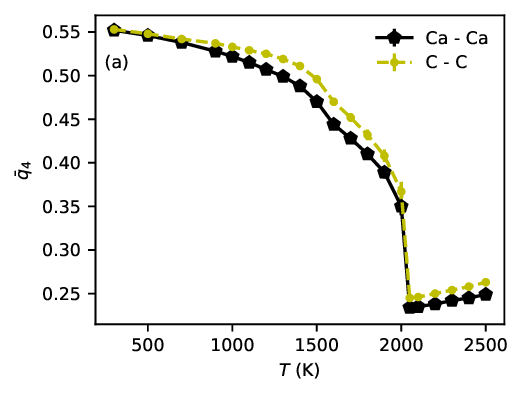}\\
    \includegraphics{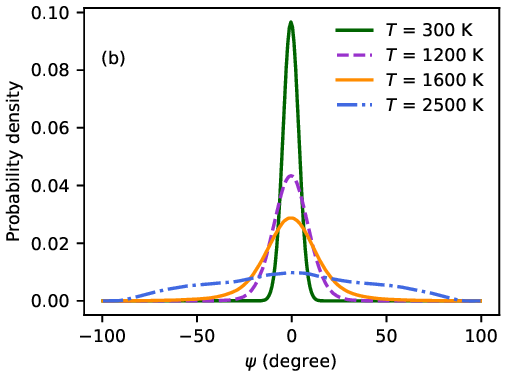}
    \caption{ (a) Temperature dependence of the $\bar{q_4}$ parameter averaged over the number of particles, calculated between calcium ions (black line), and between carbon atoms (green line). (b) Probability density of the libration angle $\psi$ at the four temperatures indicated in the legend. }  
    \label{fig:q4_psi}
\end{figure}

\begin{figure}[h!]
    \centering    
    \includegraphics{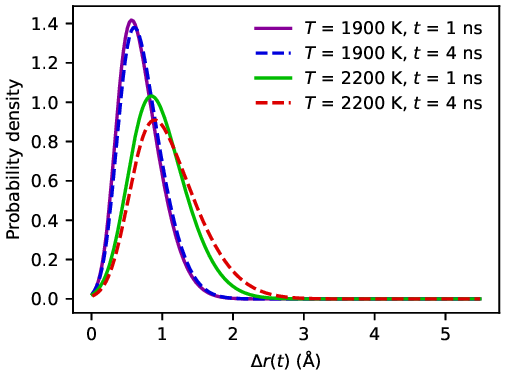}
    \caption{ Probability distribution of the displacements, $\Delta r(t)$, of carbon atoms from their initial positions at $t_0=$ 0 ns to the positions at $t=$ 1 ns (solid lines) and $t=$ 4 ns (dashed lines). Data are shown in purple and blue for $T=$ 1900 K, and in green and red for $T=$ 2200 K.}  
    \label{fig:delta_r}
\end{figure}

The Gibbs free energy $G$ was calculated at three different temperatures to represent the three phases of interest, and is shown in Fig.~\ref{fig:F_m_T}~(b). At $T=$ 300 K, the free energy has one minimum at 0.35 which corresponds to the calcite symmetry. The free energy landscapes retrieved at $T=$ 1200 K and 1600 K also have only one free energy minimum, corresponding to the phases CaCO$_3$-IV and CaCO$_3$-V respectively. In summary, $G$ has one minimum which slowly moves toward lower values of $\bar{q_6}$ as temperature increases, consistently with Fig.~\ref{fig:F_m_T}~(a). 
The continuous trend of $H$, $V$ in Fig.~\ref{fig:cp_4_5}~(a) and (b), and of $\bar{q}_6$ in Fig.~\ref{fig:F_m_T}~(a), lead us to characterize the phase transformation between CaCO$_3$-IV and CaCO$_3$-V as a typical second-order transition. Furthermore, the unique minimum in the free energy landscape at each presented temperature is another crucial feature of second-order transitions, in contrast to the multiple competing minima observed in typical first-order transitions. 
In a second-order phase transition, fluctuations in the order parameter drive the change without absorption or release of energy as latent heat. Moreover, the system exhibits critical behavior near the transition point, with some properties diverging. Our findings are consistent with previous experimental and modeling works, where continuous changes in thermodynamics variables, such as volume, cell parameters, and angles are also observed.\cite{order_disorder_2, order_disorder_2_2009, Ishizawa_2013, Ishizawa_2014} 

Next, we consider the transition between phase V and the `disordered CaCO$_3$' at $T=$ 2000 K. To analyze structural variation during this transition, we calculated the $\bar{q_4}$ Steinhardt’s order parameter between calcium (Ca-Ca) and carbon (C-C) atoms. Both mentioned $\bar{q_4}$ parameters drop abruptly from 0.55 to 0.25 after the transition into the `disordered' state as illustrated in Fig.~\ref{fig:q4_psi}~(a). Thus, the transition is characterized by a significant structural change and order decrease. In Fig.~\ref{fig:q4_psi}~(b), the probability density of the libration angle $\psi$ spreads from one distinct maximum to a nearly uniform distribution between +75$^{\circ}$ and -75$^{\circ}$, showing that the oxygen sublattice undergoes a transition to an even more disordered state than in phase V. 

We observed that this transition is reversible. Moreover, the limited diffusion in the MSD plot (Fig.~\ref{fig:msd} in the supplementary material) leads us to exclude a standard liquid state from the possible interpretations. 

Fig.~\ref{fig:delta_r} shows the probability density of displacements of carbon atoms from their initial position at time $t_0 = 0$ after times $t=$ 1 ns and $t=$ 4 ns, $\Delta r(t) = |r(t)-r(t_0)|$. Before the transition, at $T=$ 1900 K, the $\Delta r(t)$ distribution peaks at $\simeq 0.56$, and its tail does not change over time. After the transition, at $T =$ 2200 K, $\Delta r(t)$ distribution has a maximum at $\simeq 0.93$ \AA{}, and its tail moves over distances of $\simeq 0.3$ \AA{} at later times.
Overall, the increased mobility of carbon atoms, and the discontinuous change in many properties during this transition, i.e. $V$, $\bar{q}_4$, $\psi$ angle, support the idea of a first-order phase transition into a very slow evolving structure. 

\subsection{Pressure-driven CaCO\texorpdfstring{$_3$}{} transitions.}\label{subsec:press}

The second part of this research deals with the pressure-driven phase transitions. We start with a bulk calcium carbonate sample at $T=$ 1600 K and $P=$ 0 GPa as the initial configuration and explore its changes under pressure up to 13 GPa. Such a range of temperatures and pressures was chosen to explore a new calcium carbonate phase CaCO$_3$-Vb, which we will refer to as Vb, first-time resolved by X-ray diffraction pattern by Druzhbin et al.. \cite{Druzhbin}

As before, we first investigated structural changes in samples of three different sizes: small ($N_p=$ 3360), medium ($N_p=$ 9900), and large ($N_p=$ 14529). Ten independent trajectories were collected for the medium sample and five for the small and large samples. 
The heat capacity calculated at different pressures for the three samples revealed a discrepancy in the position of the peaks between the small and medium samples compared to the larger sample (see Fig.~\ref{fig:cp_4_5b_with_l} in the supplementary material). After rigorous analysis of various structural parameters, including CN, $\bar{q}_6$, and CSP, we established that the formation of defects affected the position of the second peak for the large sample. For more details, see Section~\ref{subsec:defect} in the supplementary material. 
Further research is needed to determine the exact formation mechanism of the defects observed in the large sample. This topic will be explored in our next study, as it requires the application of computational methods beyond MD.

\begin{figure}[h!]
    \centering
     \includegraphics{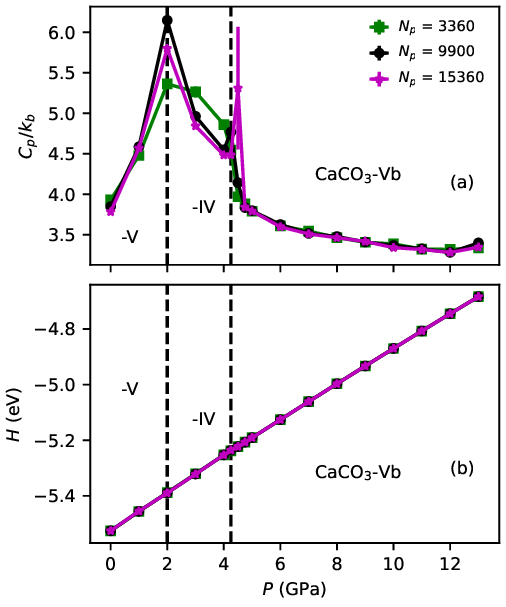}
    \caption{(a) Pressure dependence of the heat capacity $C_p$ and (b) enthalpy $H$ per particle for the three samples, i.e., $N_p=$ 3360 (green line), $N_p=$ 9900 (black line), and $N_p=$ 15360 (purple line), at $T$ = 1600 K and pressures up to 13 GPa. The vertical black dashed lines highlight the positions of the peaks.} 
    \label{fig:cp_4_5b}
\end{figure}

By considering the appropriate geometries, we built another sample with $N_p=$ 15360, which exhibited in the $C_p$ all the characteristic peaks at the same pressures as the small and medium samples (see Fig.~\ref{fig:cp_4_5b}(a)). Hereafter we refer to the latter as the large sample. The corresponding pressure dependence of the enthalpy is shown in Fig.~\ref{fig:cp_4_5b}(b). 

Two peaks are evident in the heat capacity plot: one at $P=$ 2.000 $\pm$ 0.007 GPa and another at $P=$ 4.250 $\pm$ 0.006 GPa. Both peaks were also observed in the reverse path, from $P=$ 13 GPa to $P=$ 0 GPa. The rest of this section refers to the statistics for the medium sample, as no significant differences were found between the results obtained for the large and medium samples. 

As established in the previous section through structure analysis and $\theta$-angles distribution, the phase assumed at $P=$ 0 GPa and $T =$ 1600 K is the phase V. The radial and partial distribution functions (RDF and PDFs), presented in Fig.~\ref{supp-fig:rdf_5b} of the supplementary material, do not show significant changes at that first peak pressure position, i.e. $P=$ 2 GPa. However, in the $\theta$ probability density in Fig.~\ref{supp-fig:press_theta_a} of the supplementary material, six small maxima at $P=$ 1 GPa, typical of phase V, turn into three maxima at $P=$ 2 GPa, consistent with the interpretation of a phase transition from phase V to phase IV(Section~\ref{subsec:temp}). Furthermore, from the XRD pattern analysis of our MD simulation structures, (see Section~\ref{sec:xrd_pdf}, Fig. \ref{fig:XRD_P3} of the supplementary material), we found a unit cell with point group symmetry R$\bar{3}$c and lattice parameters $a=$ 4.88 \AA{}, $c=$ 17.64 at $P=$ 3 GPa, which confirmed the transition V-IV at $P=$ 2 GPa. The transition at $P =$2 GPa is the reverse of the CaCO$_3$-IV to CaCO$_3$-V phase transition we discussed in the previous Section. This observation shows that the transition is reversible, confirming its continuous character.  
 
In Fig. \ref{fig:XRD_P6} of the supplementary material, the synthetic XRD pattern analysis at $P=$ 6 GPa is shown. The analysis revelead a monoclinic unit cell of P$2_1$/m point group symmetry with parameters $a=$ 6.17 \AA{}, $b=$ 4.92 \AA{}, $c=$ 3.93 \AA{}, and $\beta=106.3^{\circ}$. This symmetry is consistent with the CaCO$_3$-Vb structure resolved experimentally by Druzhbin et al..~\cite{Druzhbin} The unit cell parameters recovered from our XRD analysis are very similar to those reported in their work, with a maximum deviation of approximately 0.1 \AA{} for the parameter $a$. This comparison led us to identify the peak at $P=$ 4.25 GPa as the transition between phases IV and Vb.
The appearance of phase IV is consistent with another simulation study by Kawano et al.~\cite{Kawano_2009} in a different range of pressures and temperatures. 

To further characterize the IV to Vb phases and transition, we analyzed both structures with PDFs, CNs, and CSPs calculated between different types of atoms. Here, we present parameters calculated specifically for calcium atoms. An analysis involving other types of atoms is provided in the supplementary material.

\begin{figure*}
    \centering
    \includegraphics{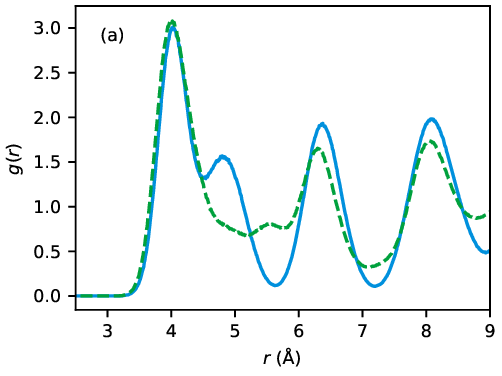}
    \includegraphics{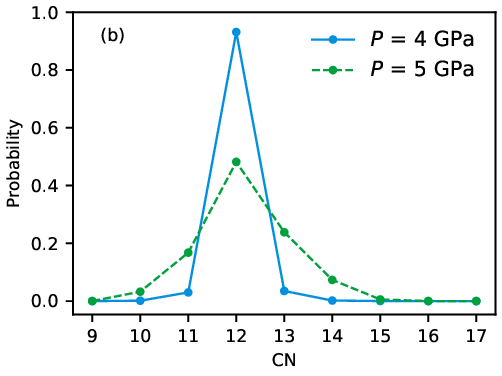}\\
    \includegraphics{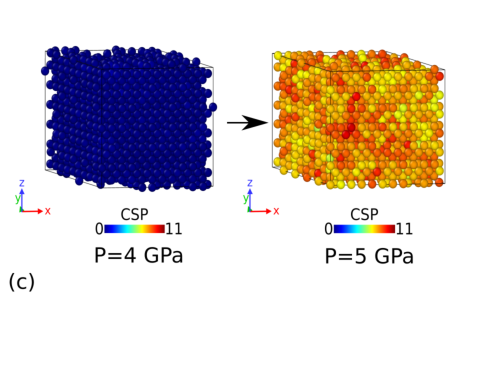}
    \includegraphics{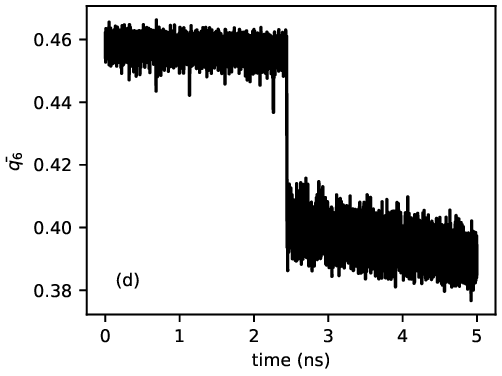}
    \caption{(a) (Ca-Ca) PDF, and (b) the probability of the (Ca-Ca) CN at $P=$ 4 GPa (blue line) and $P=$ 5 GPa (green line). (c) Simulation snapshots at $P=$ 4 GPa (left) and $P=$ 5 GPa (right) with calcium atoms only. The color legend for the CSP is provided in the color bar. (d) Time evolution of the $\bar{q_6}$ parameter for the sample continuously compressed from $P=$ 4 to 5 GPa. }  
    \label{fig:Vb}
\end{figure*}

The transition into CaCO$_3$-Vb is accompanied by an abrupt change in the Ca-Ca PDF, as shown in Fig.~\ref{fig:Vb}~(a). The second maximum at $r$ = 4.8 \AA{} for $P = $ 4 GPa (blue line) decreases and shifts to $r$ = 5.6 \AA{} at $P = $ 5 GPa (green dashed line). This indicates a rearrangement of Ca$^{2+}$ ions within the second coordination sphere following the transition identified at $P=$ 4.25 GPa. The Ca-Ca CN probability in Fig.~\ref{fig:Vb}~(b) supports the change in symmetry: CN = 12 has a probability of more than $\simeq$ 0.9 at $P = 4$ GPa (blue line), while at $P = 5$ GPa (green dashed line), after the transition, the probability for CN = 12 drops to 0.5. At this pressure, the probabilities for CNs = 11, 13, and 14 increase. The CN probability at higher pressures is provided in Section \ref{sec:pres_cn} of the supplementary material. 

The CSP for Ca$^{2+}$ ions is presented in Fig.~\ref{fig:Vb}~(c). At $P=$ 4 GPa (left snapshot), corresponding to phase IV, the CSP is approximately 0. This low CSP indicates high centrosymmetry for calcium in phase IV, as expected for a crystal with rhombohedral trigonal symmetry. After the phase transition, at $P=$ 5 GPa (right snapshot), the average CSP increases to 7, denoting a loss of central symmetry, which is typical of the CaCO$_3$-Vb phase characterized by the absence of central symmetry.


We also monitored the $\bar{q}_6$ structural parameter calculated between calcium atoms (Ca-Ca) while compressing the structure from $P=$ 4 to 5 GPa. The results are shown in Fig.~\ref{fig:Vb}~(d). An abrupt shift in the $\bar q_6$ parameter from 0.46 to 0.35 between two MD steps reveals rapid structural changes, suggesting that the entire simulation cell transitions instantaneously into the CaCO$_3$-Vb structure. Overall, from the structural analysis of the calcium sublattice, we infer that the collective motion of the atoms facilitates the spontaneous transformation of the CaCO$_3$-IV phase into the CaCO$_3$-Vb phase.
Further details on the structural analysis of this transition are presented in the supplementary material. 

\begin{figure}
    \centering
    \includegraphics{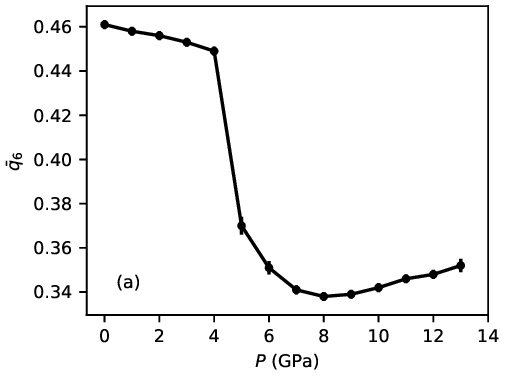}\\
    \includegraphics{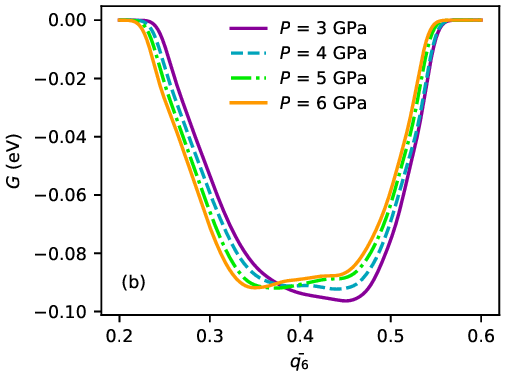}
    \caption{(a) The (Ca-Ca) $\bar{q}_6$ parameter with increasing pressure, and (b) the free energy $G$ as a function of the (Ca-Ca) $\bar{q}_6$ at the pressures given in the legend.}
    \label{fig:F_5b}
\end{figure}

The abrupt and spontaneous nature of the structural changes during the transition into the CaCO$_3$-Vb phase suggests that it may be a first-order phase transition. To verify this, we calculated the free energy $G$ at different pressures using metadynamics. We used the parameter $\bar{q}_6$, calculated for calcium within the second coordination sphere, as a collective variable (CV). In Fig.~\ref{fig:F_5b}~(a), the substantial reduction of the $\bar{q}_6$ order parameter between $P=$ 4 GPa and 5 GPa supports the occurrence of a first-order phase transition. 

Fig.~\ref{fig:F_5b}~(b) presents the free energy profile at different pressures. A clear minimum of the free energy at $\bar{q}_6 \simeq$ 0.45 at $P$ = 3 GPa (indigo line) indicates that phase IV is the equilibrium phase. As the system approaches $P=$ 4 GPa, a second local minimum appears close to $\bar{q}_6 \simeq$ 0.35 (blue line), suggesting two competing equilibrium phases, one at $\bar{q}_6 \simeq$ 0.45 (phase IV) and another at $\bar{q}_6 \simeq$ 0.35 (phase Vb). At $P$ = 5 GPa (dashed green line), we observe a unique global minimum at $\bar{q}_6 \simeq$ 0.35, which persists at $P=$ 6 GPa (orange line). The appearance of two free energy minima near the transition pressure $P=$ 4.25 GPa indicates two competing structures and supports the idea of a first-order phase transition between the CaCO$_3$-IV and CaCO$_3$-Vb structures. 

Experimental observations show that at high pressures and temperatures, calcite -- or one of its high-temperature disordered polymorphs (CaCO$_3$-IV or -V) -- undergoes a phase transformation into aragonite, \cite{Buerger} which, at pressures above 40 GPa, transforms into a post-aragonite.\cite{post-aragonite} Calcite, aragonite, and post-aragonite have different symmetries that we can resolve by XRD analysis in our simulations.
 
Here, we discuss the CaCO$_3$-Vb phase in the context of the transition from calcite, having a B1 cation array, to post-aragonite, having a B2 cation array. The B1 to B2 transition is often accompanied by more than one intermediate metastable structure.\cite{b1_b2_example_1, b1_b2_example_2} Several works demonstrated that intermediate phases can lower the potential barrier between considerable structural changes in ionic crystals.\cite{two_step_nucl, defect_speed} Various high-temperature high-pressure polymorphs, such as aragonite, aragonite-II, and CaCO$_3$-VII, have been identified as intermediate structures between calcite and post-aragonite.\cite{aragon_2_caco3_7, metastable_b1_b2} We explore whether CaCO$_3$-Vb is an example of one such structure in our simulation model. 

As shown in previous literature,~\cite{calcite_aragon_CN} the calcite-aragonite transition involves a change of the first coordination sphere of the calcium (Ca-O bond), resulting in an increase in the CN from 6 to 9. However, the Ca-Ca, C-C, and C-Ca CN do not differ between the two phases. 

In our simulation, the maximum of the C-C CN probability density is at CN = 12 at $P=$ 4 GPa and it shifts to CN = 15 at $P=$ 13 GPa (see CN probability density in Fig.~\ref{supp-fig:cn_5b_m_cc} in the supplementary material). 
At the compression rates used in our simulations (0.05-0.3 GPa ns$^{-1}$), we observed a rapid and significant increase in the CN at which the probability density becomes maximum, which allowed us to exclude the CaCO$_3$-Vb structure as an intermediate phase between calcite and aragonite. Additionally, neither of our simulations produced any aragonite-like structures.

The Ca-Ca CN probability density is at CN = 12 within the second coordination sphere in calcite. 
Using the same cutoff to calculate the CN probability density for high-pressure high-temperature structures, we obtain a CN of 19 for the aragonite-II and 18 for the CaCO$_3$-VII structure. The C-C CN, calculated within the second coordination sphere of calcite, which is also equal to 12, rises to 19 in aragonite-II and 20 in CaCO$_3$-VII. In our simulations, we observe an increase of the Ca-Ca CN probability density maximum at the position of CN = 12 in calcite to CN = 13 in the CaCO$_3$-Vb structure (see Fig.~\ref{fig:cn_5b_m_caca} in the supplementary material) and an increase of the C-C CN probability density maxima of the position at CN = 12 in calcite to CN = 15 in phase Vb (see Fig.~\ref{fig:cn_5b_m_cc}) in the supplementary material). These findings suggest that in our model the Vb structure, with P$2_1$/m symmetry, might gradually transform into a phase with P$2_1$/c symmetry or post-aragonite at very high pressures. Based on those discussions, we argue that the Vb structure could be considered an intermediate phase between calcite and post-aragonite in the model used.

\section{Conclusions}\label{sec:conclusions}

In this paper, we have analyzed selected structural transitions of calcium carbonate across varying temperatures and pressures, using the MD simulation technique in the LAMMPS engine with Raiteri's potential model. Starting with a calcite sample, our simulations revealed three transitions as the temperature was gradually increased from 300 K to 2500 K at 1 bar, and two transitions as the pressure was gradually raised from 0 to 13 GPa at a constant temperature of 1600 K.


The temperature-driven transition from calcite to the CaCO$3$-IV phase has been identified using the rotation angle $\theta$ between the covalent bonds C-O and $a$-axis of symmetry. The transition was further confirmed through
crystallographic analysis provided by the GSAS-II software. Our results reproduce previous experimental and modeling findings.

The temperature-driven transition between CaCO$_3$-IV and CaCO$_3$-V has also been revealed. A peak in the heat capacity suggested this transition, and the thermal expansion coefficient showed a good correlation with experimental findings.\cite{order_disorder_2_2009} Again, from the analysis of the $\theta$-angles, we determined that the full rotational disorder of carbonate triangles in the $ab$ plane causes this transition. A free energy analysis with metadynamics determined the CaCO$_3$-IV to CaCO$_3$-V phase transition to be a second-ordered order-disorder transition.
Our analysis of both transitions was confirmed through crystallographic analysis provided by the GSAS-II software, and agrees well with previous experimental findings.\cite{Ishizawa_2013, Ishizawa_2014} 

Upon increasing the temperature beyond 2000 K, we found a transition into a new `disordered CaCO$_3$' phase. This phase has not been reported in previous simulations or experimental results, appearing to be an artifact of the model. Analyzing both the libration angle $\psi$ of the carbonate ions relative to the $ab$ plane of symmetry, and the Ca-Ca, C-C $\bar{q_4}$ parameters, we showed that the new phase is more disordered than the ones at temperature $ T < 2000 K $.
Additionally, the gradual change in the tail of the displacement of carbon atom's probability densities shows a slow progression of the structure towards an even more disordered, liquid-like state.
 
For the pressure-driven phase transitions, we have found in certain geometries, a significant size effect caused by the formation of extended defects, o disordered interfaces among domains. Further research is required to understand the exact defect formation mechanism.

The first pressure-induced phase transition we found was between phases CaCO$_3$-V and -IV, the reverse of the CaCO$_3$-IV to -V temperature-driven transition discussed above. 

At higher pressures, we have detected another transition, into phase Vb, which has been resolved in the recent experimental work of Druzhbin et al..\cite{Druzhbin} Calculated free energy landscapes around the transition pressure show two competing free energy minima. This, together with abrupt changes of several structural parameters at the transition characterizes it as a first-ordered phase transition. Analysis of the PDFs, CNs, CSPs, and $\bar{q_6}$ order parameter demonstrated that the transition occurred through instantaneous propagation of the CaCO$_3$-Vb phase without a nucleus formation.
From the discussion about the structural properties of different high-temperature, high-pressure calcium carbonate polymorphs, we concluded that phase Vb might be an intermediate phase in the B1$\rightarrow$B2 calcite to post-aragonite phase transition. 

Despite the experimental results predicting a transition into aragonite structure after phase Vb within the pressures we studied, we have not observed any aragonite-like structure in our model. The empirical force field nature and the model we used probably prevented this transition from occurring. 

In conclusion, we have studied calcium carbonate over a wide range of temperatures and pressures. Several structural transitions were found and analyzed, including two high-temperature phase transitions: one into an artificial `disordered CaCO$_3$' structure, and the other into the CaCO$_3$-Vb phase, seen here for the first time in a simulation study. The temperature disordering of carbonate ions was thoroughly explored and compared with experimental findings, through which a strong agreement was found. We have applied metadynamics to explore free energy profiles at different temperatures and pressures.
This together with analysis of various structural and thermodynamic parameters has allowed us to distinguish the character of certain phase transitions as first-order or second-order. 
Despite the intrinsic limitations of the empirical force field we applied here, our results qualitatively reproduce many transitions across a wide range of temperatures and pressures -- including at extremes that are often difficult to reach in experiments. Finally, we discussed preliminary findings on domain and defect formation in incommensurate CaCO$_3$ phase transitions. Our preliminary analyses, not fully presented here, indicate that the incommensurate relations between the starting simulation cell and the final configuration resulted in the formation of domain walls and defects.

Additional work can surely be done to verify the ideal aspect ratio between parent and child cells, since it is well known that the transformation could involve the entire sample, as in a concerted mechanism, or only part of it. For example, the transition of the whole sample to the new phase can be greatly impaired by the formation of domains with different orientations with respect to the threefold axes. Future research will focus on a detailed investigation of the defects, including their nature and formation mechanisms. We will also work on developing supercells designed to avoid biases towards one structure over another due to commensurability issues. There is much work to be done to better understand and model the natural transitions into all possible structures that characterize the complex and rich phase diagram of calcium carbonate.

\section*{Supplementary Material}
The readers can access the supplementary material. 

\section*{AUTHOR DECLARATIONS}
The authors do not have any conflicts of interest to disclose.


\newpage

\bibliography{articles}

\end{document}



\centerline{\bf \Huge Supplementary Information on:}
\title{\bf \LARGE Structural Transitions of Calcium Carbonate by Molecular Dynamics Simulation} 



\author{Elizaveta Sidler}
\email{elizaveta.sidler@ntnu.no.}
\author{Raffaela Cabriolu}
\email{raffaela.cabriolu@ntnu.no.}

\affiliation{Department of Physics, Norwegian University of Science and Technology (NTNU), H\o gskoleringen 5, Trondheim, 7491, Norway.}




\maketitle 


\section{Potential implementation tests}\label{ImpleTest}

To validate our implementation of Raiteri's potential in the \textsc{lammps} software, we conducted a series of preliminary simulations for a system with $N_p$ = 3360. The properties obtained, listed in Table~\ref{table:1}, are consistent within the margin of error with those reported in the original article on Raiteri's model for calcium carbonate. \cite{Raiteri_2015}

\begin{table}[h!]
\centering
\begin{tabular}{|c c c c c|} 
 \hline & \\[-4.5ex]
 & a (Å) & c (Å) & $\alpha_a$ ($10^{-6}$ $K^{-1}$)& $\alpha_c$ ($10^{-6}$ $K^{-1}$)\\ 
 \hline\hline
 This research &~~~ 4.9377 $\pm$ 0.0049 &~~~ 17.2297 $\pm$ 0.0191 &~~~ -1.6325 $\pm$ 0.0008 & ~~~ 26.6189 $\pm$ 0.0003 \\ 
 From ref. \cite{Raiteri_2015} & 4.935 & 17.221 & -2.1 & 26.2 \\ [1ex] 
 \hline\hline
\end{tabular}
\caption{The unit cell parameters $a$ and $c$ correspond to average values obtained from a 4 ns NPT simulation at 300 K and 1 bar; The linear thermal expansion coefficients $\alpha_a$ and $\alpha_c$ have been calculated performing 5 ns long NPT simulations at 5 different temperatures between 100 K and 500 K. The results from reference \cite{Raiteri_2015} are also shown for comparison.}
\label{table:1}
\end{table}

\section{Temperature-driven transitions}
In this section, we present supplementary graphs on the transitions observed as temperature increases from 300 K to 2500 K, at a constant pressure of 1 bar. As the samples of three different sizes did not show any significant differences within the studied properties, we will only consider the medium sample with $ N_p = $ 9900. 

\subsection{Radial and pair distribution functions}
\begin{figure}
    \centering
    \includegraphics[width=0.75\textwidth]{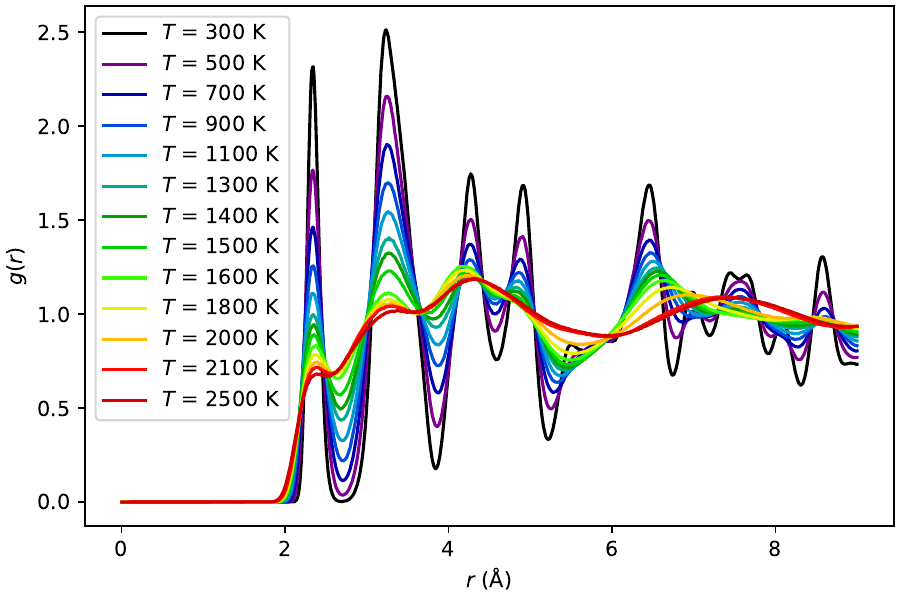}
    \caption{RDF at several temperatures as listed in the legend.}
    \label{fig:rdf_T}
\end{figure}

All the radial distribution functions (RDF) and the pair distribution functions (PDF) were identical for the three system sizes we investigated across the studied temperature ranges. The RDF for the medium sample obtained at different temperatures is shown in Fig.~\ref{fig:rdf_T}.

\newpage

\subsection{Mean squared displacement}\label{subsec:MSD}

\begin{figure}[h!]
    \centering
    \subfigure[C]{\includegraphics[width=0.32\textwidth]{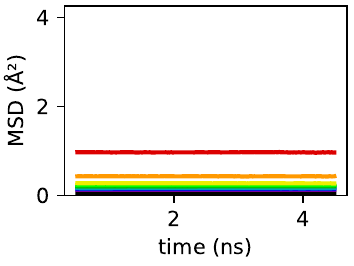}}
    \subfigure[Ca]{\includegraphics[width=0.32\textwidth]{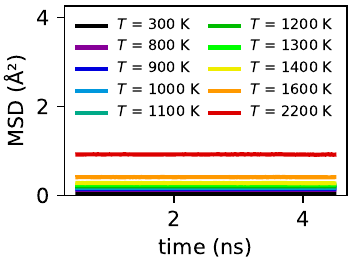}}
    \subfigure[O]{\includegraphics[width=0.32\textwidth]{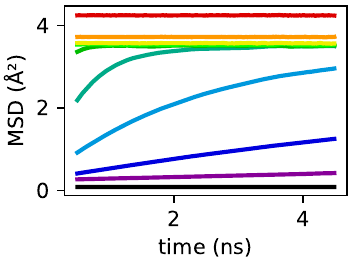}\label{MSD-O}}
    \caption{MSD of the different types of atoms at different temperatures as given in the legend.}
    \label{fig:msd}
\end{figure}

We report the mean squared displacement (MSD) plots in Fig.~\ref{fig:msd}. In Fig.~\ref{MSD-O}, increased mobility of the oxygen particles is visible at $T \approx$ 1000 K, indicating the transition of calcite to the CaCO$_3$-IV phase. 

\subsection{\texorpdfstring{$\theta$}{}-angle of the carbonate ions}

\begin{figure}[h!]
    \centering
    \subfigure[]{\includegraphics{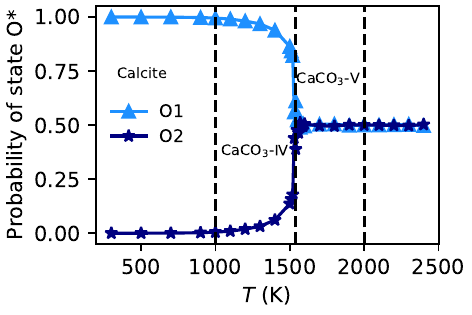}\label{fig:P_T}}
    \subfigure[]{\includegraphics{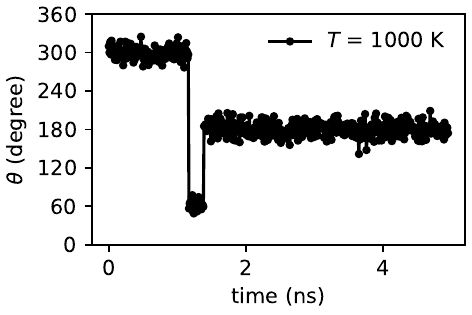}\label{fig:1theta_time}}
    \caption{(a) The probability of the angle $\theta$ for C-0 bonds, introduced in the main article in Section \ref{main-subsec:temp}, within the odd layers, to be in the O1 (light blue) and O2 (dark blue) states at different temperatures. (b) $\theta$ time evolution of one C-O covalent bond at $T$ = 1000 K.}
    \label{fig:1theta_P}
\end{figure}

The temperature dependence of the normalized probability of the $\theta$ angles in the carbonate ions appearing in states O1 or O2 within the odd layers is shown in Fig.~\ref{fig:P_T}.
As shown in Fig.~\ref{main-fig:angles} (c) of the main paper, at $T < 1000$ K the $\theta$-angle remains essentially constant over 5 ns. This is reflected in Fig.~\ref{fig:P_T}, where, at $T < 1000$ K, the probabilities of $\theta$ being in the O1 and O2 states are approximately 1.0 and 0.0, respectively. At the transition into phase IV, which occurs a $T = 1000$ K, the $\theta$-angle starts to rotate (see Fig.~\ref{fig:1theta_time}), marking the disorder within the same layer. In the temperature range between $ 1540K > T > 1000$ K, the probability of the angle being in the O1 state continuously decreases, while the probability of it being in the O2 increases. Finally, at $ T \geq 1540$ K, the probability of the $\theta$ angle being in either state reaches 0.5, marking the transition to phase V, which is characterized by fully disordered $\theta$ angles.

\newpage
\subsection{Order parameter in metadynamics simulations}\label{q_6}

\begin{figure}[h!]
    \centering
    \subfigure[]{\includegraphics{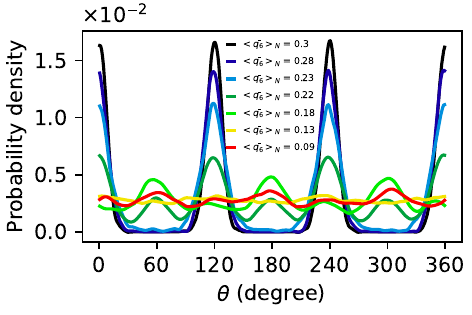}\label{fig:theta_q}}
    \subfigure[]{\includegraphics{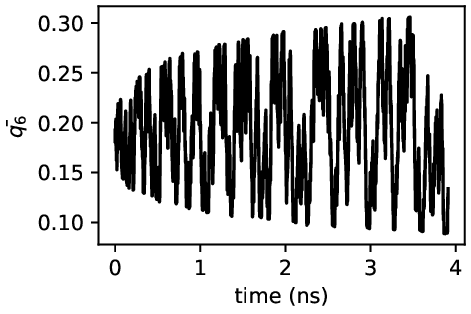}\label{fig:q_time}}
    \caption{Analysis from a metadynamics simulation at T = 1500 K: (a)$\theta$ probability density distribution with corresponding $\bar{q_6}$ parameters, and (b) $\bar{q_6}$ versus time.}
    \label{fig:theta_q_time}
\end{figure}

The graphs in this subsection show that the local Steinhardt's bond orientational order parameter $\bar{q_6}$, calculated between oxygen atoms (O-O) is a suitable collective variable (CV) for our metadynamics simulations to distinguish between calcite, CaCO3 -IV, and CaCO3 -V phases, facilitating the exploration of the
free energy profile. As in the paper, $\bar{q_6}$ in the graph, refers to the order parameter calculated between oxygen atoms (O-O) and averaged over the number of particles. 
Fig.~\ref{fig:theta_q} shows the correlation between the order parameter and the $\theta$ probability density at $T = 1500$ K. Using different bias potentials, our metadynamics simulations can sample all the possible $\theta$ values for which, the correspondent order parameter span from 0.1 to 0.3, covering the ranges of calcite, CaCO3-IV,
and CaCO3-V structures, as shown in Fig.~\ref{main-fig:F_m_T}~(a) of the main article. Fig.~\ref{fig:q_time} shows the variability of the $\bar{q_6}$ values during the metadynamics simulation at that same temperature over 4 ns.
\\
\\

\subsection{X-Ray diffraction analysis }\label{XRD_T}

\begin{figure}[h!]
    \centering
    \subfigure[$T=$ 300 K]{\includegraphics[width=0.45\textwidth]{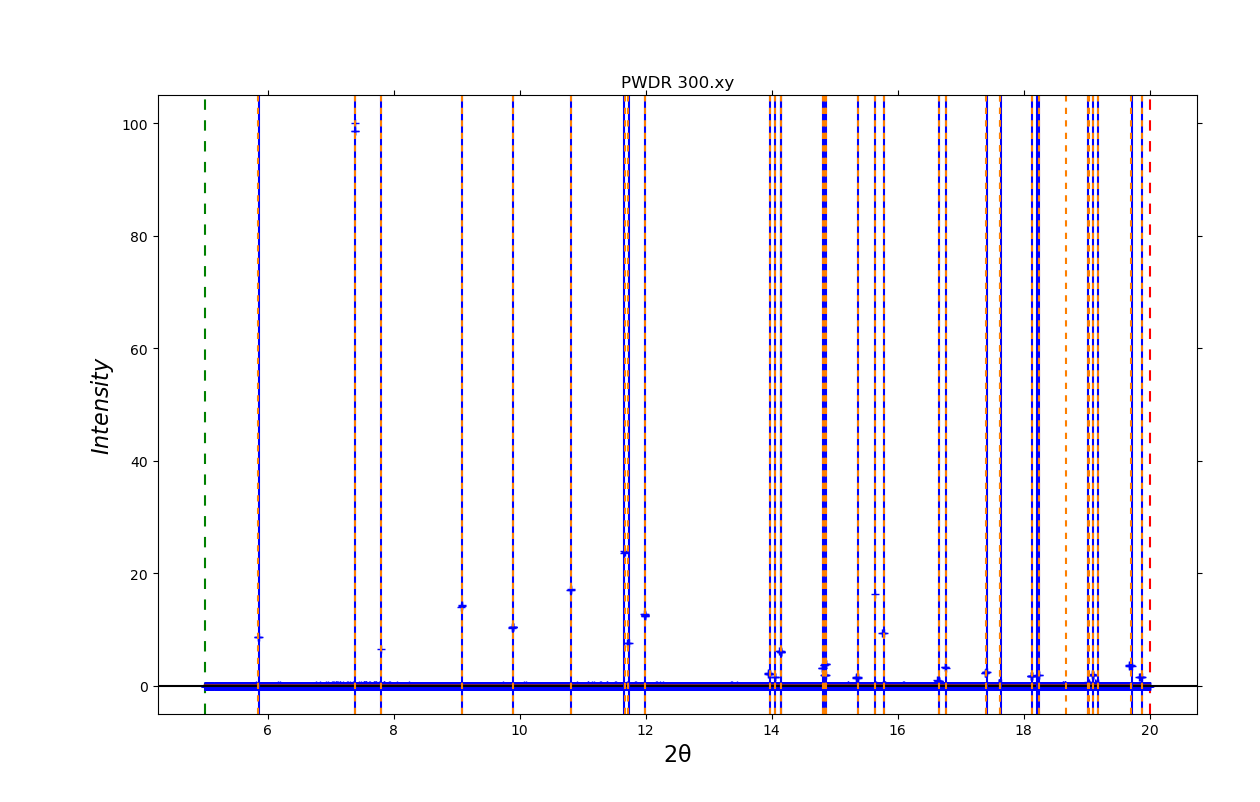}}
    \subfigure[$T=$ 1200 K]{\includegraphics[width=0.45\textwidth]{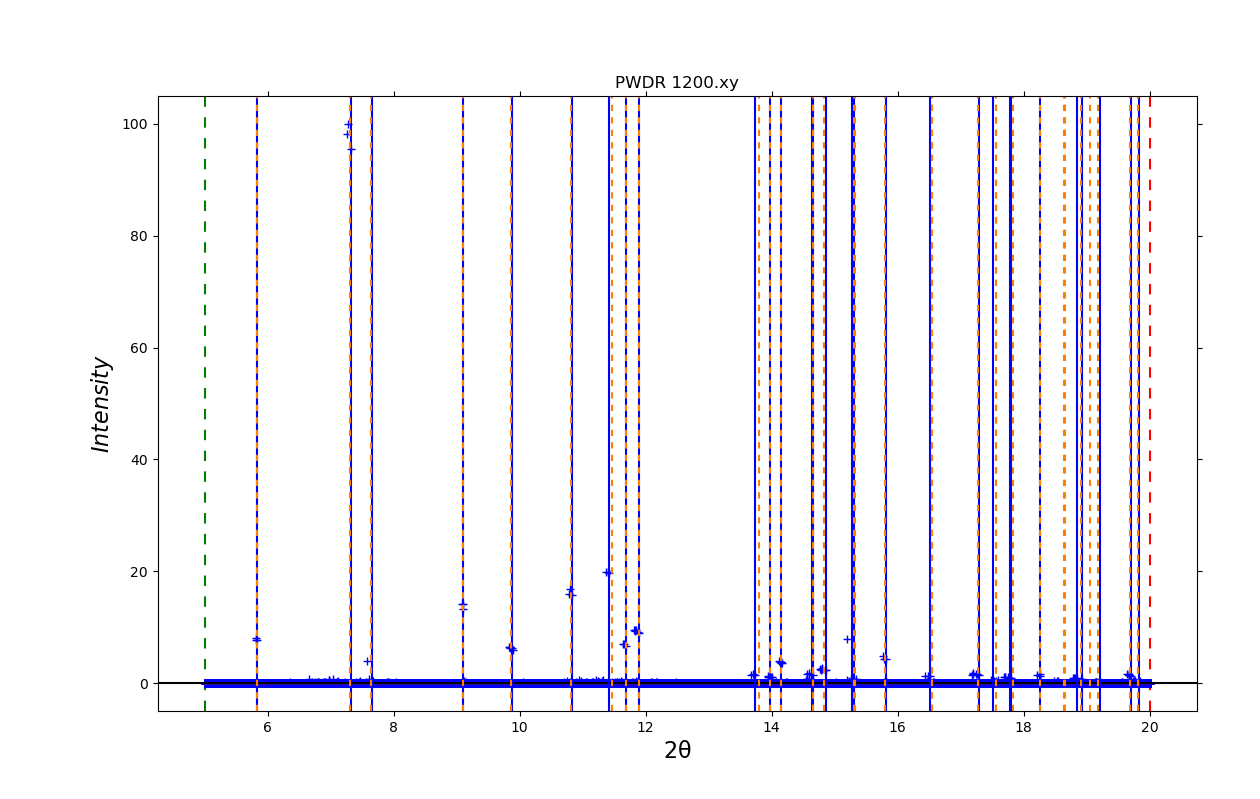}}
    \subfigure[$T=$ 1600 K]{\includegraphics[width=0.45\textwidth]{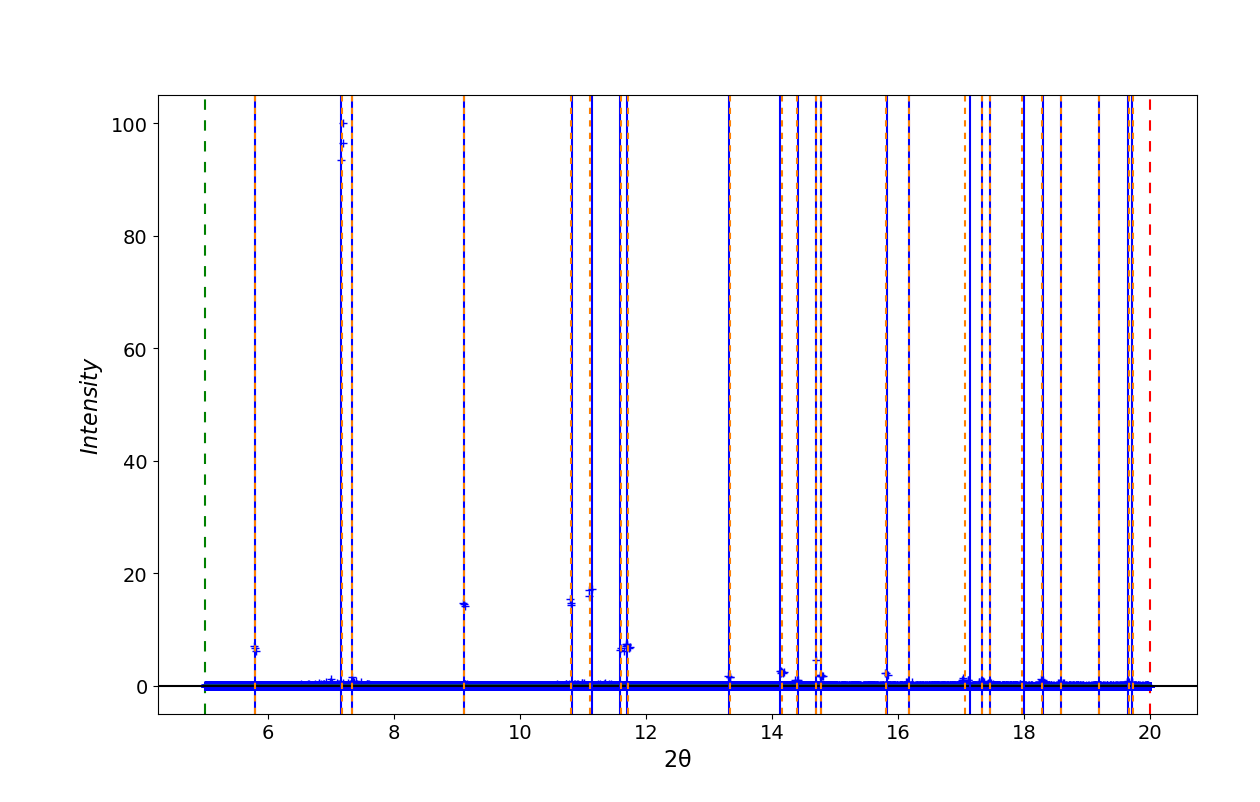}}
    \subfigure[$T=$ 1900 K]{\includegraphics[width=0.45\textwidth]{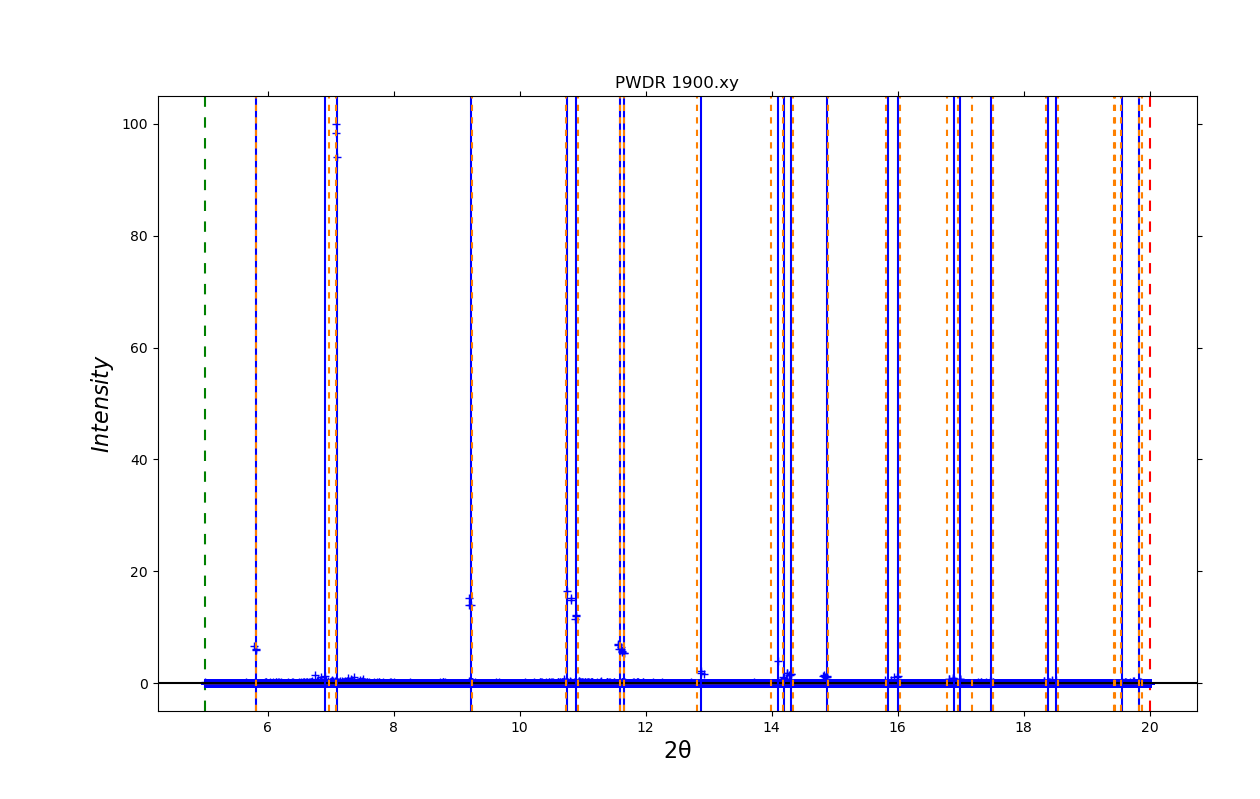}}
    \caption{XRD plots comparison at 4 different temperatures, as specified. The blue points represent synthetic XRD data generated from our simulation data; The blue vertical lines indicate the positions where diffraction peaks would appear on the 2$\theta$ axis, based on the synthetic XRD data from the MD simulations. The orange lines represent the positions of the diffraction peaks according to the unit cell model provided by \textsc{gsas-ii}. The images were generated with \textsc{gsas-ii}.}
    \label{fig:XRD_GSAS}
\end{figure}

X-ray diffraction (XRD) analysis was performed using a combination of the \texttt{pymatgen} python package, and the \textsc{gsas-ii} (General Structure Analysis System II) software.
In Fig.~\ref{fig:XRD_GSAS}, the blue points represent synthetic XRD data generated using the \texttt{pymatgen} python package, based on our simulation data. \textsc{gsas-ii} generated the blue vertical lines in the same figure, and using the `unit cell indexing' option, the software suggested several model unit cells as best fits for the synthetic data. Among those, we chose the best unit cell model with the parameters given in the main paper. The orange lines in Fig.~\ref{fig:XRD_GSAS} represent the positions of the diffraction peaks for the chosen models. 
A nearly perfect match between the model unit cell and the XRD from the simulation can be observed for all the structures investigated: calcite (see fig. \ref{fig:XRD_GSAS}(a) at T = 300K), CaCO$_3$-IV (see fig. \ref{fig:XRD_GSAS}(b) at T = 1200K) and CaCO$_3$-V (see fig. \ref{fig:XRD_GSAS} (c) and (d) at T = 1600K and T = 1990 K). 
        
\clearpage
\section{Pressure-driven CaCO\texorpdfstring{$_3$}{} transitions}\label{sec:pressure}

\subsection{Radial and pair distribution functions}\label{subsec:PDF-pressure}

\begin{figure}[h!]
    \centering
    \subfigure[RDF]{\includegraphics[width=0.4\textwidth]{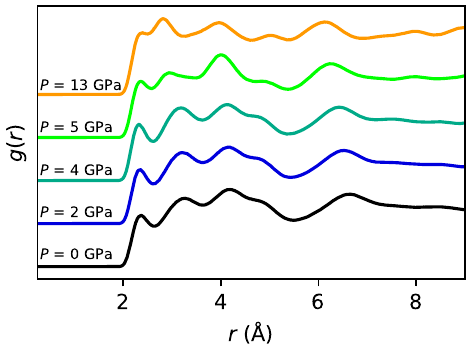}\label{fig:rdf_5b_all}}
    \subfigure[PDF Ca - Ca]{\includegraphics[width=0.4\textwidth]{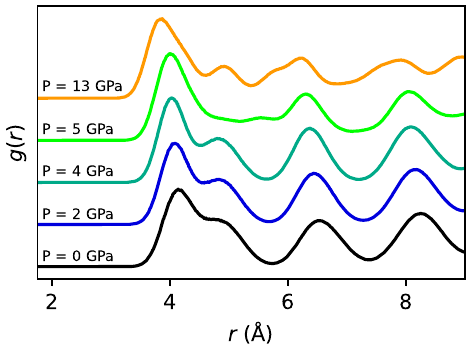}\label{fig:rdf_5b_caca}\label{fig:rdf_5b_ca_ca}}
    \subfigure[PDF C - Ca]{\includegraphics[width=0.4\textwidth]{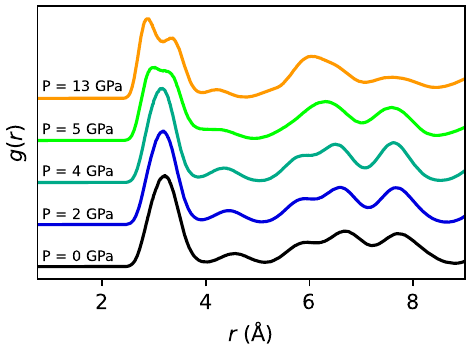}\label{fig:rdf_5b_c_ca}}
    \subfigure[PDF C - C]{\includegraphics[width=0.4\textwidth]{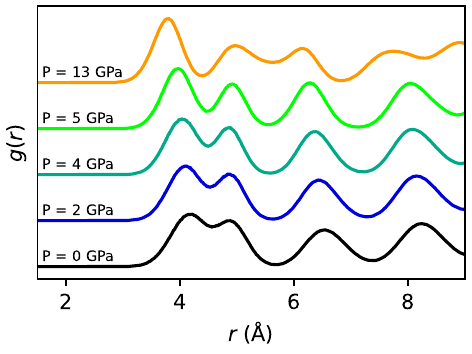}\label{fig:rdf_5b_c_c}}
    \caption{RDF and PDFs at different pressures, as marked on the figures, for the medium sample: (a) the RDF considering distances between all atoms; (b) the Ca-Ca PDF considering distances between calcium atoms; (c) the C-C PDF considering distances between carbon atoms; (d) the C-Ca PDF considering distances between carbon and calcium atoms.}
    \label{fig:rdf_5b}
\end{figure}

Radial and partial distribution functions (RDFs and PDFs) for the medium sample at $T=$ 1600 K and different pressures are shown in Fig.~\ref{fig:rdf_5b}. Fig.~\ref{fig:rdf_5b_all} represents the RDF, while Figs.~\ref{fig:rdf_5b_ca_ca} and ~\ref{fig:rdf_5b_c_ca} show the PDF for Ca-Ca and C-Ca respectively. 
Noticeable structural changes are observed between $P=$ 4 and 5 GPa. In Fig.~\ref{fig:rdf_5b_all}, the second maximum of the RDF shifts from $r=$ 3.19 at $P=$ 4 GPa to $r=$ 2.95 at $P=$ 5 GPa and becomes lower than the third maximum. In Fig. \ref{fig:rdf_5b_ca_ca}, the second maximum of the PDF for Ca-Ca flattens at $P=$ 5 GPa. In Fig. \ref{fig:rdf_5b_c_ca}, the first maximum at $P=$ 4 GPa splits into two maxima at $P=$ 5 GPa. No apparent structural rearrangement follows from the PDF C-C in Fig.~\ref{fig:rdf_5b_c_c}.
 
Furthermore, Fig.~\ref{fig:rdf_5b} is used to define our first and second coordination spheres based on the positions of the first and second minima in the PDF at $P=$ 0 GPa, respectively. For the C-Ca bonds, the first shell is defined at the coordination sphere distance $r_{cut} = 3.86$ \AA{}, while for Ca-Ca and C-C bonds in our model, the second shell is defined at the coordination sphere distance of $r_{cut} = 5.65$ \AA{}. 

\newpage
\subsection{\texorpdfstring{$\theta$}{}-angles in the carbonate ions}

\begin{figure}[h!]
    \centering
    \subfigure[]{\includegraphics[width=0.45\textwidth]{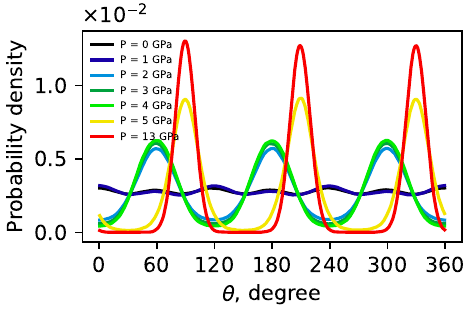}\label{fig:press_theta_a}}
    \subfigure[P = 13 GPa]{\includegraphics[width=0.41\textwidth]{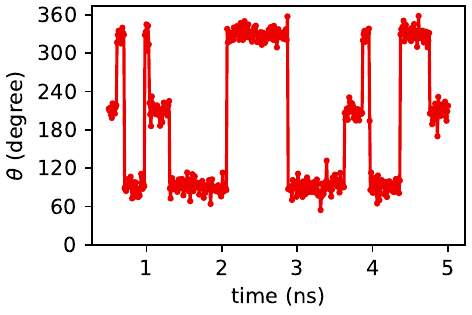}\label{fig:press_theta_b}}
    \caption{(a) $\theta$ probability density at different pressures given in the legend and (b) $\theta$ time evolution of one C-O bond at $P=$ 13 GPa. }
    \label{fig:press_theta}
\end{figure}

Fig.~\ref{fig:press_theta_a} shows the $\theta$ probability density for the medium sample at $T=$ 1600 K at different pressures. The six nearly flat peaks observed at $P\sim$ 0 GPa and $P$ = 1 GPa merge into three distinct peaks at $P$ = 2 GPa, marking the transition into phase IV. After the transition into phase Vb, occurring at $P=$ 4.25 GPa, the three maxima shift from the typical oxygen-ordered state in phase IV to a different ordered arrangement. This $\theta$ configuration persists up to $P=$ 13 GPa with a probability that increases with the pressure. Fig.~\ref{fig:press_theta_b} shows that, in the Vb phase, the $\theta$angles exhibit a behavior similar to that observed in the CaCO$_3$-IV phase -- flipping between discrete states by $120^{\circ}$. 

\subsection{X-Ray diffraction analysis}\label{sec:xrd_pdf}

\begin{figure}[h!]
    \centering
    \subfigure[$P=$ 3 GPa]{\includegraphics[width=0.45\textwidth]{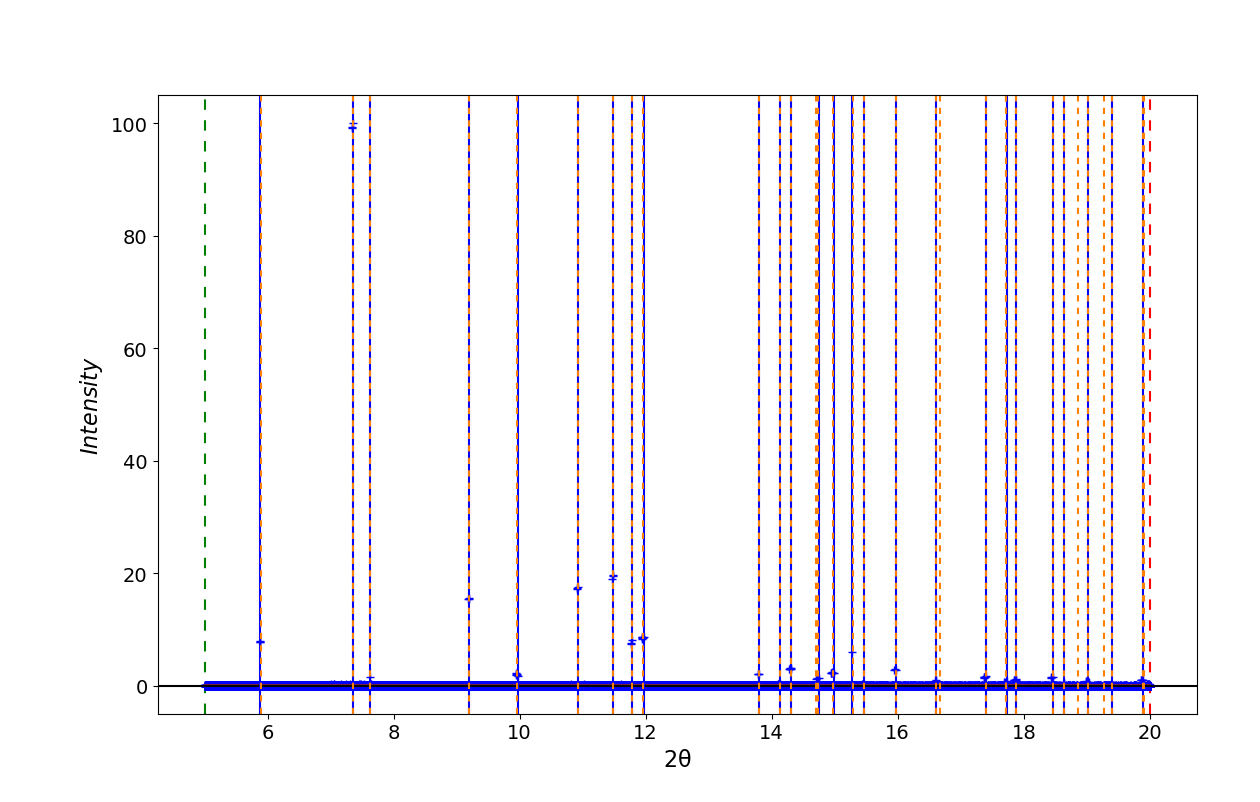}\label{fig:XRD_P3}}
    \subfigure[$P=$ 6 GPa]
    {\includegraphics[width=0.45\textwidth]{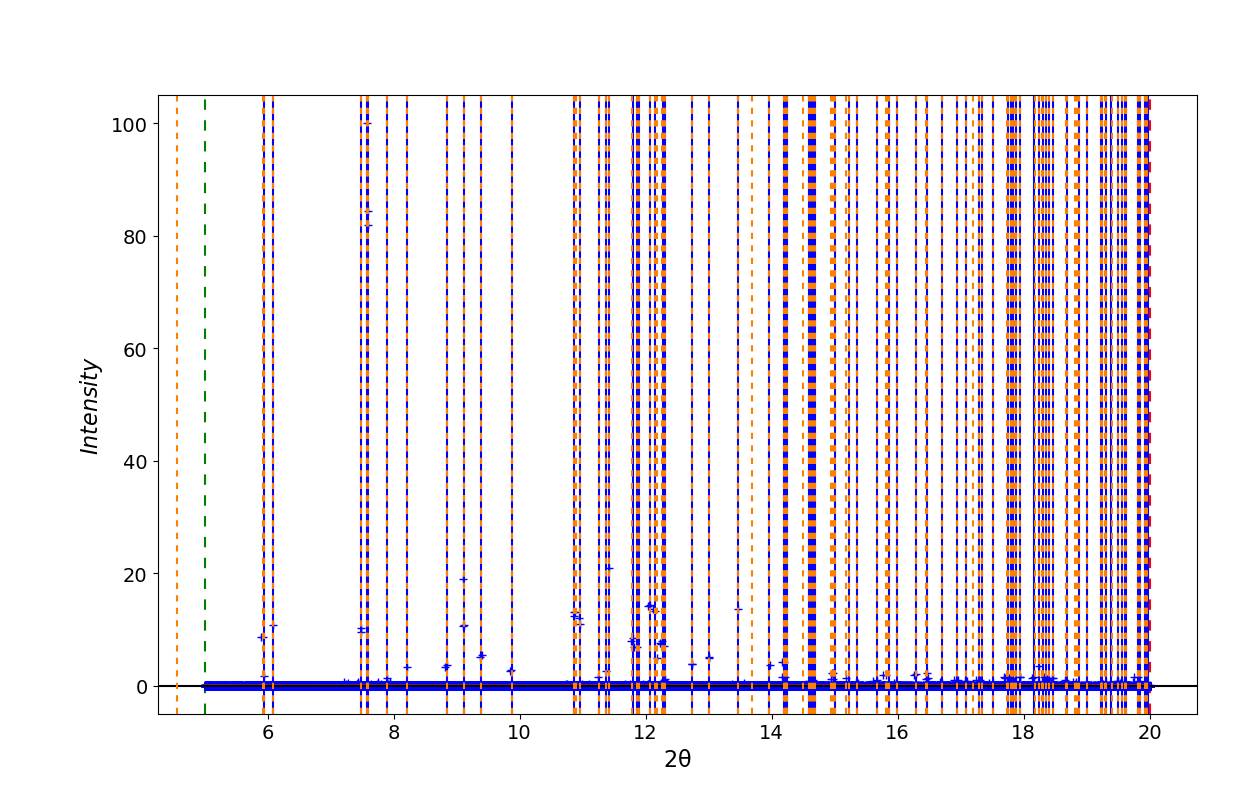}\label{fig:XRD_P6}}
    \caption{XRD plots comparison at two different pressures, as specified. The blue points represent synthetic XRD data generated from our simulation data; The blue vertical lines indicate the positions where diffraction peaks would appear on the 2$\theta$ axis, based on the synthetic XRD data from the MD simulations. The orange lines represent the positions of the diffraction peaks according to the unit cell model provided by \textsc{gsas-ii}. The images were generated with \textsc{gsas-ii}.}
    \label{fig:xrd_unit_cell_5b}
\end{figure}

Following the same XRD analysis procedure described in Section \ref{XRD_T}, we extracted the unit cells at $P=$ 3 GPa and $P=$ 6 GPa. The correspondence in the diffraction peak positions relative to the simulation data, blue lines, and model unit cell, orange lines, is shown in Fig.~\ref{fig:XRD_P3} for phase IV at $P=$ 3 GPa, and in fig. \ref{fig:XRD_P6} for phase Vb at $P=$ 5 GPa. 

\newpage 
\subsection{Coordination number analysis}\label{sec:pres_cn}

\begin{figure}[h!]
    \subfigure[C-Ca]{\includegraphics[width=0.32\textwidth]{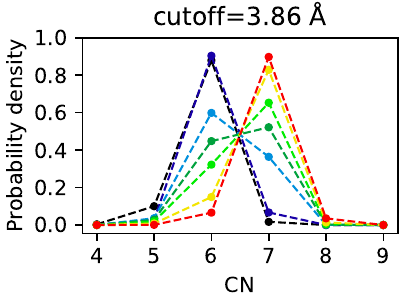}\label{fig:cn_5b_m_cca}}
    \subfigure[Ca-Ca]{\includegraphics[width=0.32\textwidth]{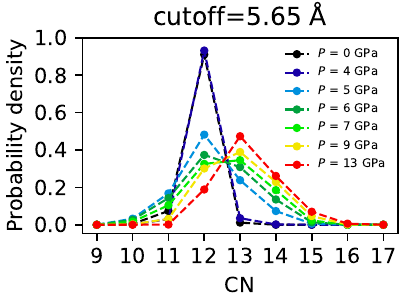}\label{fig:cn_5b_m_caca}}
    \subfigure[C-C]{\includegraphics[width=0.32\textwidth]{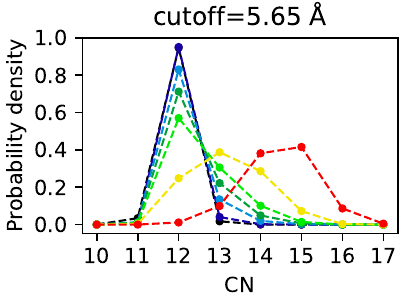}\label{fig:cn_5b_m_cc}}
    \caption{(a) C-Ca, (b) Ca-Ca and (c) C-C CN probability density at different pressures indicated in legend at $T=$ 1600 K.}
    \label{fig:cn_5b_m}
\end{figure}

The coordination number (CN) probability density is presented in Fig.~\ref{fig:cn_5b_m} for various pressures for the medium sample. Fig. \ref{fig:cn_5b_m_cca} concerns the first coordination shell, while Figs. \ref{fig:cn_5b_m_caca} and Figs. \ref{fig:cn_5b_m_cc} refer to the second coordination shell. The cutoff radii shown in the figure used to define the first and second coordination shells are indicated in the figures and have been obtained by the analysis of the PDFs as explained at the end of subsection \ref{subsec:PDF-pressure}.

In Fig. \ref{fig:cn_5b_m_cca}, the maximum CN probability density for C-Ca shows an abrupt drop between $P=$ 4 GPa and 5 GPa at CN = 6, indicating the transition from phase IV to Vb. From previous literature, a CN = 6 is expected for the IV phase group symmetry, while after the transition, higher CNs become populated reflecting the higher density packing of the Vb polymorph. 
As the pressure increases, higher CNs become more populated, with the probability density maximum shifting to CN = 7 for $P \geq$ 6 GPa, and very high probability densities are observed from $P =$ 9 GPa onward. Fig. \ref{fig:cn_5b_m_caca} conveys a similar message for the Ca-Ca coordination: the CN probability density also drops abruptly between $P=$ 4 GPa and 5 GPa at CN = 12, followed by a gradual increase in the population of other CNs at higher pressures. A CN = 12 of Ca-Ca in the second coordination shell is expected for the IV phase group symmetry, reflecting a highly ordered arrangement of the ions. After the transition, the probability of higher CNs to be populated increases, as the higher density packing of the Vb polymorph requires. 
The probability density maximum shifts to CN = 13 for $P \geq $ 7 GPa, with a gradual increase as the pressure continues to rise. In Fig. \ref{fig:cn_5b_m_cc}, there is a moderate drop in the CN probability density maximum from $P=$ 4 GPa to $P=$ 7 GPa at CN = 12, followed by a continuous increase in the CN population as the pressure rises, with the CN = 15 assuming the maximum probability density at $P=$ 13 GPa. 

\subsection{Central symmetry parameter analysis}

\begin{figure}[h!]
    \centering
    \includegraphics{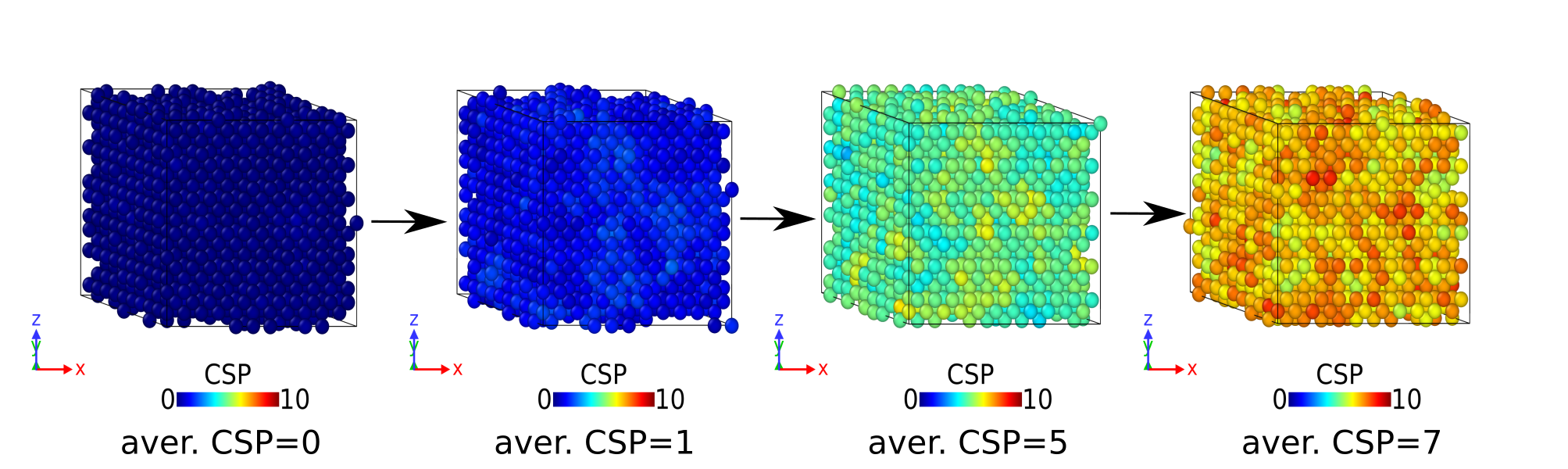} 
    \caption{CSP analysis of four selected snapshots during the compression at $T =$ 1600 K from $P=$ 4 GPa to $P=$ 5 GPa at a rate of 0.2 GPa/ns. The starting configuration ($P \approx$ 4.19 GPa) with an average CPS of 0 corresponds to phase IV, the last configuration ($P \approx$ 4.31 GPa ) corresponds to phase Vb, and the intermediate configurations reflect the CSP during the transition between phases IV and Vb. The pressure step between adjacent snapshots is 0.04 GPa.}
    \label{fig:ovito_4_5_m}
\end{figure}

\begin{figure}[h!]
     \centering
    \subfigure[$P=$ 4 GPa; CSP = 0]{\includegraphics[width=0.23\textwidth]{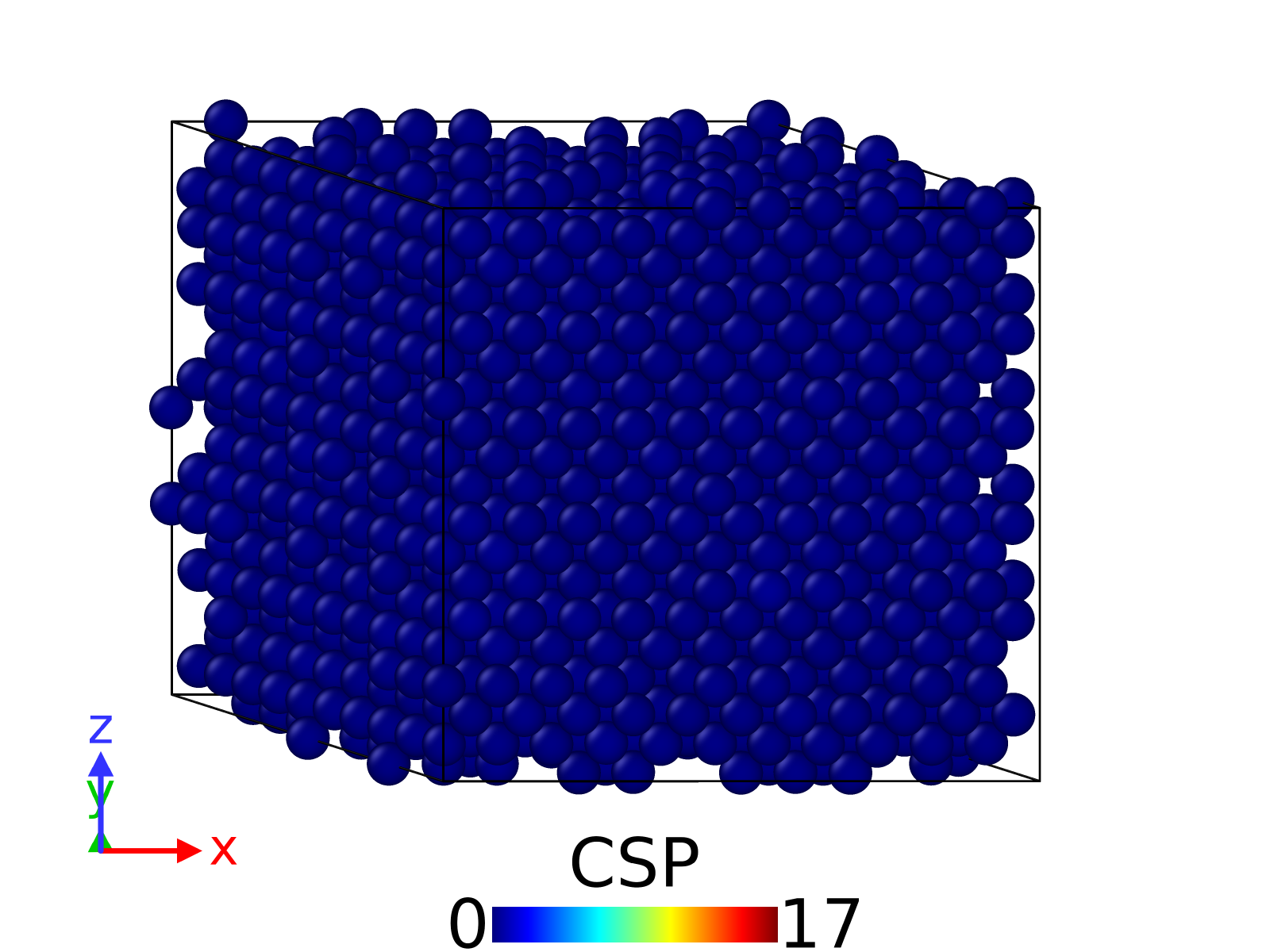}\label{fig:ovito_6_7_m_a}} 
     \subfigure[$P$ = 5 GPa; CSP = 8]{\includegraphics[width=0.23\textwidth]{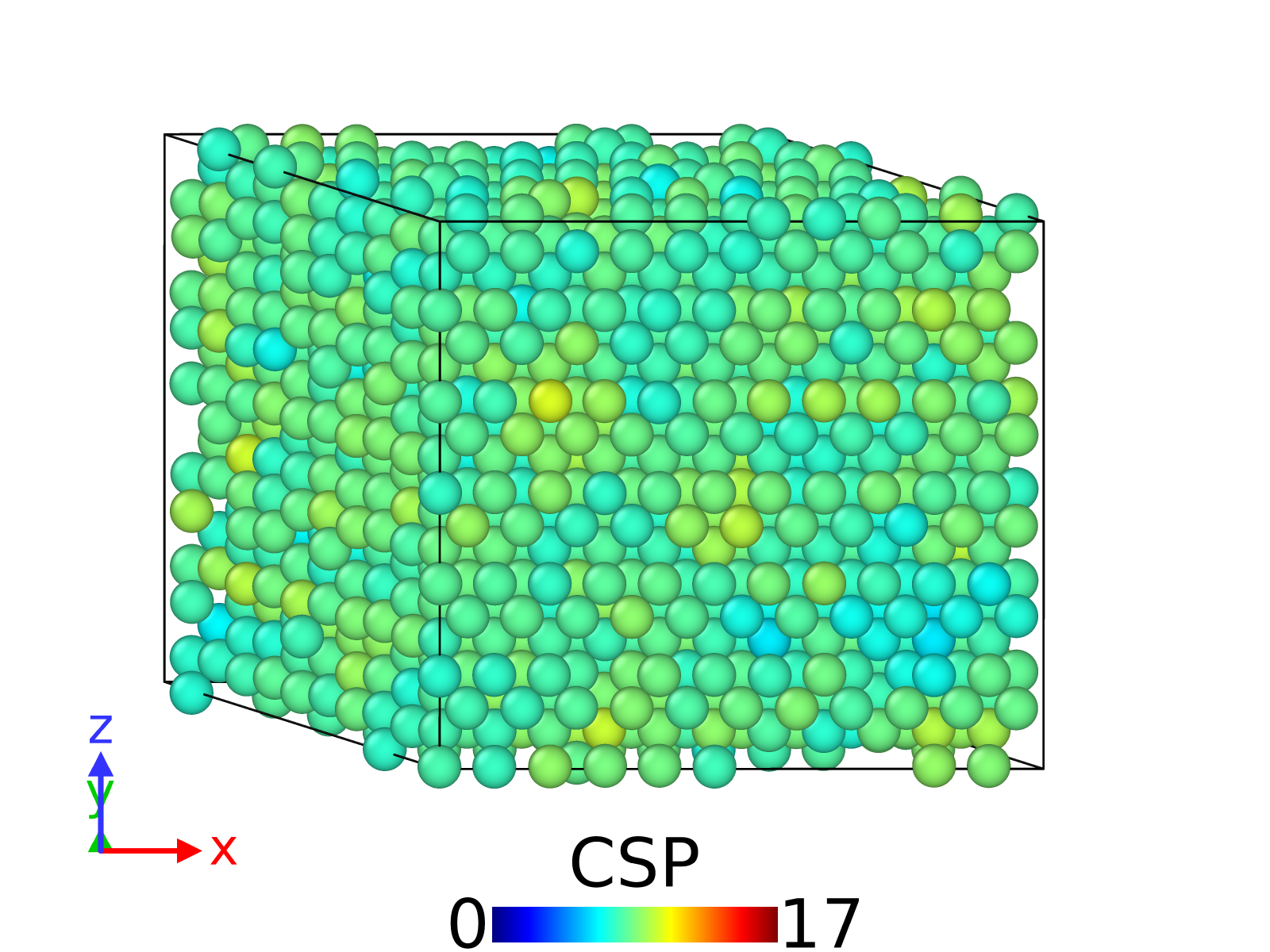}\label{fig:ovito_6_7_m_f}}
     \subfigure[$P$ = 6 GPa; CSP = 11]{\includegraphics[width=0.23\textwidth]{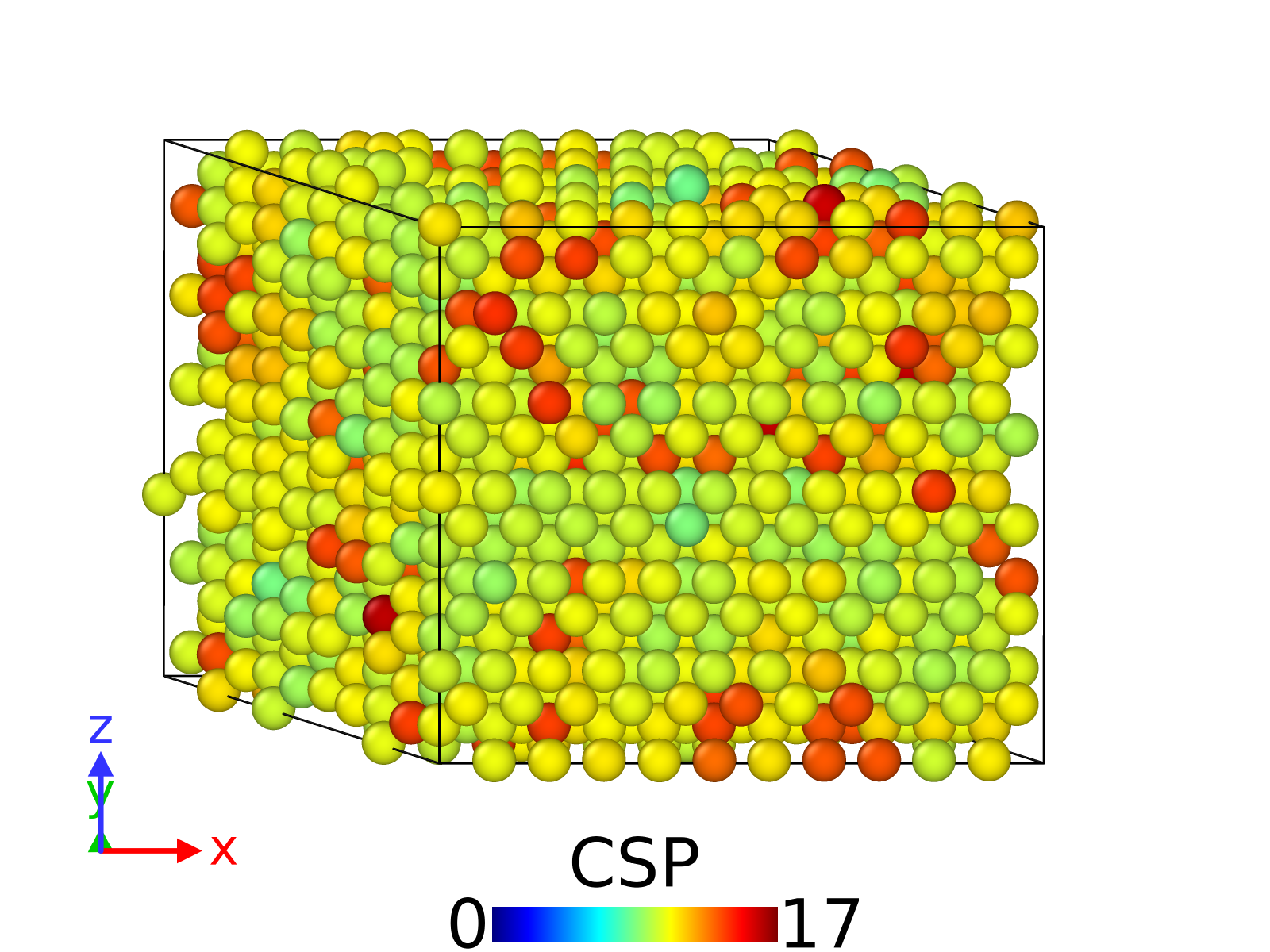}}
     \subfigure[$P$ = 7 GPa; CSP = 15]{\includegraphics[width=0.23\textwidth]{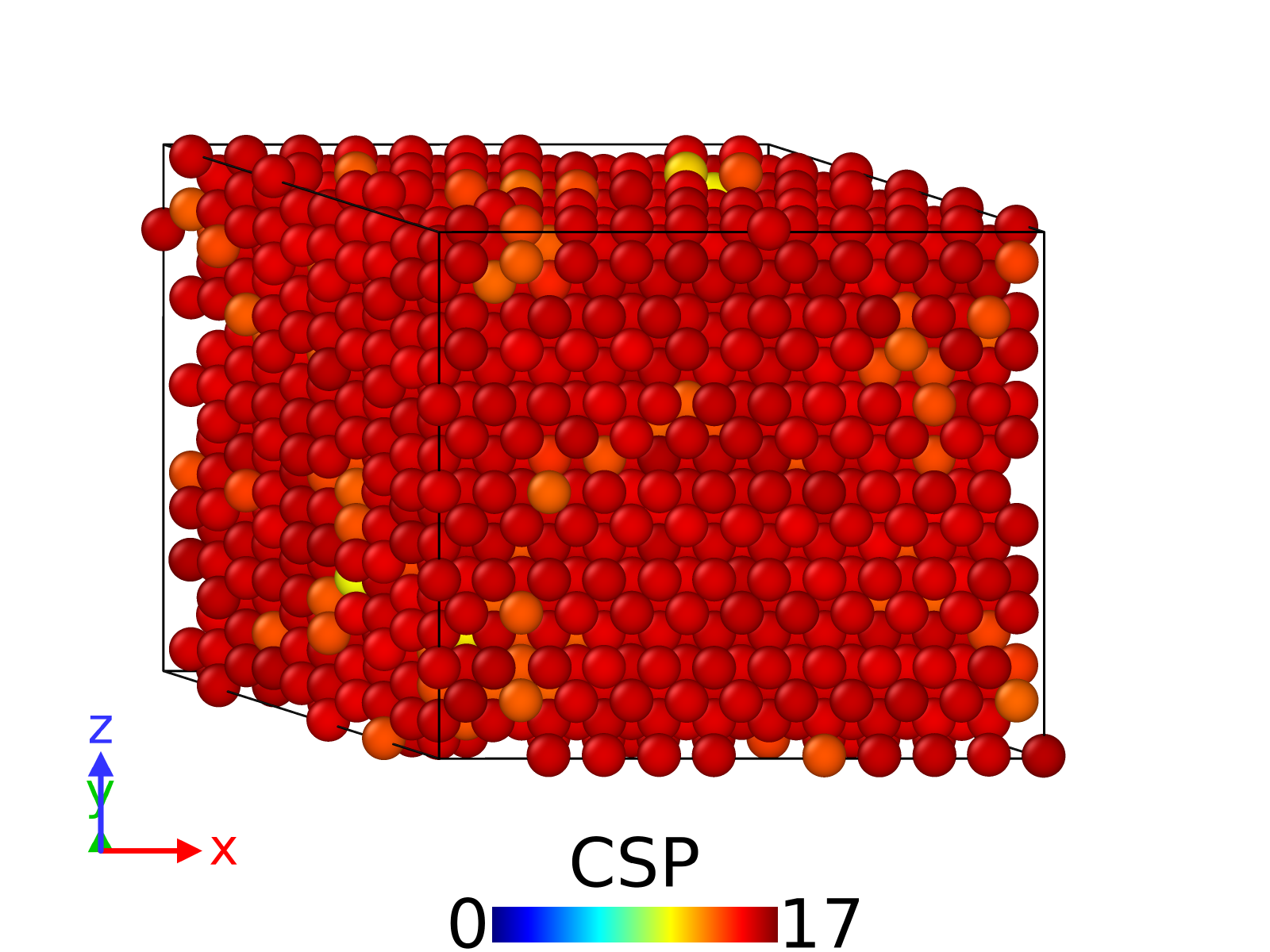}}
     \subfigure[$P$ = 9 GPa; CSP = 15]{\includegraphics[width=0.23\textwidth]{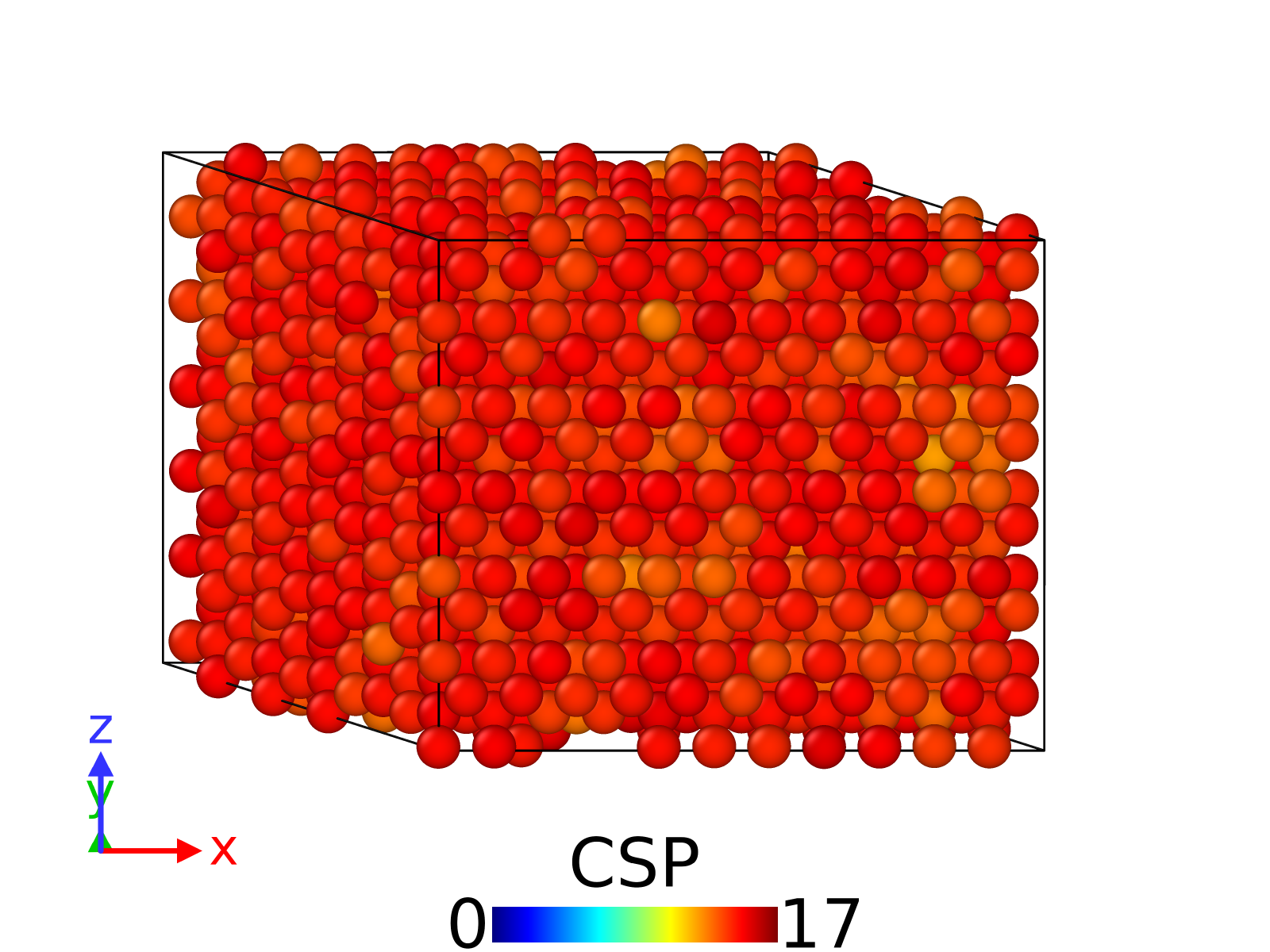}}
     \subfigure[$P$ = 10 GPa; CSP = 13]{\includegraphics[width=0.23\textwidth]{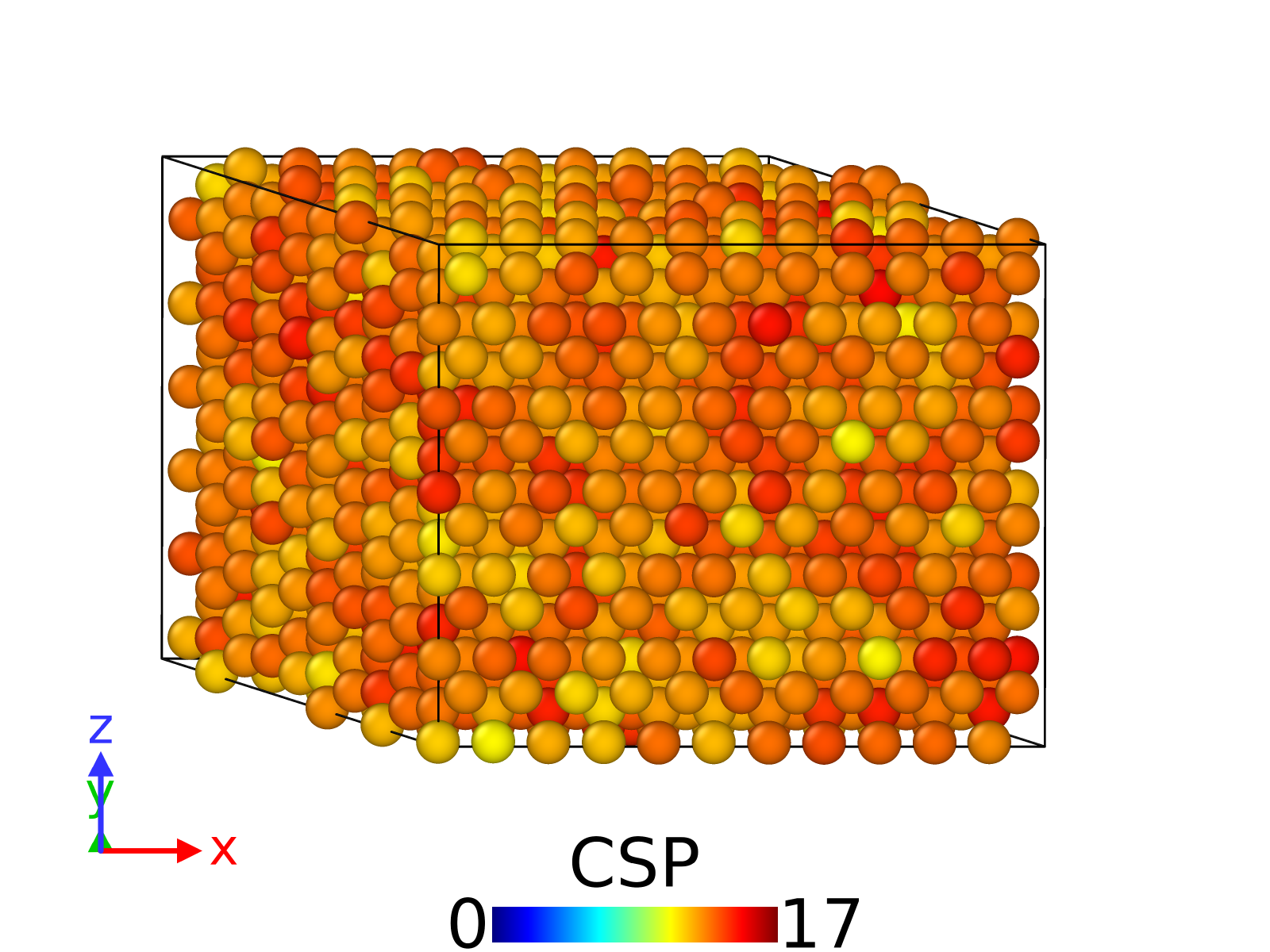}}
     \subfigure[$P$ = 11 GPa; CSP = 11]{\includegraphics[width=0.23\textwidth]{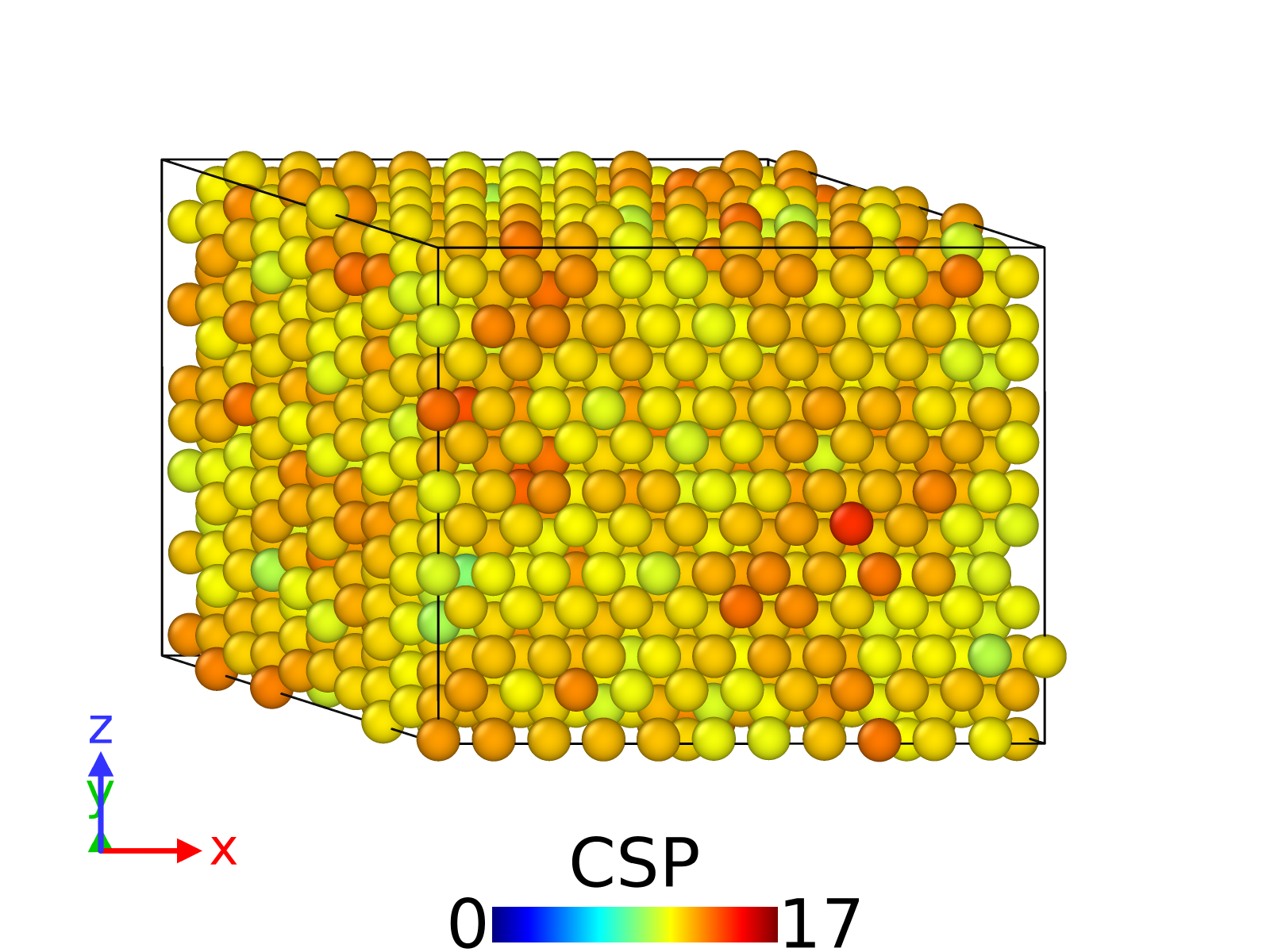}}
     \subfigure[$P$ = 13 GPa; CSP = 9]{\includegraphics[width=0.23\textwidth]{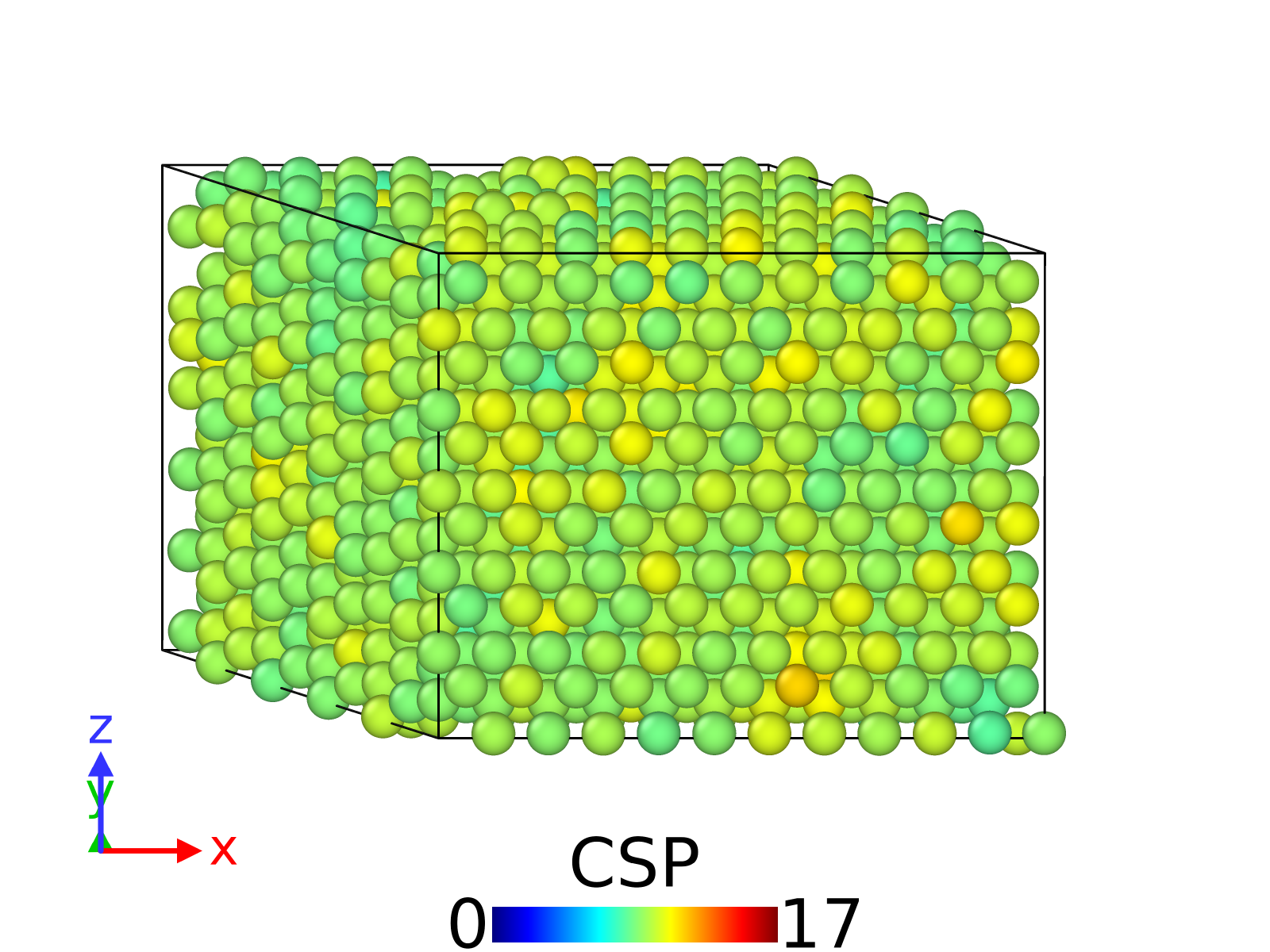}}
     \caption{CSP analysis at $T =$ 1600 K compressed from $P=$ 4 GPa to $P=$ 13 GPa, along with the average CSP per snapshot.}
     \label{fig:ovito_7_9_11_13_m}
 \end{figure}

The central symmetry parameter (CSP) analysis between calcium ions at $T =$ 1600 K at different pressures is shown in Fig. \ref{fig:ovito_4_5_m} and Fig. \ref{fig:ovito_7_9_11_13_m}. 
It is important to note that calcite and calcite IV exhibit typical rhombohedral trigonal symmetry characterized by high central symmetry. For such structures, the CSP parameter is known to be approximately zero, indicating strong central symmetry. On the other hand, phase Vb belongs to the P$2_1$/m space group and the monoclinic crystal system, which is characterized by a lack of central symmetry. Consequently, the CSP of phase Vb takes high values, reflecting the absence of central symmetry.

For our analysis, the medium sample was compressed at a rate of 0.2 GPa per ns from $P =$ 4 GPa to $P=$ 5 GPa. 

Four representative snapshots are presented in Fig. \ref{fig:ovito_4_5_m} to illustrate the variation in the average CSP around the transition from IV to Vb, which occurs at $P=$ 4.25 GPa. These snapshots are approximately 0.04 GPa apart, with the first snapshot being at $P \approx$ 4.19 GPa and the last at $P \approx$ 4.31 GPa. These snapshots do not show any nucleus formation but highlight a sudden change in the central symmetry at the transition pressure, with average CSP rising from 0 to 7 in the short pressure range of 4.19-4.31 GPa.

In Fig.\ref{fig:ovito_7_9_11_13_m}, snapshots at pressures from $P=$ 4 GPa to $P=$ 13 GPa, along with their average CSP, illustrate that after the sudden rise associated with the transition IV to Vb, the CSP gradually increases up to 15 at $P=$ 7 GPa and decreases to an average value of 9 at $P$ = 13 GPa.

\clearpage
\section{Domains and defects formation}\label{subsec:defect}

\begin{figure}[h!]
    \centering
     \includegraphics[width=0.4\textwidth]{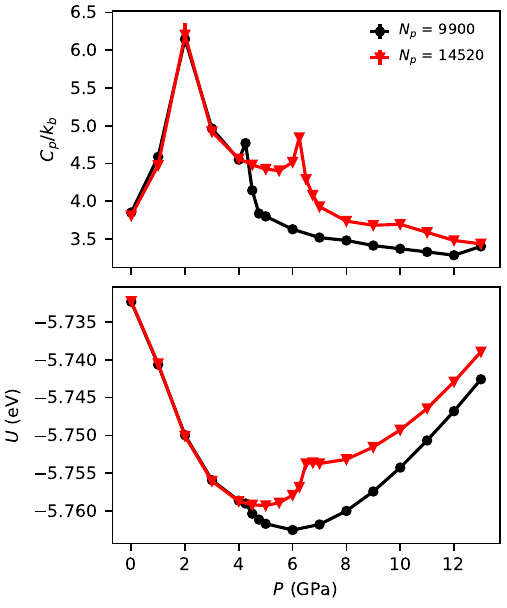}
    \caption{(a) $C_p$ and (b) $U$ plots for the medium (black line) and large (red line) samples at $T=$ 1600 K during the compression from $P=$ 0 Gpa to $P=$ 13 GPa.}
    \label{fig:cp_4_5b_with_l}
\end{figure}

As mentioned in Section~\ref{main-subsec:press} of the main article, discrepancies were observed in the pressure dependence of the heat capacity ($C_p$) for the sample with $N_p =$14520 particles,  starting at $P=$ 4.25 GPa, compared to the medium and small samples with $N_p =$3360 and 9900 particles, respectively. For clarity, only the $C_p$ and energy ($U$) for the large and medium samples are shown in Figs.~\ref{fig:cp_4_5b_with_l}(a) and (b). Fig.~\ref{fig:cp_4_5b_with_l}(a) shows that the two $C_p$ curves depart from each other at $P=$ 4.25 GPa, and that the peak for the larger sample (red curve) occurs at $P=$ 6.25 GPa instead of $P =$ 4.25 GPa as for the medium sample (black curve). Fig.~\ref{fig:cp_4_5b_with_l} (b) indicates that the large sample also exhibits excess energy starting from $P=$ 4.25 GPa compared to the medium sample. Further investigations have led us to link this excess of energy to the appearance of structural defects in the sample with $N_p =$14520 particles, referred to for this discussion as a ``large defective sample" or simply ``large sample". 

It is worth mentioning that no structural changes nor peaks in the heat capacity for the large sample around $P=$ 4.25 GPa were observed in several simulations lasting over 50 ns. This led us to conclude that the discrepancy between the large and the medium samples in the heat capacity is not due to a poor statistic or limited equilibration of our sample. 


\subsection{Central symmetry parameter analysis}

\begin{figure}[h!]
    \centering
    \subfigure[$P$ = 6,17 GPa]{\includegraphics[width=0.2\textwidth]{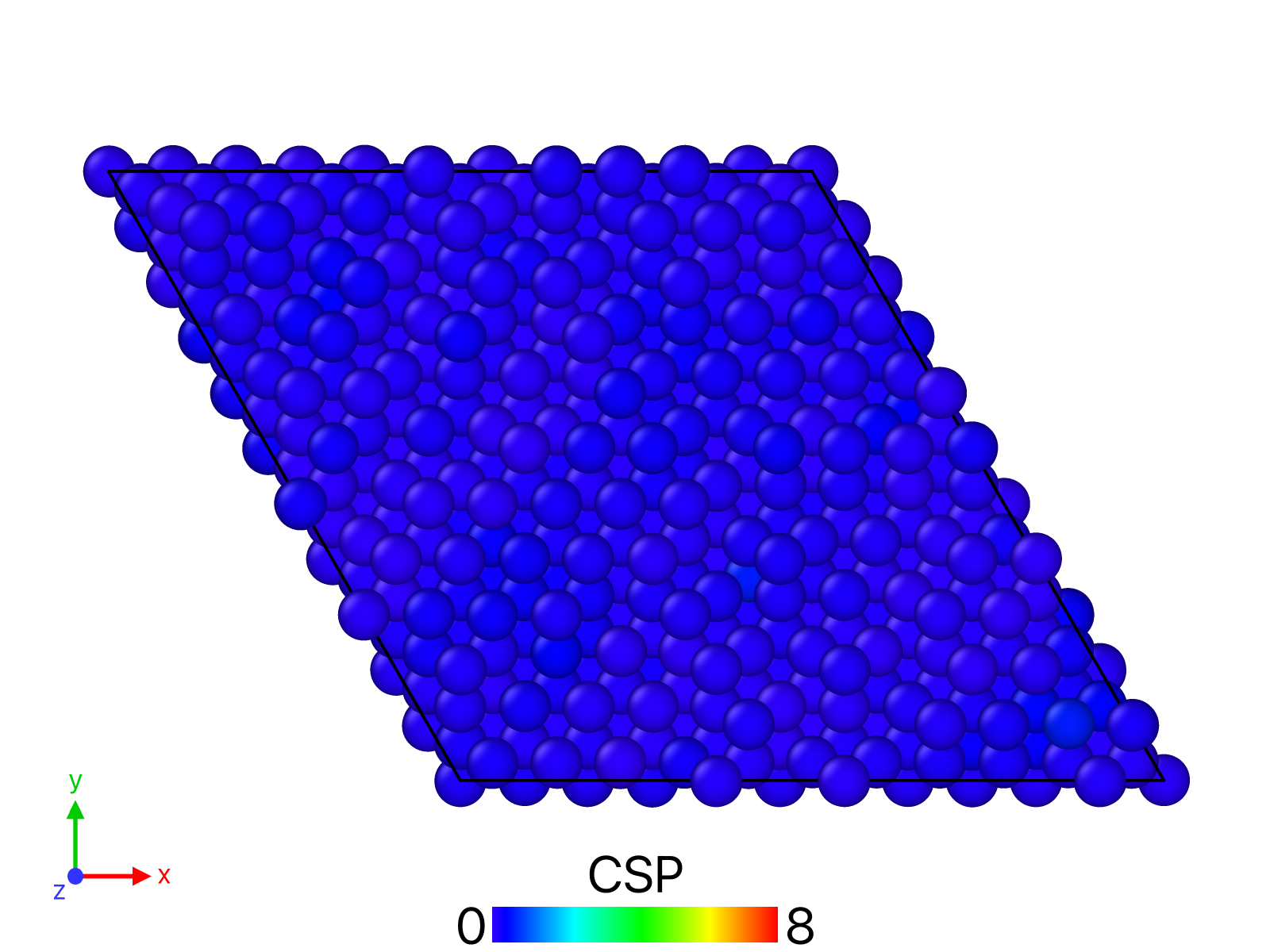}} 
    \subfigure[$P$ = 6,17 GPa]{\includegraphics[width=0.2\textwidth]{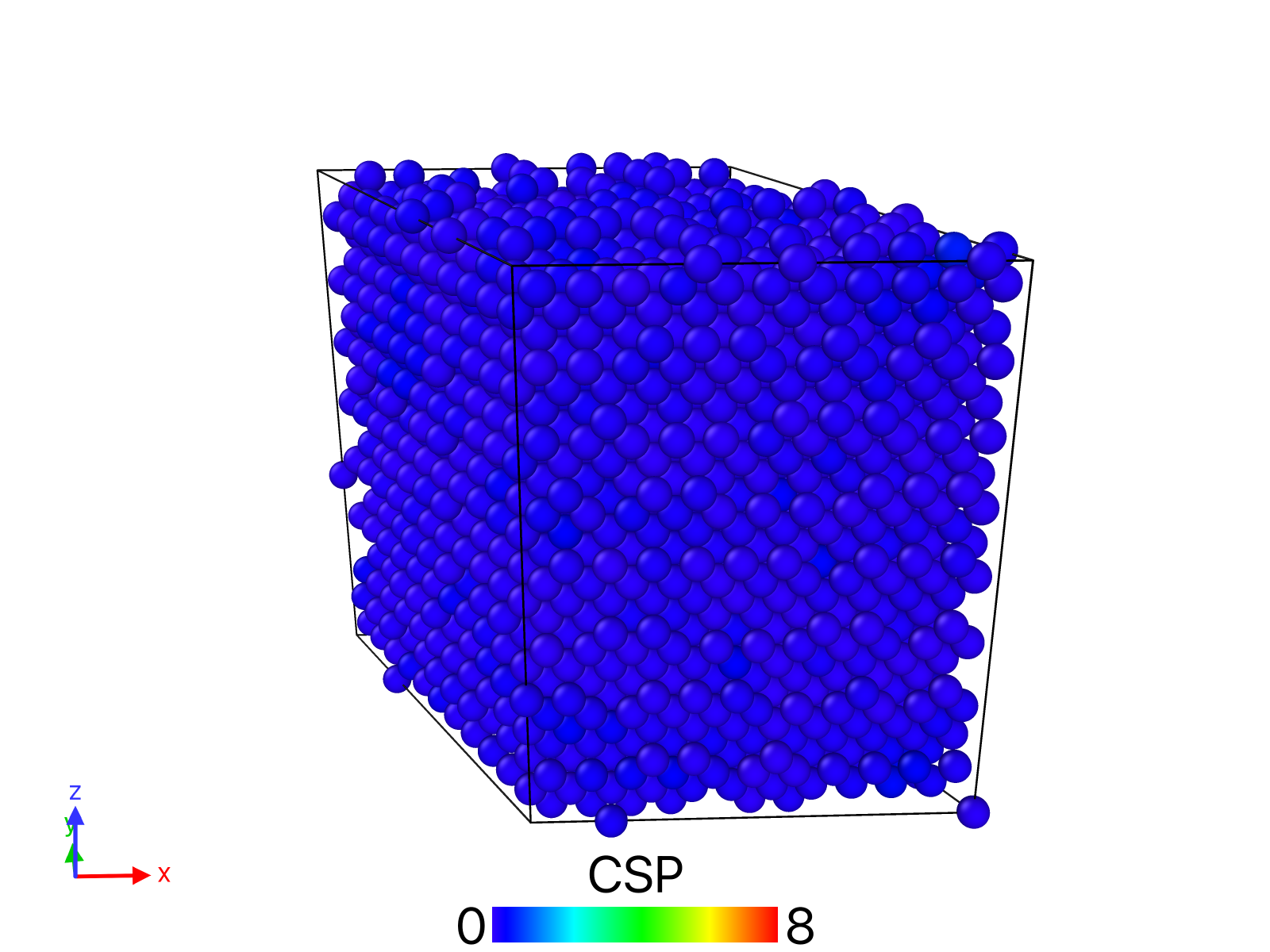}}    
    \subfigure[$P$ = 6,47 GPa]{\includegraphics[width=0.2\textwidth]{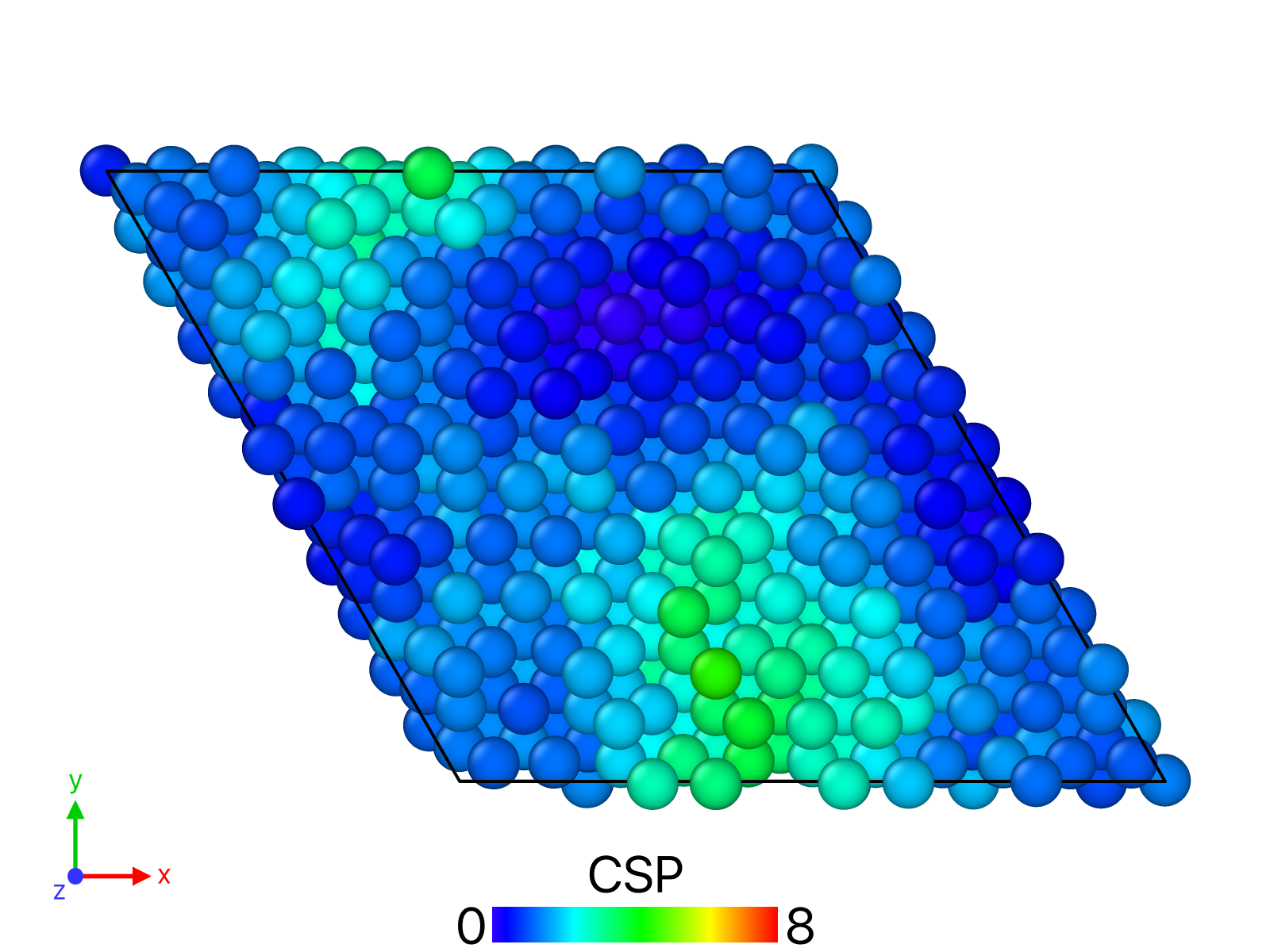}} 
    \subfigure[$P$ = 6,47 GPa]{\includegraphics[width=0.2\textwidth]{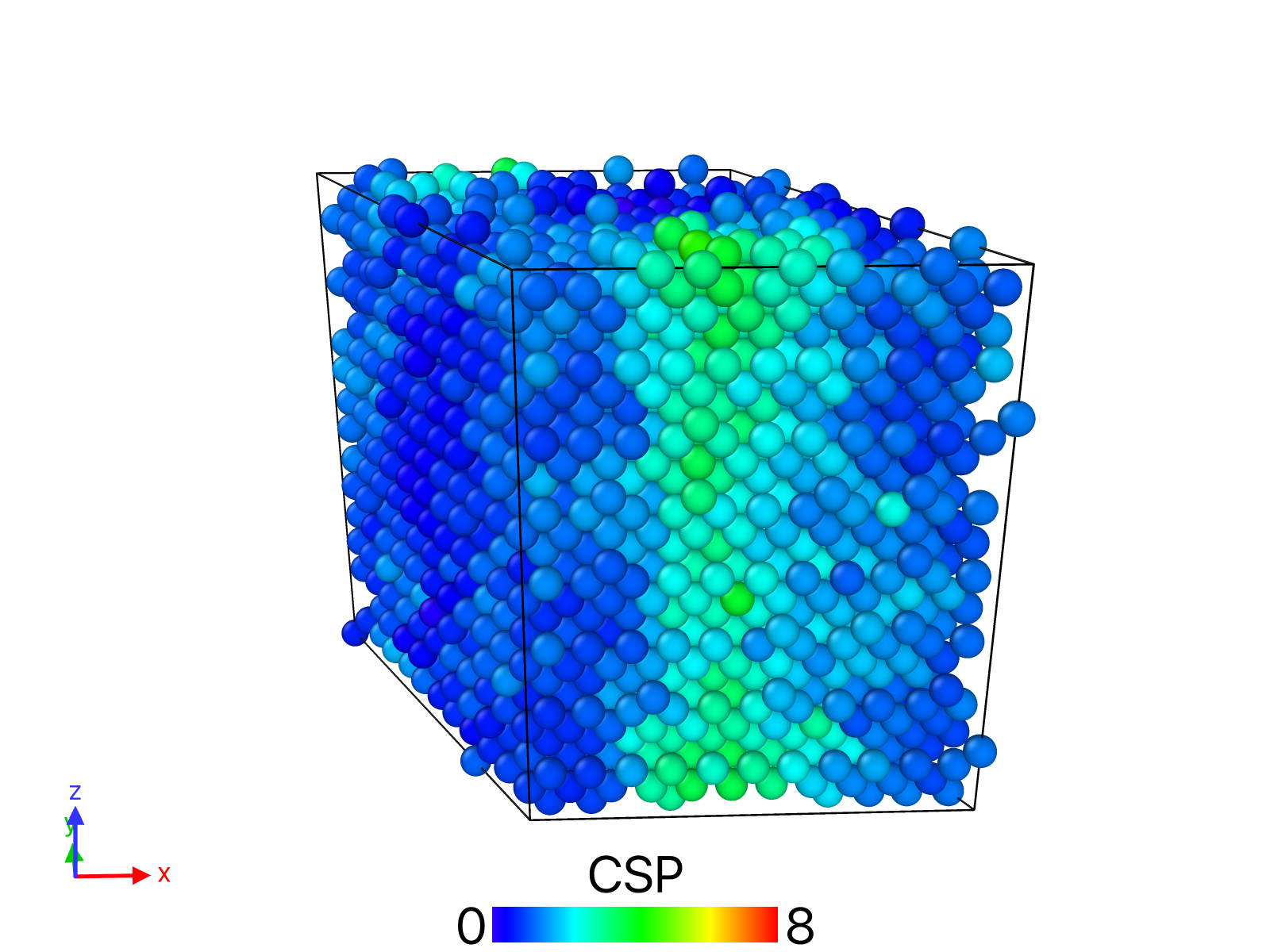}}
    \subfigure[$P$ = 6,77 GPa]{\includegraphics[width=0.2\textwidth]{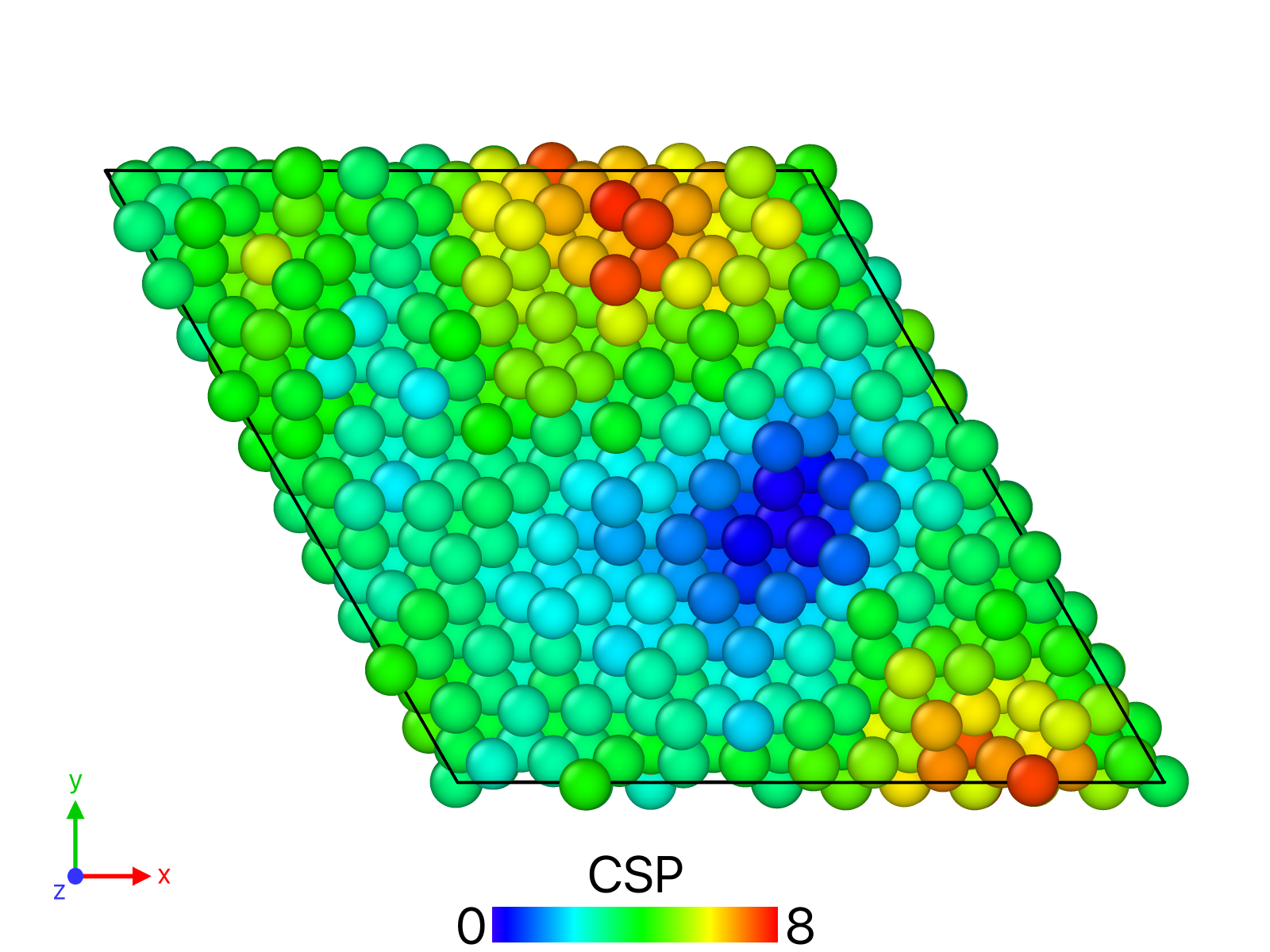}} 
    \subfigure[$P$=6,77 GPa]{\includegraphics[width=0.2\textwidth]{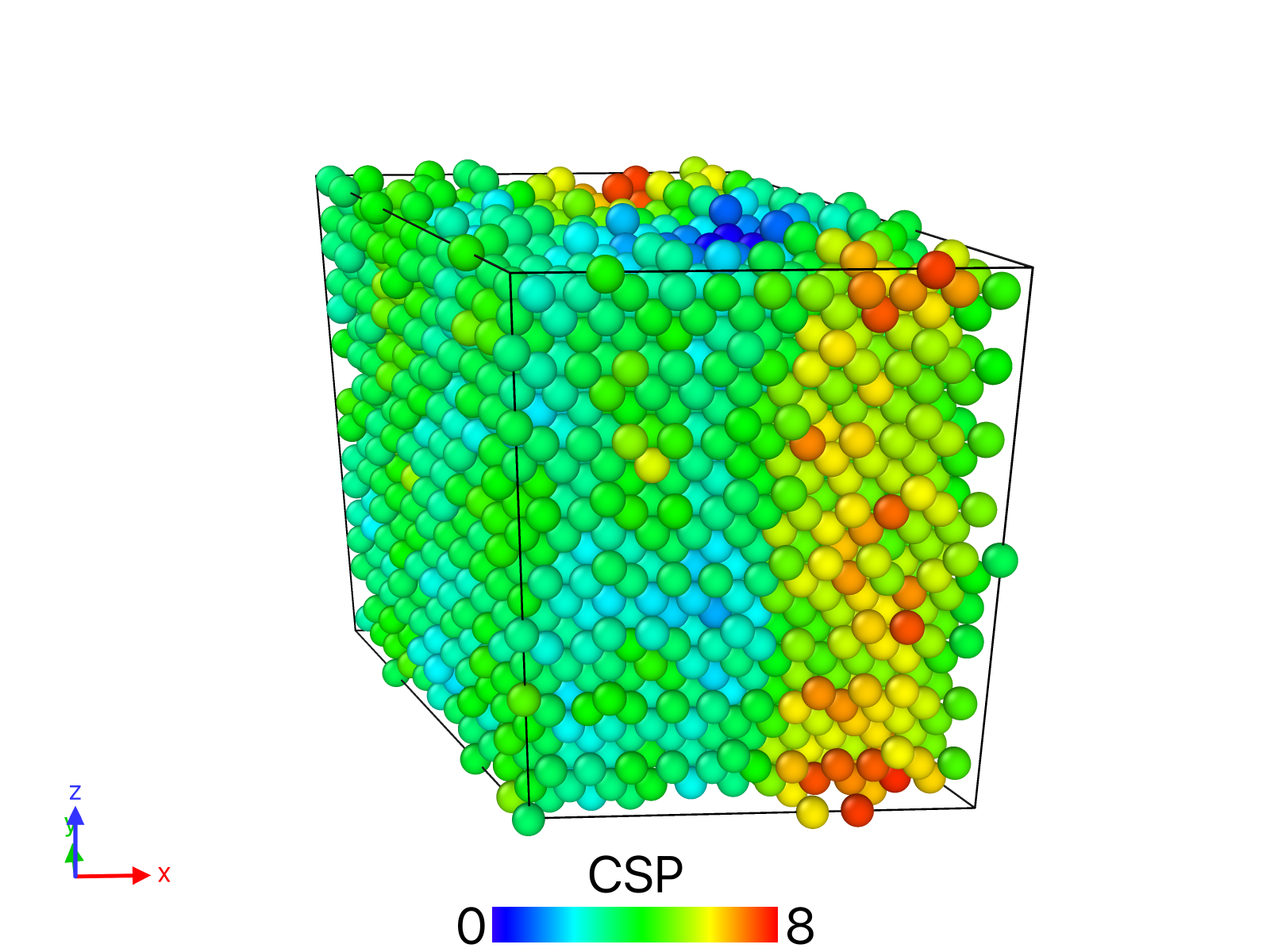}}
    \subfigure[$P$=7,00 GPa]{\includegraphics[width=0.2\textwidth]{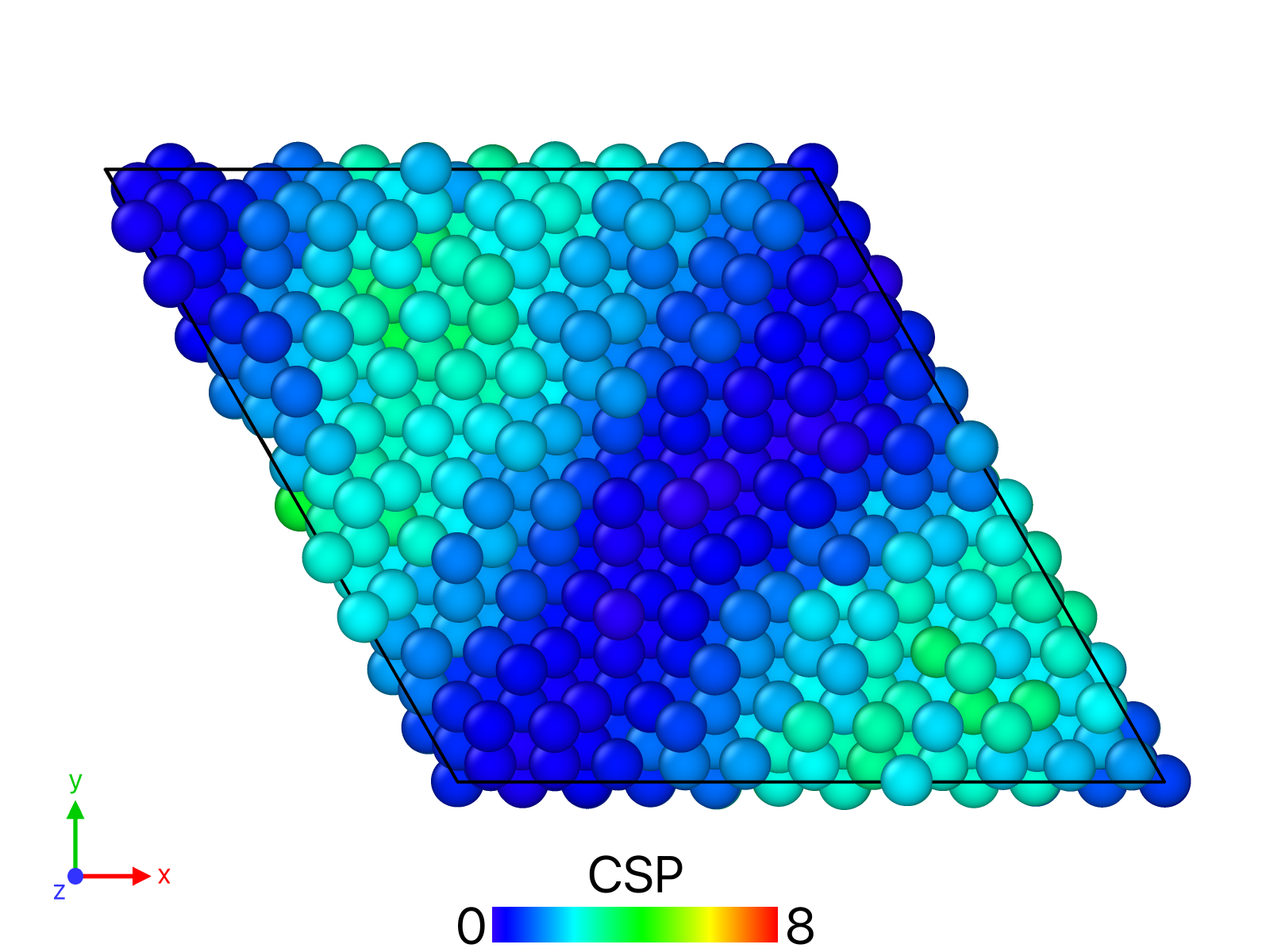}} 
    \subfigure[$P$=7,00 GPa]{\includegraphics[width=0.2\textwidth]{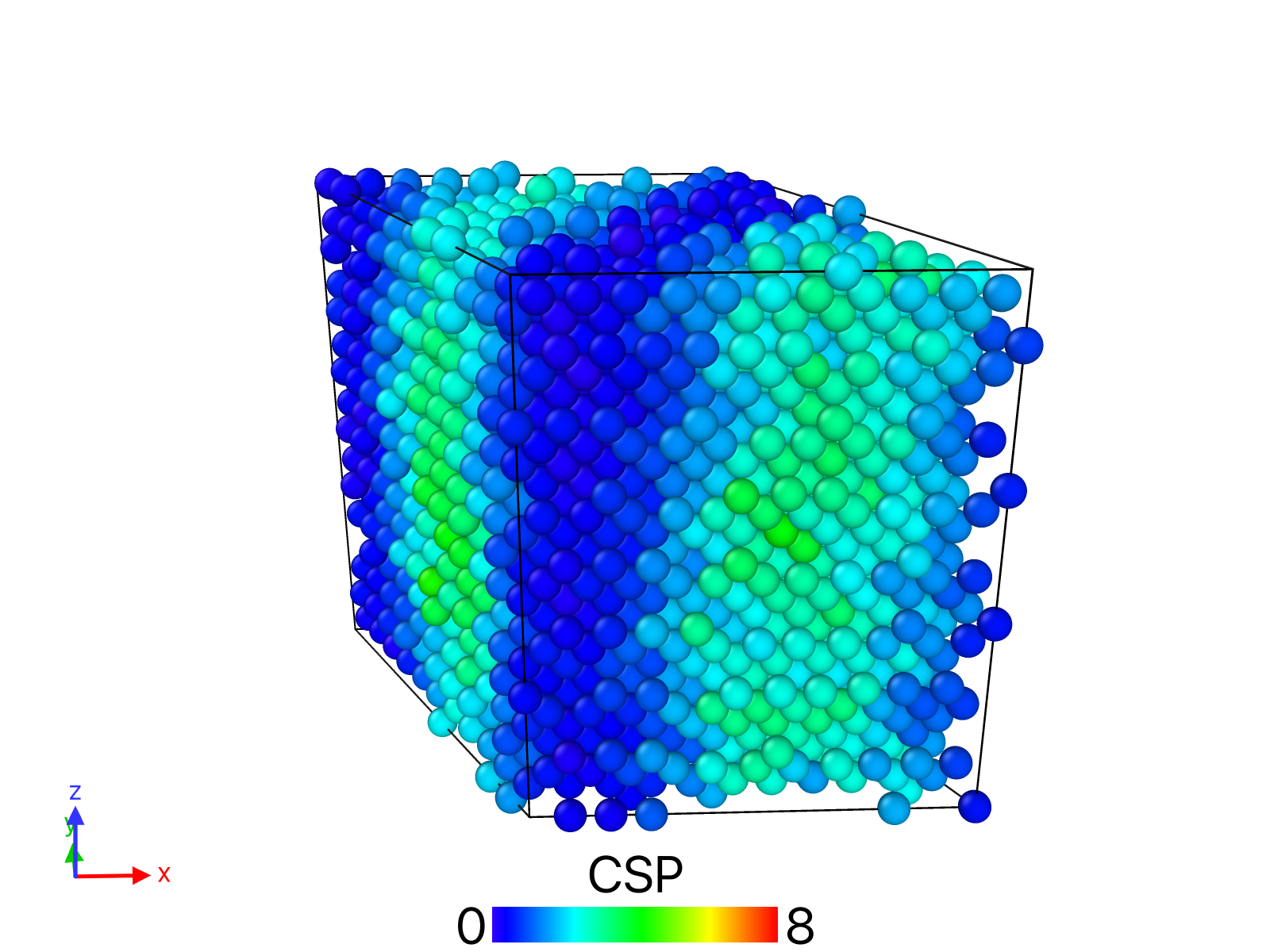}} 
    \caption{CSP analysis of four selected snapshots during the compression at $T =$ 1600 K from $P=$ 6 GPa to $P=$ 7 GPa at a rate of 0.2 GPa/ns. The starting configuration ($P \approx$ 6.17 GPa) with an average CPS of 0 corresponds to phase IV, the last configuration ($P=$ 7 GPa ) shows the coexistence of phases IV and Vb, and the intermediate configurations ($P=$ 6,47 and 6,77 GPa) reflect the CSP during the transition in between.}
    \label{fig:ovito_6_7_top}
\end{figure}

To investigate the structural changes occurring in the large sample, we calculated the centrosymmetry parameter (CSP) for the Ca$^{2+}$ ions at different pressures. Selected snapshots of the simulation during the compression from 6 GPa to 7 GPa are presented in Fig.~\ref{fig:ovito_6_7_top} in the side and top view at four different pressures. The analysis of the CSP for pressures $P <$ 6 GPa is not shown here because no significant variation from the zero value, typical of the calcite and phase IV symmetry, is measured. At $P=$ 6.17 GPa, snapshots in Fig.~\ref{fig:ovito_6_7_top}(a) (side view) and (b) (top view), right before the $C_p$ peak at $P =$6.25, 
the average CSP value fluctuates between 0 and 0.5. After $P =$6.25, (see snapshots (c)-(h)), we measure non-homogenous values of CSP within the same sample. In these snapshots, and within the same sample, we measure areas with CSP approximately equal to 0, indicating local regions with high central symmetry, and areas with CSP approximately equal to 7/8, indicating low central symmetry as seen in the Vb phase. Such heterogeneities in the CSP values of the same sample reveal that the $C_p$ peak at $P=$ 6.25 GPa is associated to the formation of domains with different symmetries. All the observed domains are perpendicular to the $ab$-plane, as shown in the snapshots. 

We have also observed three pressure regimes for the appearance of the domains. The domains fluctuate in the pressure range 6.25 GPa $\leq P \leq $ 7 GPa, where regions with CSP of approximately 0 and regions with CSP of approximately 7/8 coexist in the same sample (see snapshots (c)-(h) in Fig. \ref{fig:ovito_6_7_top})). 

\begin{figure}[h!]
    \centering
    \subfigure[$P =$ 8 GPa]{\includegraphics[width=0.2\textwidth]{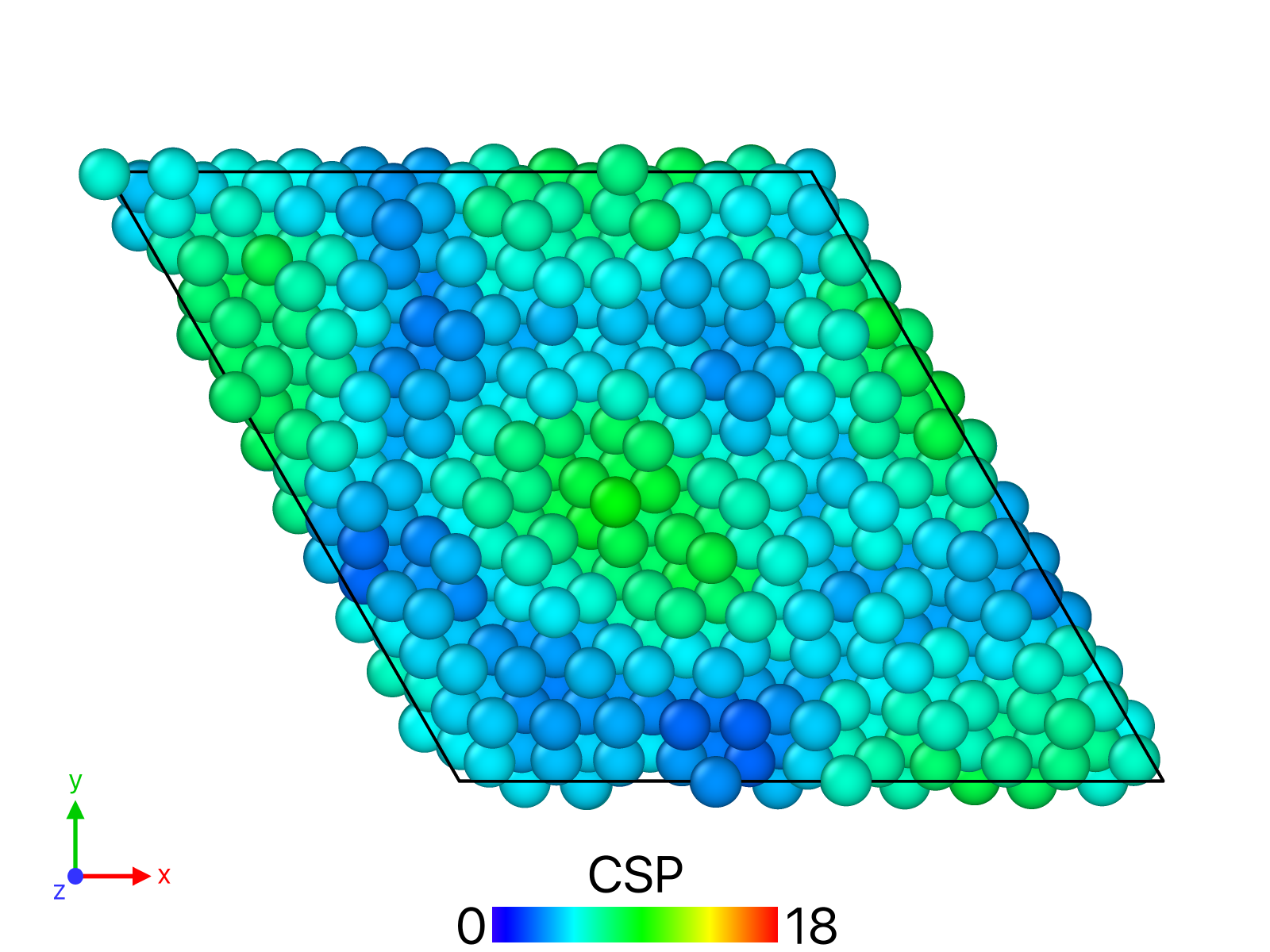}}
    \subfigure[$P =$ 8 GPa]{\includegraphics[width=0.2\textwidth]{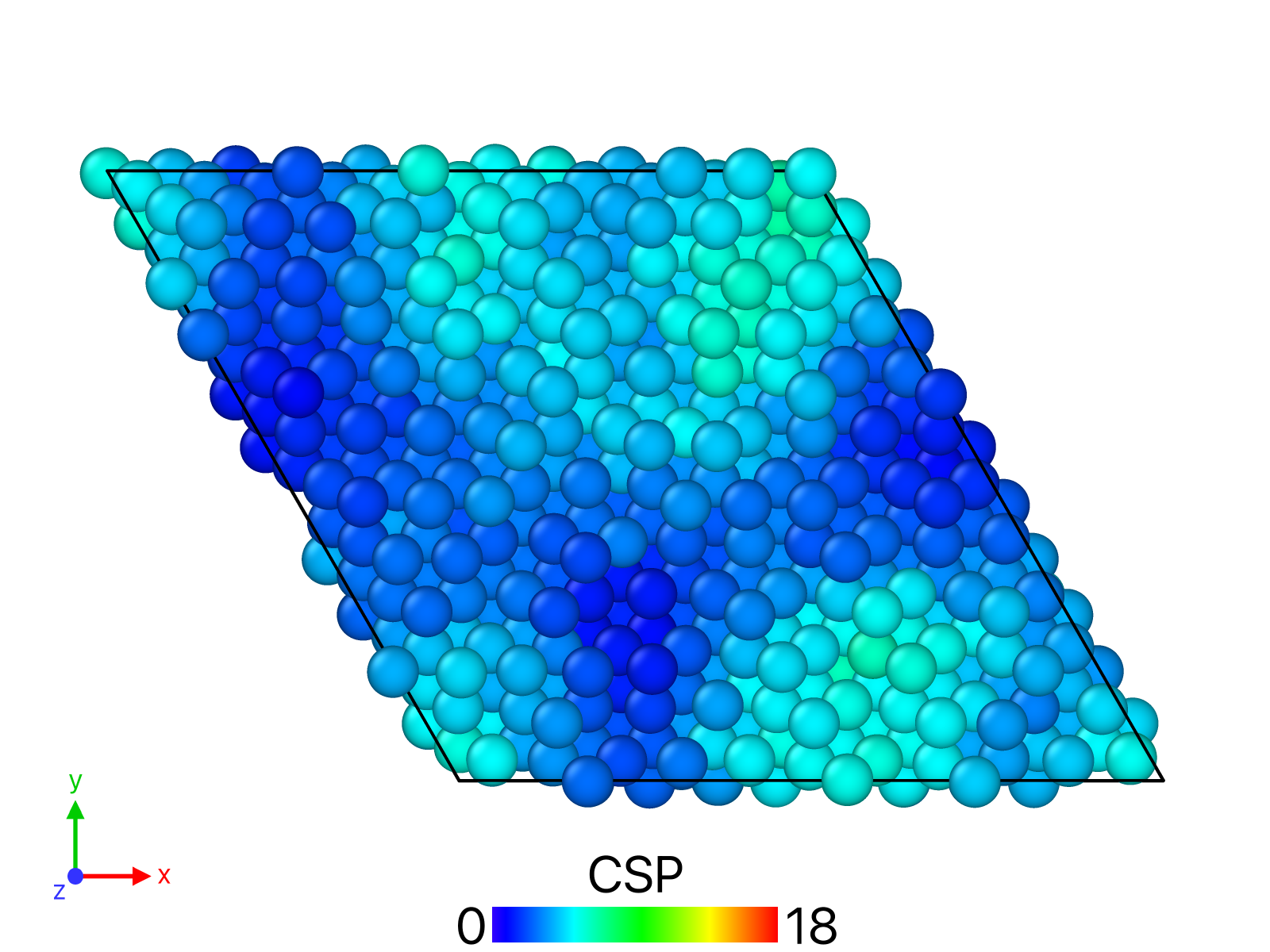}} 
    \subfigure[$P =$ 10 GPa]{\includegraphics[width=0.2\textwidth]{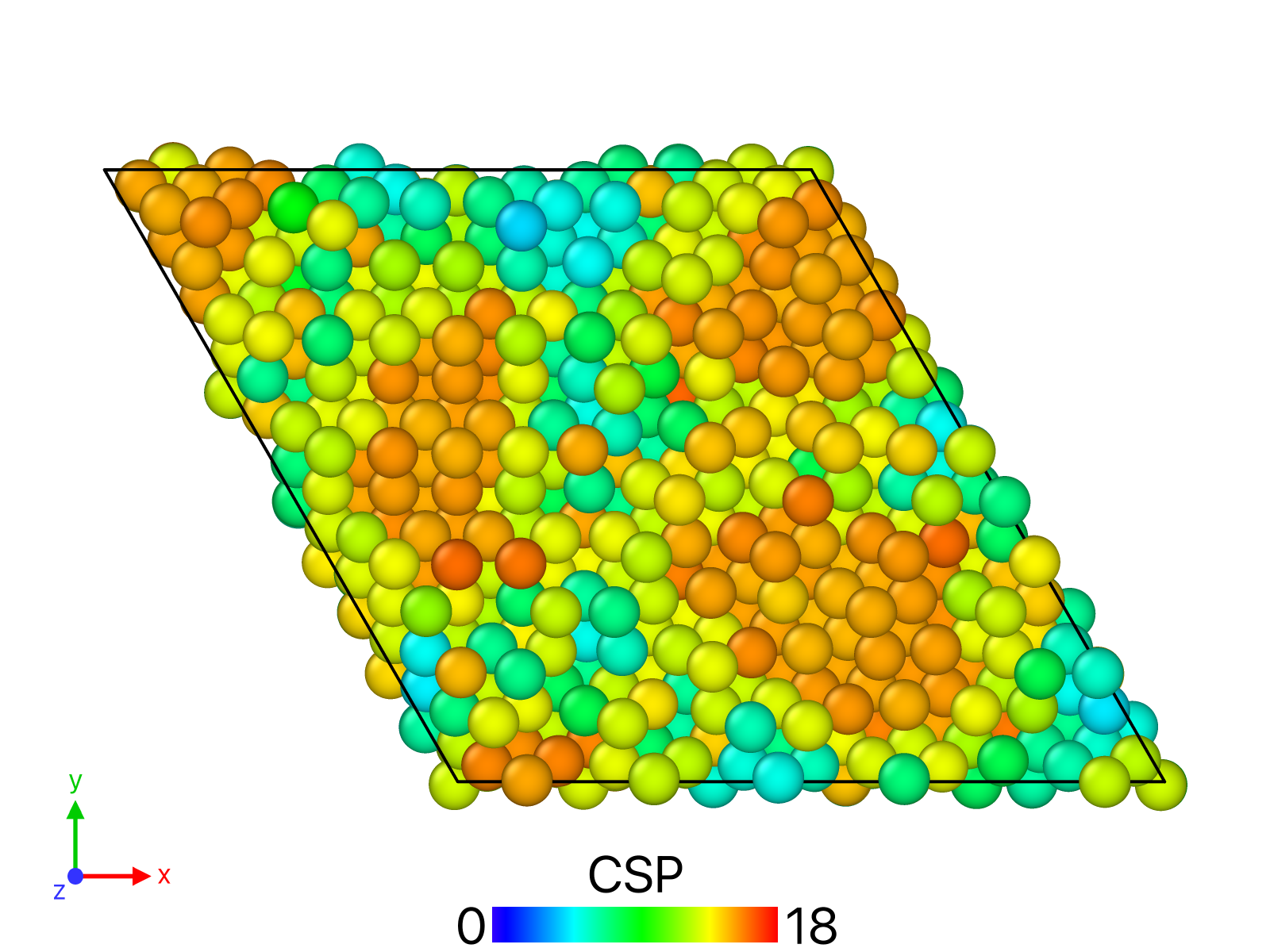}}
    \subfigure[$P =$ 10 GPa]{\includegraphics[width=0.2\textwidth]{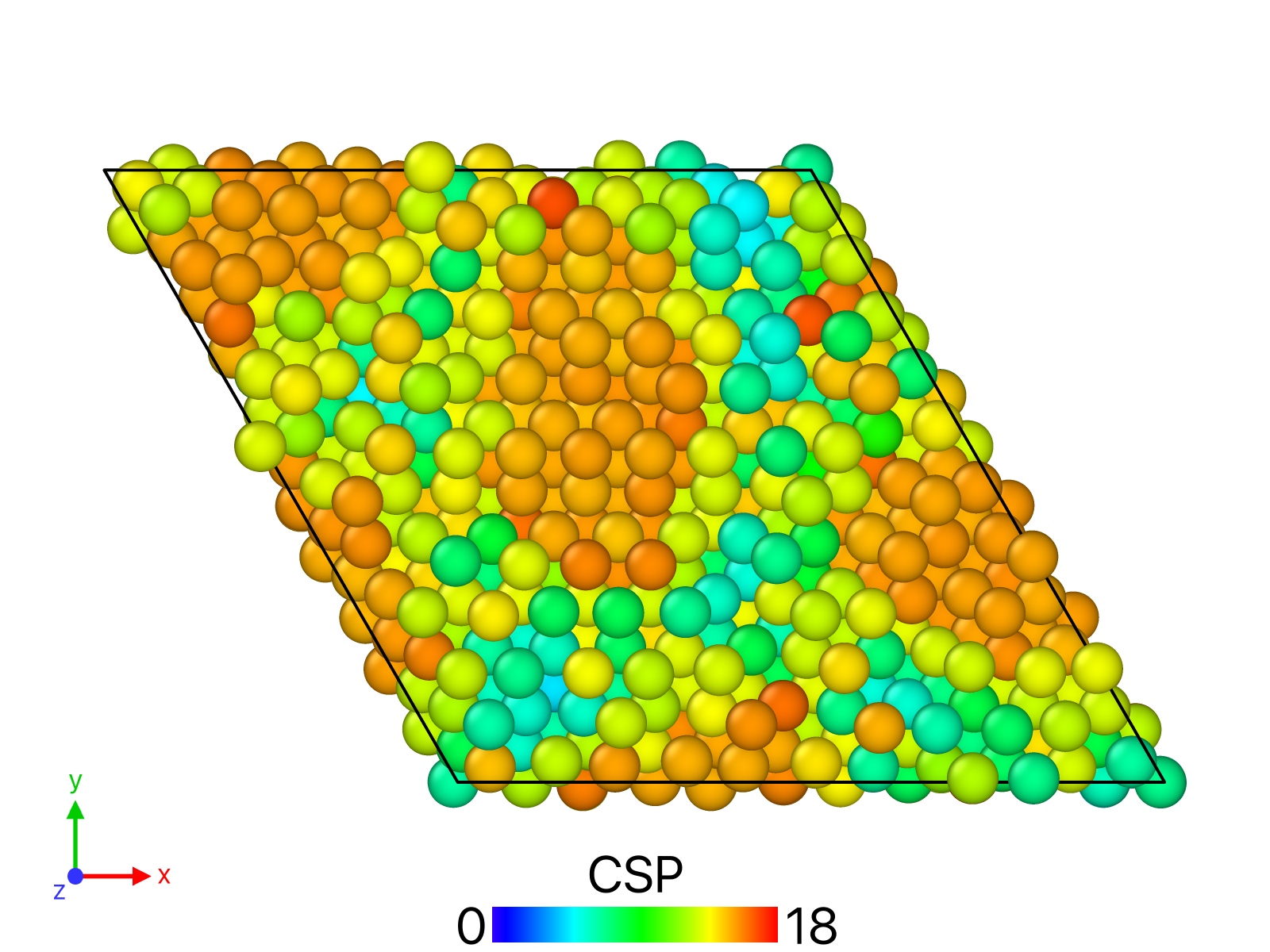}}
    \caption{Snapshots from the simulation of the large-defective sample at pressure 10 GPa $\geq P \geq$ 8 GPa as shown in the legend. In this pressure range, rearranging domains are observed.}
    \label{fig:ovito_8_10}
\end{figure}

\begin{figure}[h!]
    \centering
    \subfigure[$P$ = 11 GPa]{\includegraphics[width=0.2\textwidth]{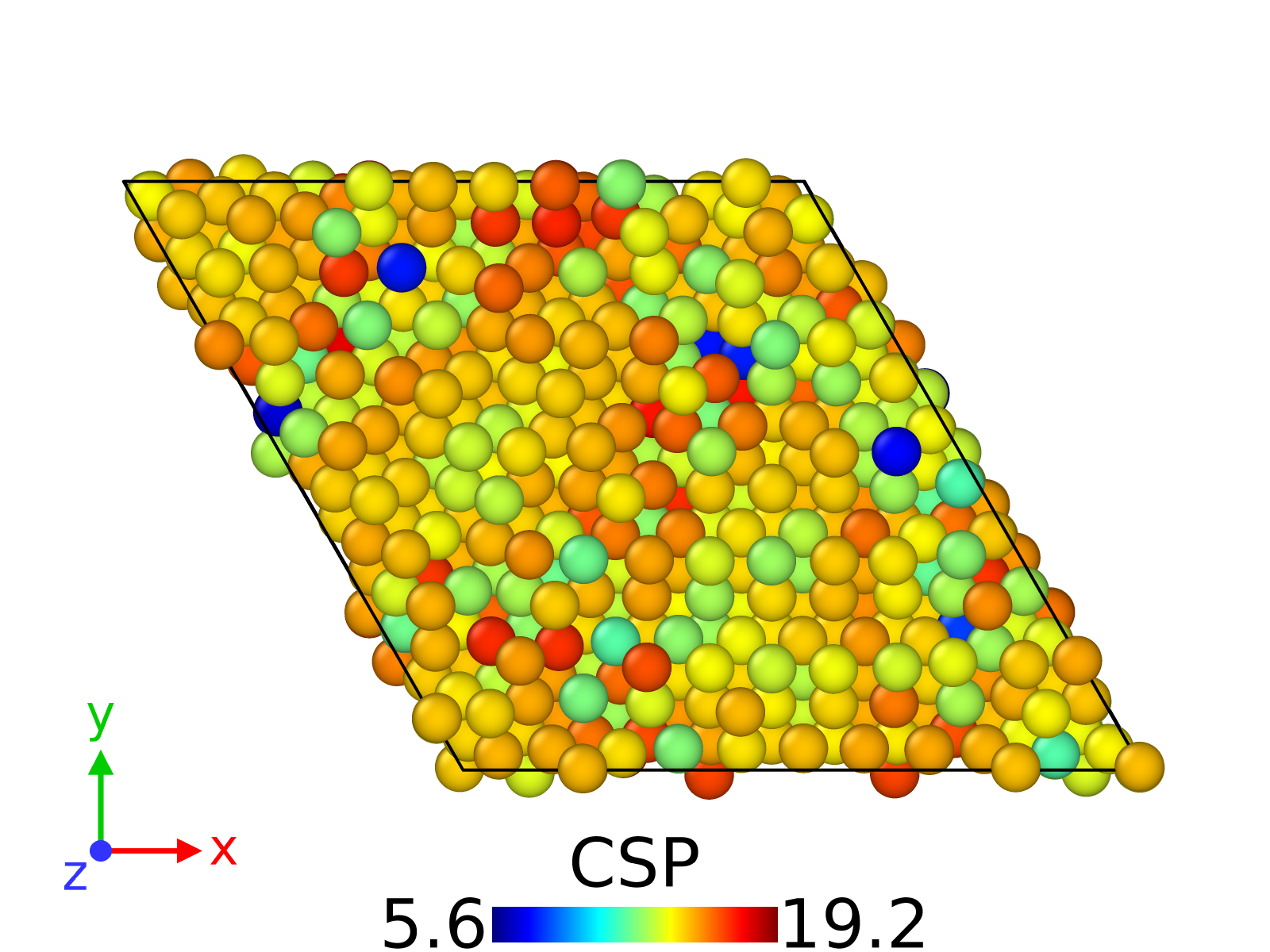}}
    \subfigure[$P$ = 12 GPa]{\includegraphics[width=0.2\textwidth]{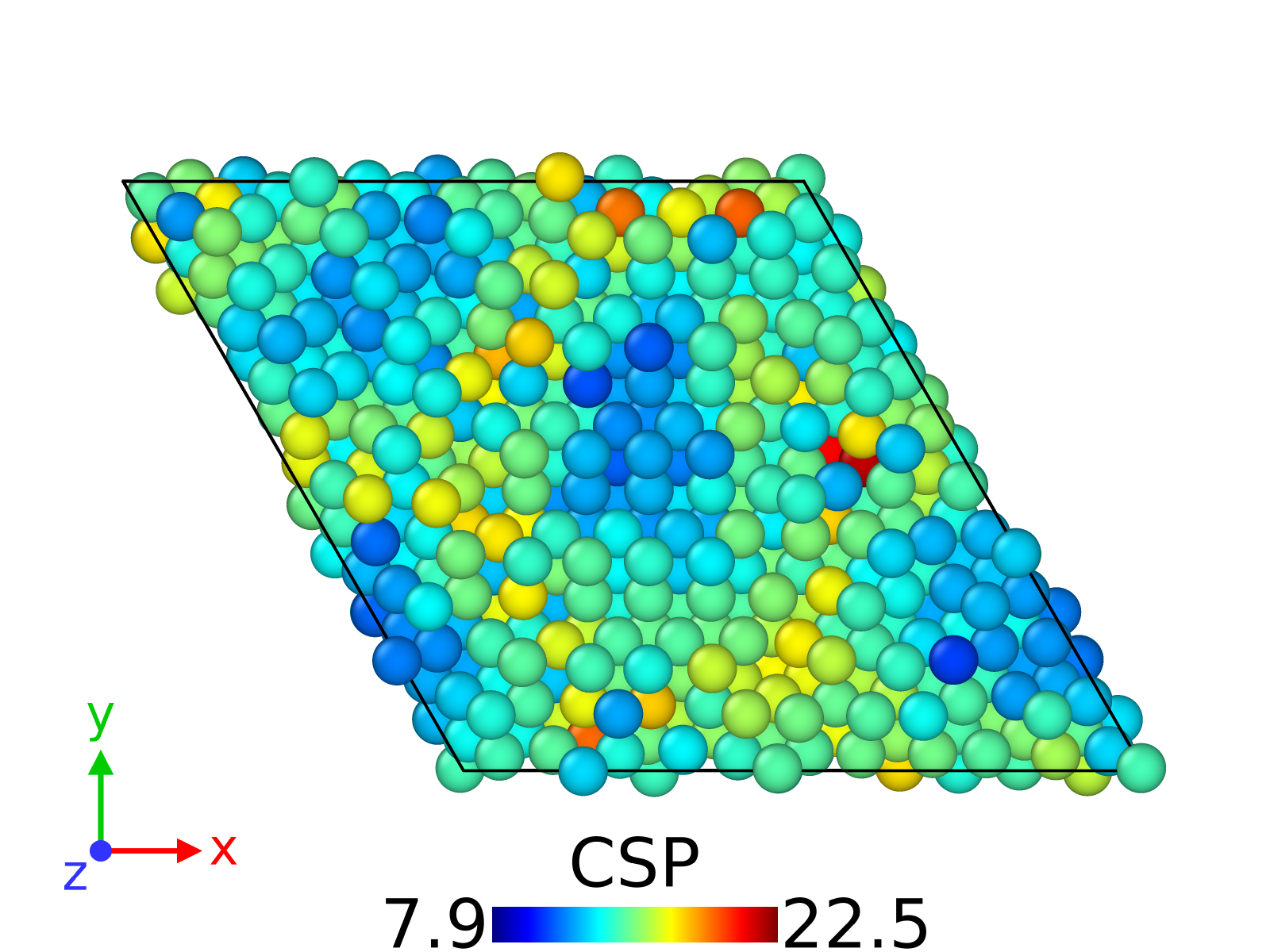}}
    \subfigure[$P$ = 13 GPa]{\includegraphics[width=0.2\textwidth]{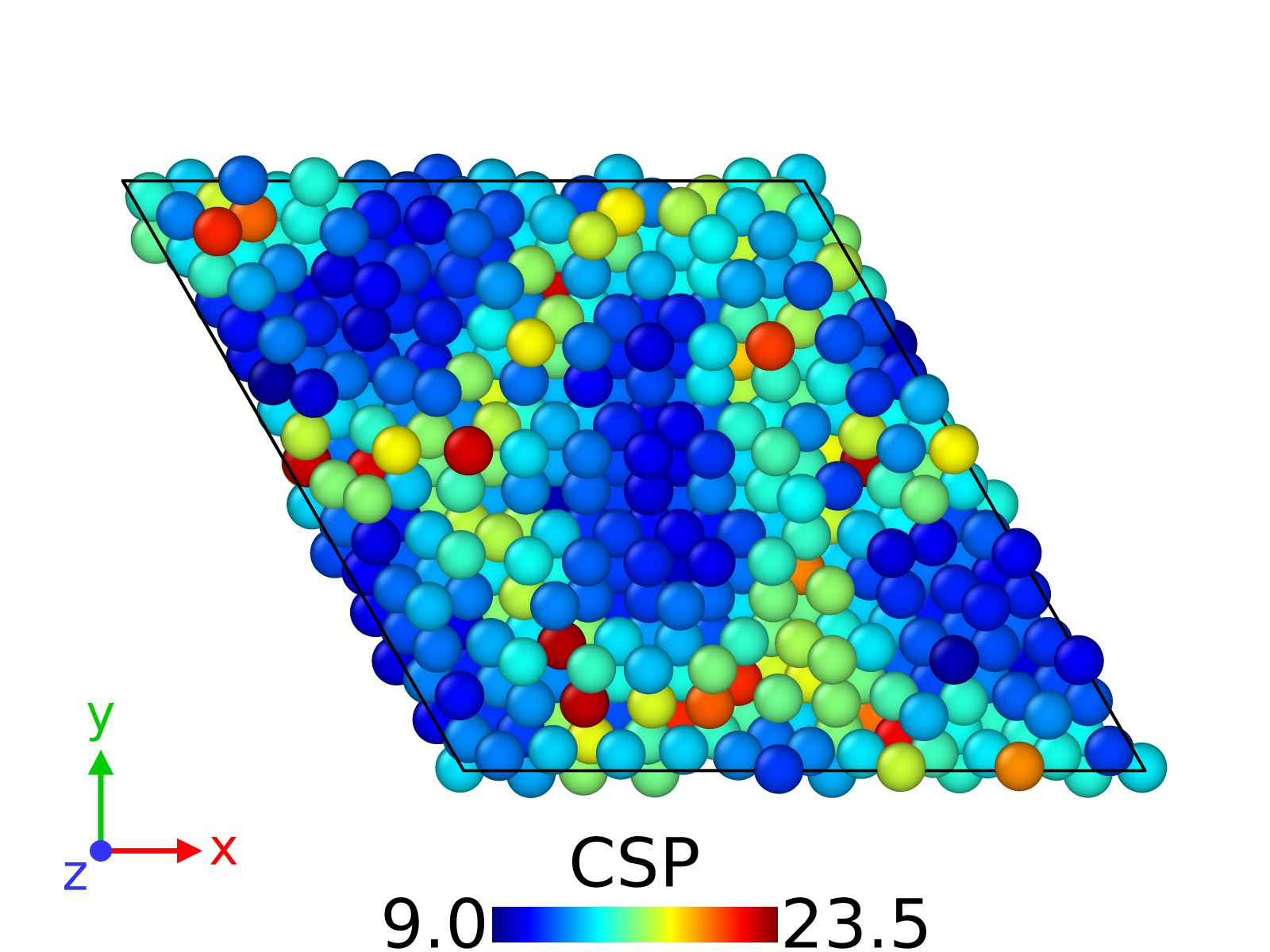}}
    \caption{Snapshots from the simulation of the large-defective sample at pressure  $P \geq$ 11 GPa as shown in the legend. Defects are stable and limited by grain boundaries.} 
    \label{fig:ovito_11_13}
\end{figure}

In the range between 8 GPa and 11 GPa, the domains rearrange but never completely disappear. Additionally, the regions with CSP approximately equal to 0 tend to shrink, and defects with local regions with very high CSP appear(see snapshots (a)-(d) in Fig.~\ref{fig:ovito_8_10})). For pressures  $P>$ 11 GPa, the domains become stable, do not rearrange, and appear to be delimited by grain boundaries or defects. In this regime regions characterized by CSP of approximately 0 are absent, and only regions with CSP higher than 7 are present (see snapshots (a)-(c) in Fig.~\ref{fig:ovito_11_13})).

\subsection{Pair distribution functions}

\begin{figure}[h!]
    \centering
    \subfigure[Ca - Ca]{\includegraphics[width=0.46\textwidth]{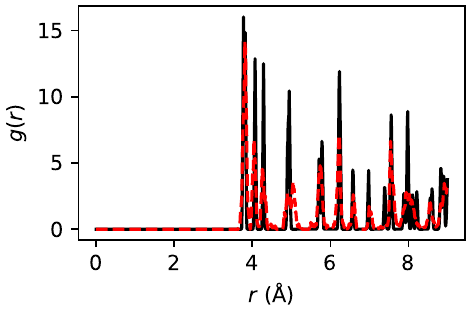}}
    \subfigure[C - C]{\includegraphics[width=0.46\textwidth]{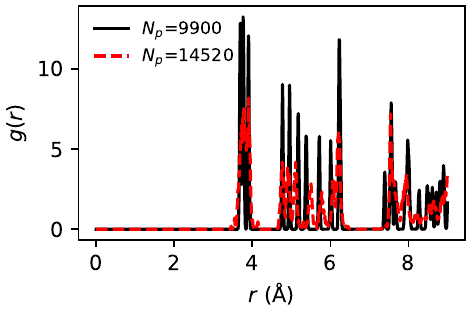}}
    \caption{PDF Ca-Ca (a) and C-C (b) inside the domains in the large sample and the middle sample at P=13 GPa. All the plots were calculated using ovito software.}
    \label{fig:rdf_ml}
\end{figure}

As seen by the CSP snapshot, at very high pressure, i.e. $P=$ 13 GPa in Fig. \ref{fig:ovito_11_13}(c), the large sample appears to be constituted by stable domains delimited by grain boundaries. At this stage, the PDFs for Ca-Ca and C-C compared with the PDF in the medium sample at the same pressure appear very similar, as presented in Fig.~\ref{fig:rdf_ml}. Picks at the same distances and of similar heights are visible, attesting that, as the pressure rises, the differences due to the domain formation tend to disappear.

\subsection{Coordination number analysis}

\begin{figure}[h!]
    \centering
    \subfigure[C-Ca]{\includegraphics[width=0.32\textwidth]{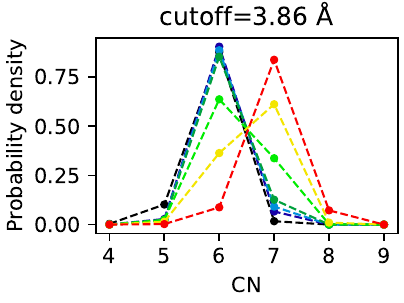}}
    \subfigure[Ca-Ca]{\includegraphics[width=0.32\textwidth]{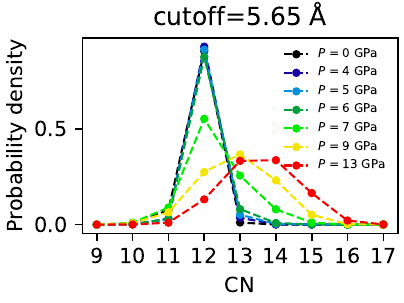}}
    \subfigure[C-C]{\includegraphics[width=0.32\textwidth]{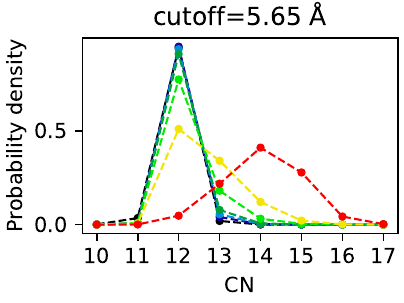}}
    \caption{CN at different pressure between different types of atoms for the large sample.}
    \label{fig:cn_5b_l}
\end{figure}

The study of the CN probability density shown in Fig.~\ref{fig:cn_5b_l} also supports the idea that the second peak for the larger sample in the $C_p$ of Fig.~\ref{fig:cp_4_5b_with_l} (a)  corresponds to the structural transition from phase IV to phase Vb in our model.  
Fig.~\ref{fig:cn_5b_l} (a) illustrates the CN probability density for the first coordination shell, while Figs.~\ref{fig:cn_5b_l} (b) and ~\ref{fig:cn_5b_l} (c) represent the second coordination shell. These graphs are similar to the corresponding graphs for the CN probability density of the medium sample presented in Fig.~\ref{fig:cn_5b_m}.
In Fig. ~\ref{fig:cn_5b_l}(a), the pressure drop of the maximum CN probability density for C-Ca at CN = 6 appears between $P=$ 6 GPa and $P=$ 7 GPa, rather than between $P=$ 4 GPa and $P=$ 5 GPa as observed for the medium sample (see Fig.\ref{fig:cn_5b_m_cca}). The maximum of the probability density shifts to CN = 7 for $P\geq$ 9 GPa and reaches a very high value at P = 13 GPa.  

In Fig. ~\ref{fig:cn_5b_l}(b), the CN probability density drops
abruptly between $P=$ 6 GPa and $P=$ 7 GPa at CN = 12, rather than between $P=$ 4 GPa and $P=$ 5 GPa as observed for the medium sample (see Fig.\ref{fig:cn_5b_m_caca}). Finally, the moderate drop of the CN probability density at CN = 12 between $P=$ 6 GPa and $P=$ 7 GPa in Fig. ~\ref{fig:cn_5b_l}(b) occurs between $P=$ 4 GPa and $P=$ 5 GPa in Fig.~\ref{fig:cn_5b_m_cc} for the medium sample. As the pressure increases, the same higher CNs become more populated also for the larger sample as in Fig.~\ref{fig:cn_5b_m} for the medium sample.   

\subsection{\texorpdfstring{$\bar {q_6}$}{} order parameter}

\begin{figure}[h!]
    \centering
    \includegraphics[width=0.45\textwidth]{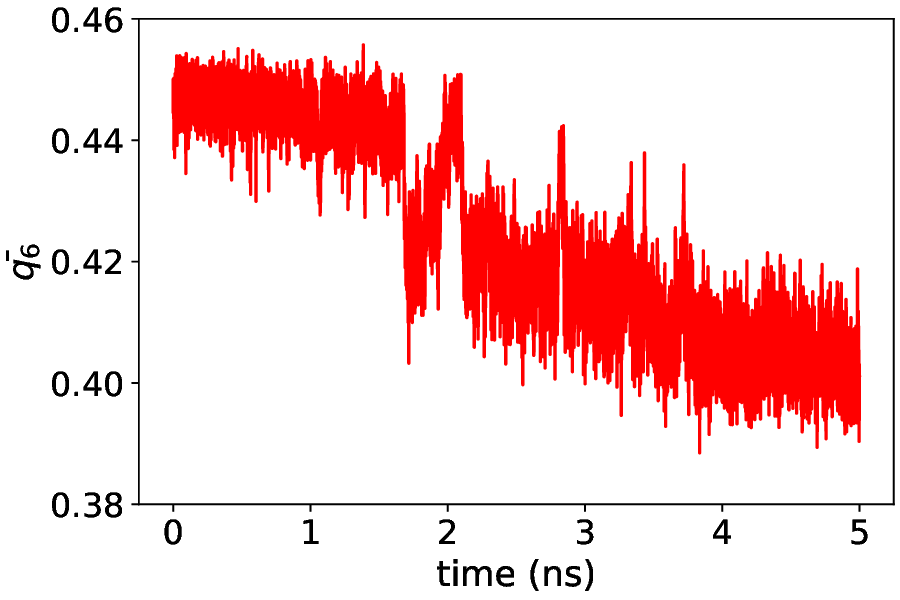}
    \caption{$\bar{q_6}$ (Ca-Ca) parameter in the large sample between $P=$ 6 and 7 GPa.}
    \label{fig:q6_t_l}
\end{figure}

Inspired by the structural investigations of the phase IV and Vb in the medium sample, we present in Fig.~\ref{fig:q6_t_l} the time evolution of the $\bar{q_6}$ (Ca-Ca) parameter for the large sample during the compression from $P=$ 6 GPa to $P=$ 7 GPa. The $\bar{q_6}$ parameter decreases continuously in the first 1.5 ns from approximately 0.46 to 0.43, values consistent with the one observed for phase IV (see Fig. \ref{main-fig:Vb} (d) and Fig. \ref{main-fig:F_5b} of the main text). A sudden drop of the order parameter suggests the transient formation of structures with order parameters close to the value characteristic of the Vb structure, i.e. $\bar{q_6}$ = 0.41. Around 2 ns, the curve fluctuates and rises to values typical of the IV structure again. After 2.3 ns the curve shows an overall decreasing trend with very large fluctuations. This irregular behavior and the significant fluctuations of the order parameter reflect the formation of transient regions with IV and Vb characteristics within the same sample during the time evolution. These observations confirm the coexistence of IV and Vb structures on a short-lived basis, with their appearance and disappearance contributing to the observed fluctuations.

\newpage
\subsection{Commensurability in the IV to V transition}

Studying structural transformation using molecular modeling is challenging due to the intrinsic microscopic scales of the simulation cell. Larger samples tend to capture better long-range interactions that may be essential to describing the formation of the new phase. However, the right dimension is not known in advance, nor the ideal geometries of the simulation cells. Numerous simulations with nine different sample sizes have been performed to ensure that the pressure-driven transitions investigated were not affected by the finite size of the samples. Our preliminary studies support that the mismatch in the $C_p$ between small, medium, and large samples is not a ``size effect". 

All the systems were heated to $T$ = 1600 K and then compressed up to $P$ = 10 GPa with steps 2 GPa or 1 GPa around the transition pressure. Among all those cases, several samples with a total number of particles greater than the ones in the large domain sample did not show any domain/defect formation or the mismatch of the $C_p$ graph at the pressure transition. To complete our investigations, we mentioned in the main text a sample with $N_p = 15360$ particles that did not show any mismatch in the heat capacity, demonstrating that the observed mismatch is not a matter of the sample size.

\begin{table}[h!]
\centering
\caption{\label{tab:unit_cells} Cell parameters and symmetry group details of the structures observed in the simulations of the medium $N_p$ = 9900 at different pressures and temperatures. }
\begin{tabular}{cccccccccc}
phase & $T$ (K) & $P$ (GPa) & space group & $a$ & $b$ & $c$ & $\alpha$ & $\beta$ & $\gamma$ \\
\hline
IV & 1200 & 0 & R$\bar{3}$c & 4.93 & 4.93 & 17.62 & 90 & 90 & 120 \\
V & 1600 & 0 & R$\bar{3}$m & 4.92 & 4.92 & 9.17 & 90 & 90 & 120 \\
V & 1900 & 0 & R$\bar{3}$m & 4.86 & 4.86 & 9.65 & 90 & 90 & 120 \\
IV & 1600 & 3 & R$\bar{3}$c & 4.88 & 4.88 & 17.64 & 90 & 90 & 120 \\
Vb & 1600 & 6 & P$2_1$/m & 6.17 & 4.92 & 3.93 & 90 & 106.3 & 90  
\end{tabular}
\end{table}


In table \ref{tab:unit_cells} the unit cell parameters of our simulation sample are presented. The details in this table have been extracted through the X-ray analysis with \texttt{pymatgen} and the \textsc{gsas-ii} software, as explained above. 

As seen in the CSP analysis snapshots, all domains or defects propagate continuously along the three-fold $c$ direction but show a discontinuous shape in the $ab$ plane. The phase IV unit cell is rhombohedral, typical of a trigonal crystal system, characterized by non-orthogonal axes. In contrast, the phase Vb unit cell is monoclinic, with two orthogonal angles and a third angle slightly different from 90 degrees. To achieve a smooth transition from phase IV to Vb, the parent supercell's lattice parameters should be multiples of those in the child phase, allowing a smooth atomic mapping. Additionally, the crystal lattice periodicity in the child phase should match or be a multiple of the parent phase's periodicity. We aimed to let the samples evolve under pressure and temperature range to observe structural transitions without bias towards any particular structure. 

However, it is known that commensurate phase transitions, or transitions between phases that have commensurate structural relationships, proceed smoothly to the child phase that fits perfectly within the structure of the original, parent phase, without any mismatches. In contrast, incommensurate transitions have non-integer ratio periodicities, leading to complex structures with domain walls and/or defects. Our preliminary analyses, not fully presented here, indicate that the discrepancy observed in the heat capacity graph arises from the incommensurate relations between our fortuitously chosen starting simulation cell, that, in turn resulted in the formation of domain walls and defects. 

Additionally, depending on the aspect ratio between parent and child cells, the transformation can involve the entire sample, as with the small and medium samples in our case, or only part of it, as with the domain-forming large sample. For example, the transition of the whole sample to the new phase can be greatly impaired by the formation of domains with different orientations with respect to the threefold axes.

Future research will focus on a detailed investigation of the defects, including their nature and formation mechanisms. We will also work on developing supercells designed to avoid biases towards one structure over another due to commensurability issues. Another area of interest is isolating the effects of the model potentials from commensurability issues to better understand how to observe transitions into all possible structures that characterize the complex and rich phase diagram of calcium carbonate.
\newpage

\bibliographystyle{unsrt}
\bibliography{articles.bib}